\documentclass[pra,aps,superscriptaddress,notitlepage,longbibliography,nofootinbib]{revtex4-2}
\usepackage[margin=1.54cm]{geometry}
\usepackage{wrapfig}
\usepackage{graphicx}
\usepackage{mathtools}
\usepackage{bm}
\usepackage{amssymb}
\usepackage{epstopdf}
\usepackage{siunitx}
\usepackage{xcolor}
\usepackage[ngerman,english]{babel}
\usepackage[T1]{fontenc}
\usepackage{amsmath}
\usepackage{bbold}
\usepackage{hyperref}
\usepackage{comment}
\usepackage[normalem]{ulem}
\usepackage{CJKutf8}
\usepackage{tikz}
\usetikzlibrary{positioning}

\DeclareUnicodeCharacter{2009}{ }

\DeclareMathOperator{\sinc}{sinc}

\usepackage{hyperref}
 \hypersetup{
     colorlinks = true,
     linkcolor  = black,
     filecolor  = blue,
     citecolor  = magenta,      
     urlcolor   = black,
     }

\usepackage{braket}

\newcommand{\sigp}{\sigma^+}
\newcommand{\sigm}{\sigma^-}

\newcommand{\sigx}{\sigma_x}

\newcommand{\Tr}[1]{\text{Tr} \left\{#1\right\}}

\definecolor{orange}{RGB}{200,100,0}

\usepackage[markup=underlined]{changes}

\newcommand{\Romannum}[1]{\MakeUppercase{\romannumeral #1}}

\usepackage{letltxmacro}
\LetLtxMacro{\ORIGselectlanguage}{\selectlanguage}
\makeatletter
\DeclareRobustCommand{\selectlanguage}[1]{%
  \@ifundefined{alias@\string#1}
    {\ORIGselectlanguage{#1}}
    {\begingroup\edef\x{\endgroup
       \noexpand\ORIGselectlanguage{\@nameuse{alias@#1}}}\x}%
}
\newcommand{\definelanguagealias}[2]{%
  \@namedef{alias@#1}{#2}%
}

\makeatother

\definelanguagealias{en}{english}
\definelanguagealias{EN}{english}
\definelanguagealias{eng}{english}
\definelanguagealias{de}{ngerman}

\begin{document}

\title{Gauge-Independent Emission Spectra and Quantum Correlations in the Ultrastrong Coupling Regime of {Open System} Cavity-QED}

\author{Will Salmon}
\thanks{These two authors contributed equally to this work: will.salmon@queensu.ca, cgustin@stanford.edu}
\affiliation{Department of Physics, Engineering Physics and Astronomy, Queen's University, Kingston, ON K7L 3N6, Canada}
\author{Chris Gustin}
\thanks{These two authors contributed equally to this work: will.salmon@queensu.ca, cgustin@stanford.edu}
\affiliation{Department of Physics, Engineering Physics and Astronomy, Queen's University, Kingston, ON K7L 3N6, Canada}
\affiliation{Department of Applied Physics, Stanford University, Stanford, California 94305, USA}
\author{Alessio Settineri}
\affiliation{Dipartimento di Scienze Matematiche e Informatiche, Scienze Fisiche e  Scienze della Terra, Universit\`{a} di Messina, I-98166 Messina, Italy}

\author{Omar Di Stefano}
\affiliation{Dipartimento di Scienze Matematiche e Informatiche, Scienze Fisiche e  Scienze della Terra, Universit\`{a} di Messina, I-98166 Messina, Italy}

\author{David Zueco}
\affiliation{Instituto de Ciencia de Materiales de	Arag\'{o}n and Departamento de F\'{i}sica de la Materia Condensada, CSIC-Universidad de Zaragoza, Pedro Cerbuna 12, 50009 Zaragoza, Spain}
\affiliation{Fundaci\'{o}n ARAID, Campus R\'{i}o Ebro, 50018 Zaragoza, Spain}

\author{Salvatore Savasta}
\affiliation{Dipartimento di Scienze Matematiche e Informatiche, Scienze Fisiche e  Scienze della Terra, Universit\`{a} di Messina, I-98166 Messina, Italy}
\affiliation{Theoretical Quantum Physics Laboratory, RIKEN Cluster for Pioneering Research, Wako-shi, Saitama 351-0198, Japan}



\author{Franco Nori}
\affiliation{Theoretical Quantum Physics Laboratory, RIKEN Cluster for Pioneering Research, Wako-shi, Saitama 351-0198, Japan}
\affiliation{Physics Department, The University of Michigan, Ann Arbor, Michigan 48109-1040, USA}

\author{Stephen Hughes}  
\affiliation{Department of Physics, Engineering Physics and Astronomy, Queen's University, Kingston, ON K7L 3N6, Canada}

\date{\today}

\begin{abstract}
   A quantum dipole interacting with an optical cavity is one of the key models in cavity quantum electrodynamics (cavity-QED). To treat this system theoretically, the typical approach is to truncate the dipole to two levels. However, it has been shown that in the ultrastrong-coupling regime, this truncation naively destroys gauge invariance. By truncating in a manner consistent with the gauge principle, we introduce master equations {for open systems} to compute gauge-invariant emission spectra, photon flux rates, and quantum correlation functions which show significant disagreement with previous results obtained using the standard quantum Rabi model. Explicit examples are shown using both the dipole gauge and the Coulomb gauge.
\end{abstract}

\maketitle


\section{Introduction}

\begin{figure}[hb]
    \centering
    \includegraphics[width=0.7\linewidth]{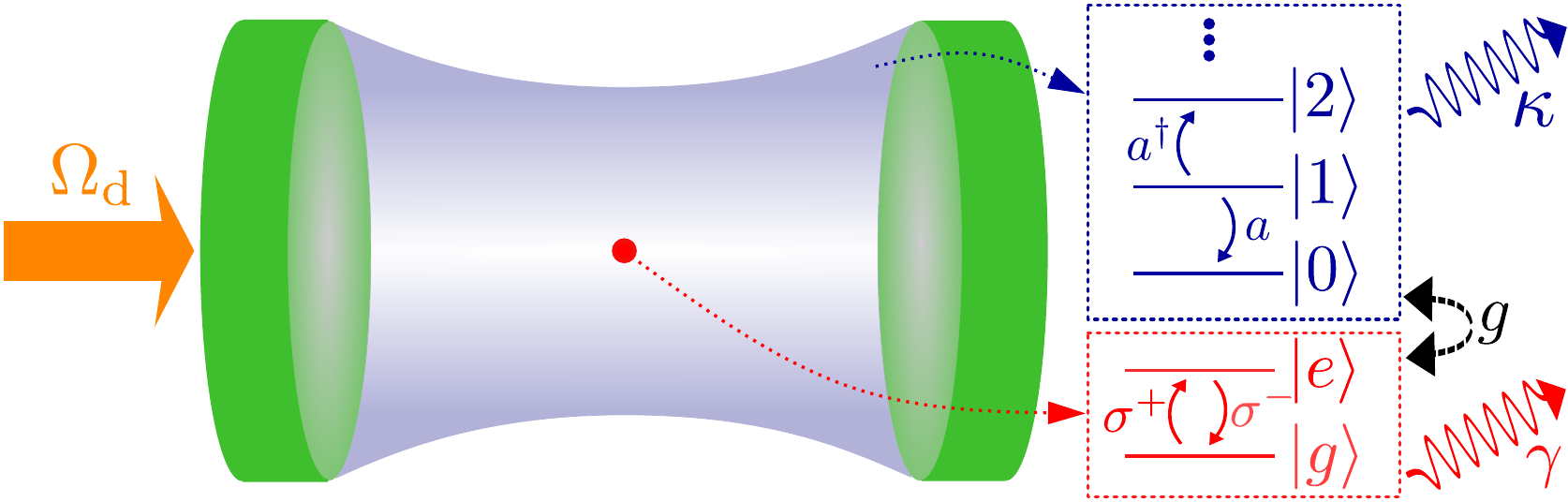}
    \caption{{Schematic of a generic cavity-QED system. The optical cavity mode has quantized energy levels (in blue), with a decay rate $\kappa$. The matter system is a truncated TLS (in red), with a possible spontaneous emission decay rate $\gamma$. The two systems have a coherent coupling strength $g$. A coherent laser (in orange) drives the system with Rabi frequency $\Omega_{\rm d}$.}}
    \label{fig:system}
\end{figure}

The intricate interactions between light and matter allow one to observe drastically different behavior depending on the relative strength of the light-matter coupling. In the weak-coupling regime, the losses in the system exceed the light-matter coupling strength, and energy in the system is primarily lost before it has the chance to coherently transfer between the matter and the light. Accessing this regime experimentally has allowed for breakthroughs in quantum technologies such as single-photon emitters~\cite{salter_entangled-light-emitting_2010,somaschi_near-optimal_2016,senellart_2017, tomm_2021}. Beyond weak-coupling, in the strong-coupling regime the rate of decoherence is smaller than the rate of excitation exchange, allowing for the observation of vacuum Rabi oscillations: the coherent oscillatory exchange of energy between light and matter. The strong-coupling regime has helped initiate a second generation of quantum technologies~\cite{buluta2011natural,georgescu2012quantum}.
{See Fig.~\ref{fig:system} for a simple schematic of a typical cavity-QED system with system-bath leakage.}

Around 2005, the ``ultrastrong-coupling'' (USC) regime was predicted for intersubband polaritons~\cite{ciuti_quantum_2005}. This regime is characterized not by still lower rates of decoherence, but by a coupling strength that is a comparable fraction of the bare energies of the system. The dimensionless parameter $\eta=g/\omega_0$ (i.e., the cavity-emitter coupling rate divided by the transition frequency) is used to quantify this coupling regime for cavity-QED. 
Typically, USC effects are expected when $\eta \gtrsim 0.1$, at which point the rotating wave approximation (RWA) used in the weak and strong regimes becomes invalid. Reported signs of USC emerged in 2009 with experiments involving quantum-well intersubband microcavities~\cite{anappara_signatures_2009}, achieving $\eta \approx 0.11$. Terahertz-driven quantum wells have also demonstrated USC effects~\cite{zaks_thz-driven_2011}, and similar effects have been exploited to achieve carrier-wave Rabi flopping with strong optical pulses~\cite{hughes_breakdown_1998,mucke_signatures_2001,ciappina_carrier-wave_2015}. To date, many different systems have exhibited USC~\cite{frisk_kockum_ultrastrong_2019,forn-diaz_ultrastrong_2019}. Recently, using plasmonic nanoparticle crystals, $\eta = 1.83$ has been achieved, with potential to lead to $\eta = 2.2$~\cite{mueller_deep_2020}.

With experiments pushing the normalized coupling strength continuously higher, the interest in USC effects also continues to grow, helping to improve the underlying theories of light-matter interactions~\cite{PhysRevA.81.042311}, even at arbitrarily high coupling strengths~\cite{PhysRevLett.126.153603}. There have also been various predictions made about what novel technologies USC will bring about, including modifications to chemical or physical properties of various systems caused by their USC to light~\cite{ciuti_quantum_2005,herrera_cavity-controlled_2016}, and the potential to create faster quantum gates and gain a high level of control over chemical reactions~\cite{frisk_kockum_ultrastrong_2019}. To push these advancements forward, it is essential to have a fundamental understanding of the physics involved with these systems and to 
accurately connect to experimental observables.

The cornerstone model in cavity-QED is a two-level system (TLS) interacting with a quantized cavity mode~\cite{scully1999quantum}. This model has been applied to atoms~\cite{miller_trapped_2005,schuster_nonlinear_2008,flick_atoms_2017,hamsen_two-photon_2017}, quantum dots~\cite{yoshie_vacuum_2004,reithmaier_strong_2004,hennessy_quantum_2007,bose_all-optical_2014}, and circuit QED~\cite{you2011atomic,beaudoin_dissipation_2011,gu_microwave_2017,mirhosseini_cavity_2019}. Outside the USC regime, this model is typically represented by the canonical Jaynes-Cummings (JC) Hamiltonian~\mbox{\cite{jaynes_comparison_1963}}, which makes a RWA and can be easily diagonalized. In the USC regime, however, it is necessary to retain counter-rotating terms, giving rise to the quantum Rabi model (QRM)~\mbox{\cite{frisk_kockum_ultrastrong_2019,forn-diaz_ultrastrong_2019,niemczyk_circuit_2010}}. By detecting resonance fluorescence of light emitted from the cavity as quantified by the first-order degree of coherence correlation function (CF), the spectral content of these cavity-QED models can be explored, while the second-order intensity CF is fundamental to understanding the photon statistics as probed by intensity interferometry.

{
The main contribution 
of 
this work
is to present a   self-consistent and unambiguous way to model
observables in the USC regime of open system 
cavity-QED. 
Apart from addressing the subtle (and unknown) effects of dissipation, and excitation, and input-output, 
we show the striking influence of
modelling experimentally relevant 
observables such as the emission spectra and quantum correlation functions. We  also show
how and why  the 
form of the system-bath interactions matters, yet the form {\it is} gauge invariant, if---and only if---treated properly (in contrast to the standard master equation approaches) using gauge invariant master equations. We show equivalence between dipole gauge and Coulomb gauge master equations, if one applies {\it gauge corrections} in a consistent way, and we also demonstrate the drastic failure of currently adopted master equations in the USC regime.
Our framework and formalism, to  the best of our knowledge, constitutes a first way to do this, and can thus be applied to a wide range of measurements in the USC regime for open systems.}

\section{Gauge invariance and system-reservoir interactions}

It  was recently  shown that extra care is needed when constructing gauge-independent theories~\cite{Settineri2021Apr}, for computing experimental observables for suitably strong light-matter interactions. This development started with a series of papers dealing with so-called gauge ambiguities in the USC regime~\cite{de_bernardis_breakdown_2018,adam_stokes_gauge_2019,di_stefano_resolution_2019}. As a $U(1)$ gauge theory, different gauges in QED manifest in different representations of the Hamiltonian of a given system, but these should be unitarily equivalent and give rise to equivalent physical observables.
Without proper care, gauge invariance of cavity-QED theories can break down when considering USC~\cite{stokes_gauge_2020}. This is due to the truncation of the matter system's formally infinite Hilbert space to the two lowest eigenstates in forming the TLS---only keeping an infinite number of energy levels formally preserves gauge invariance~\cite{Rouse2021Feb}. Consequently, previous model predictions in the USC regime can be ambiguous since the predictions are impacted by the choice of gauge. While this issue has been known in general for several decades~\cite{PhysRevA.36.2763}, only recently was this specific problem presented as rather insurmountable~\cite{stokes_gauge_2020}. However, the issue has been resolved by using a self-consistent theory at the system Hamiltonian level~\cite{di_stefano_resolution_2019,2002.02139}, restoring gauge invariance to the theory for systems with a finite Hilbert space.

Despite this, 
additional subtleties occur in the USC regime regarding the interaction of the cavity-QED system with its environment. To connect to experiments, one also requires an input-output model of dissipation from the cavity to external modes, requiring an \textit{open-system} model of cavity-QED. In the USC regime, complications arise with this input-output formalism associated with approximations typically made outside of the USC regime. These complications originate from the hybridization of light and matter that occurs in USC, and as such the quanta of excitations inside the cavity-QED system have different quasiparticle representations than the photons actually emitted from the system. Moreover, the separation of operators into light and matter components becomes highly gauge-specific in the USC regime, and proper care must be taken to ensure self-consistency. 

To fully synthesize these considerations with the restoration of gauge invariance,  we present a dissipative and gauge-invariant master equation model, which is {\it required} to properly describe experimentally-observable quantities arising from output channels of the cavity. Key experiments to probe such observables include resonance fluorescence and two-photon detection schemes, and we make a direct connection to both of these. We also show how previous
QRM master equations in the USC regime are ambiguous in
general as they produce gauge dependent results for observables, and we show how to fix such problems. Moreover, our theories can be used to explore the precise form of the system-bath interactions, which in fact yield different experimental signatures
in the USC regime. 

\section{Model}

In the dipole gauge {(namely, the multipolar gauge
in the dipole approximation)}, we can write the system Hamiltonian, using the QRM, as ($\hbar=1$)
\begin{equation}\label{eqn:QR}
    H_{\rm QR} = \omega_c a^\dagger a + \omega_0 \sigp \sigm + ig(a^{\dagger}-a)(\sigp+\sigm),
\end{equation}
where $\omega_0$ ($\omega_c$) is the TLS (cavity) transition frequency, $\sigp$ ($\sigm$) is the raising (lowering) operator for the TLS, and $a^{\dagger}$ ($a$) is the cavity mode creation (annihilation) operator; $g$ is the TLS-cavity coupling strength. We take $\omega_c=\omega_0$ throughout. In contrast to the Coulomb gauge, straightforwardly truncating the dipole in the light-matter interaction to a TLS subspace does not break gauge invariance in the dipole gauge~\cite{di_stefano_resolution_2019}. Making a RWA on Eq.~\eqref{eqn:QR} (i.e., neglecting counter-rotating terms $a^\dagger\sigp$ and $a\sigm$, which do not conserve excitation number), yields the simpler JC Hamiltonian.

Outside of the USC regime, the usual approach to include dissipation is with a Lindblad master equation~\cite{carmichael_statistical_2013},
\begin{equation}\label{eqn:ME-CQED}
    \Dot{\rho} = -\frac{i}{\hbar}[H_{\rm QR},\rho] + \mathcal{L}_{\rm bare}\rho,
\end{equation}
where $\rho$ is the reduced density matrix. The dissipation term, $\mathcal{L}_{\rm bare}\rho = \frac{\kappa}{2}\mathcal{D}[a]\rho$, is the Lindbladian superoperator where $\mathcal{D}[O]\rho = \left(2O\rho O^\dagger - \rho O^\dagger O - O^\dagger O\rho \right)$ and $\kappa$ is the cavity photon decay rate. Since dissipation is usually dominated by cavity decay, we neglect direct TLS relaxation and pure dephasing~\cite{settineri_dissipation_2018,zueco2019ultrastrongly}. {However,
the theory of how to include TLS dissipation is discussed in Appendix~\ref{sec:AppA}.4}

The Lindbladian can be derived by following the typical approach in which one neglects the TLS-cavity interaction when considering the coupling of these systems to the environment~\cite{beaudoin_dissipation_2011}. However, when moving into the USC regime, this approach fails, and the Lindbladian must be derived while self-consistently including the coupling between the subsystems. For sufficiently strong subsystem coupling, transitions occur between {\it dressed} eigenstates of the full Hamiltonian rather than between eigenstates of the individual free Hamiltonians~\cite{settineri_dissipation_2018}.

In the USC regime, the system has transition operators $\ket{j}\bra{k}$ which cause transitions between the dressed eigenstates of the system $\{\ket{j},\ket{k}\}$. To obtain these transitions for the cavity mode operator, we use dressed operators~\cite{settineri_dissipation_2018},
\begin{equation}\label{eqn:dressed_ops}
    x^+ = \sum_{j,k>j} C_{jk}\ket{j} \bra{k},
\end{equation}
and $x^- = (x^+)^{\dagger}$, where the sum is over states $\ket{j}$ and $\ket{k}$, with $\omega_k>\omega_j$, $C_{jk} = \bra{j}\Pi_{\rm C}\ket{k}$, and we neglect thermal excitation effects; $\Pi_{\rm C}$ is an operator which couples linearly to dissipation channel modes which we assume proportional to the cavity electric field operator such that $\Pi_{\rm C} = i(a^{\dagger}-a)$. We then replace $\mathcal{L}_{\rm bare}$ in Eq.~\eqref{eqn:ME-CQED} with $\mathcal{L}_{\rm dressed}\rho = \frac{\kappa}{2}\mathcal{D}[x^+]\rho$, to arrive at the dressed state (DS) master equation. One can also use a generalized master equation to capture coupling to frequency-dependent reservoirs~\cite{settineri_dissipation_2018,cao2010dynamics} 
{See Appendix~\ref{sec:AppA} for a derivation of the generalized master equation, and 
Sec.~\ref{sec:G} for an example application using an Ohmic bath.}

Beyond this dressing transformation, it has been shown that there exists a potential gauge ambiguity in the electric field operator which causes further problems when computing observables in the USC regime~\cite{di_stefano_resolution_2019}; namely, $\Pi_{\rm C}$  corresponds to the Coulomb gauge electric field, but the QRM Hamiltonian is derived in the dipole gauge. The gauge transformation from the Coulomb gauge to the dipole gauge is generated by a unitary transformation, which for the restricted TLS subspace is given by the projected unitary operator~\cite{di_stefano_resolution_2019} ${\cal U}=\exp(-i\eta(a+a^\dagger)\sigma_x)$. The photon destruction operator transforms as $a\rightarrow {\cal U} a {\cal U}^\dagger = a+i\eta\sigma_x$~\cite{Settineri2021Apr}. Thus, to ``gauge-correct'' the master equation in the dipole gauge, we conduct the dressing operation as above, but with 
\begin{equation}
x^\pm \rightarrow x^\pm_{\rm GC}= \sum_{j,k>j} C_{jk}'\ket{j}\bra{k},
\end{equation}
where we take $C_{jk}' = \bra{j} \mathcal{U}\Pi_{\rm C}\mathcal{U}^{\dagger}\ket{k} = \bra{j}\Pi_{\rm D}\ket{k}= \bra{j}i(a^\dagger-a)+2\eta\sigx\ket{k} $; see Appendix~\ref{sec:AppA}  for a derivation of the master equation in the dipole and Coulomb gauges and their equivalence.

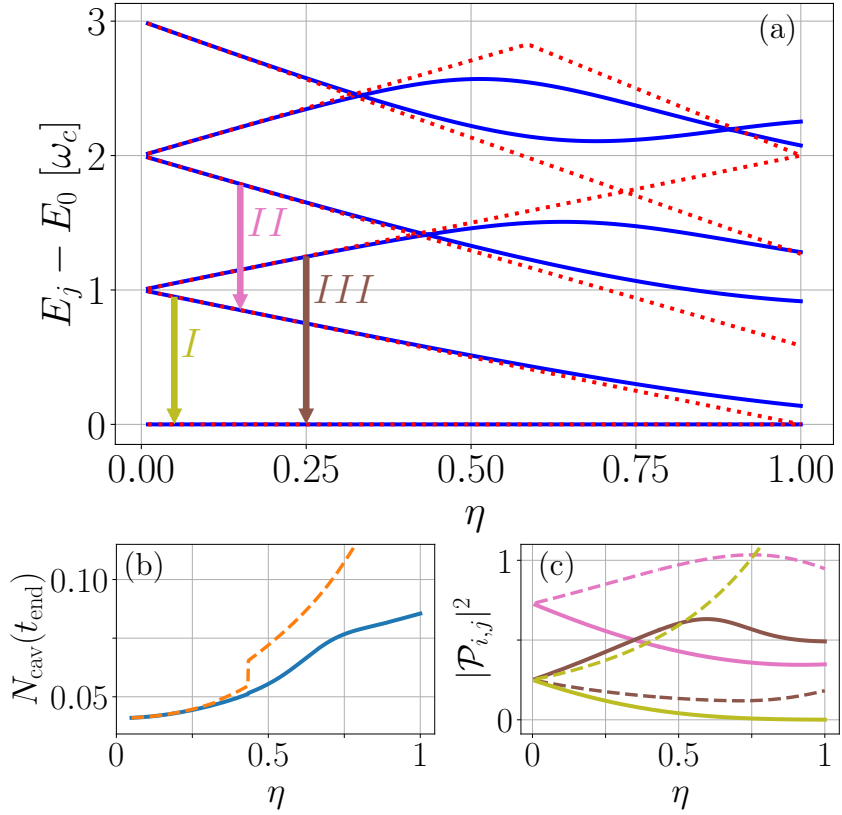
\begin{figure}[t]
    \begin{tikzpicture}
    \centering
        [inner sep=0mm]
        \node [xshift=0mm] (figa) {\includegraphics[scale=0.50,trim=-6mm 0 0 0cm]{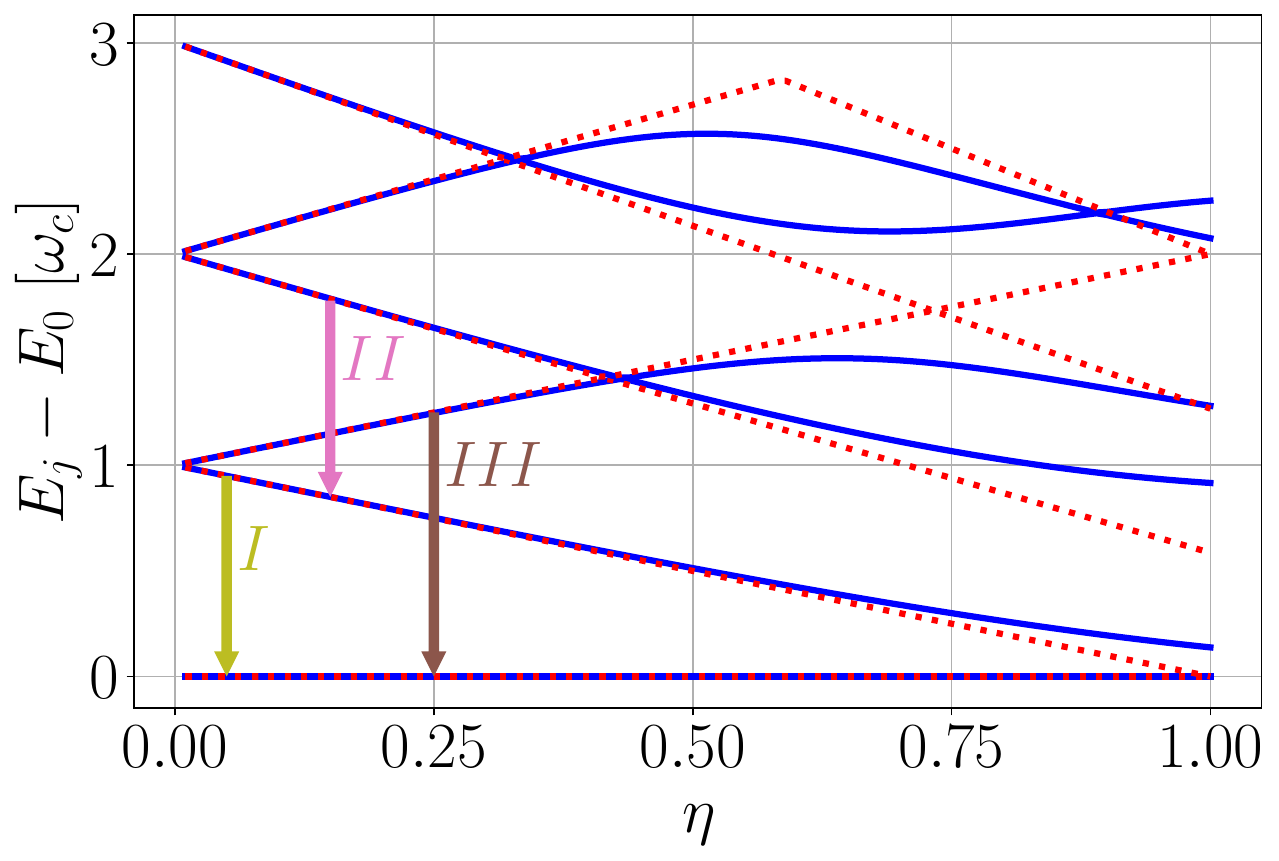}};
        \node[below=of figa,xshift=-27mm,yshift=13mm] (figb) {\includegraphics[scale=0.43]{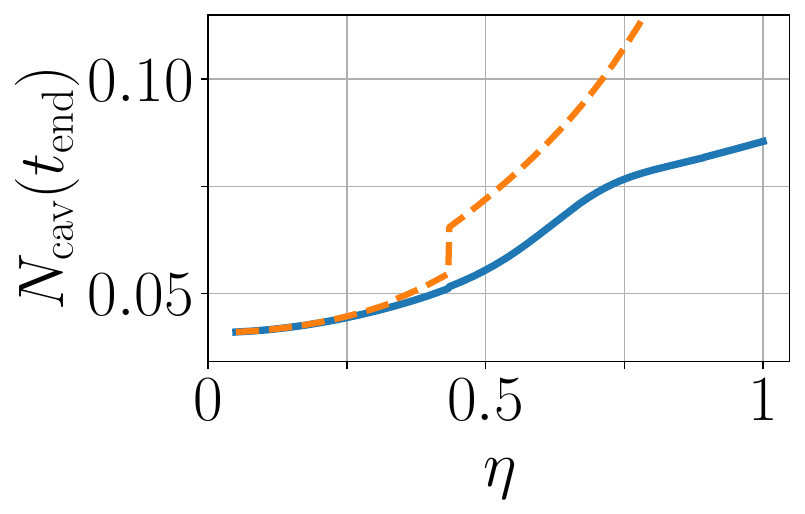}};
        \node[right=of figb, xshift=-12mm] (figc) {\includegraphics[scale=0.43]{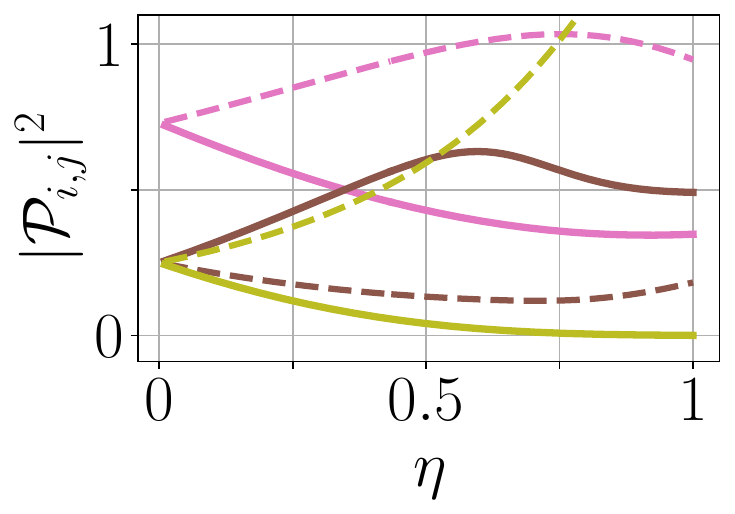}};
        \node[xshift=-10mm, yshift=-6mm] at (figa.north east) {\large(a)};
        \node[xshift=20mm, yshift=-5.3mm] at (figb.north west) {\large(b)};
        \node[xshift=16mm, yshift=-5.3mm] at (figc.north west) {\large(c)};
    \end{tikzpicture}
    \vspace{-0.2cm}
    \caption{(a) The energy eigenvalues of the six lowest states of the QRM (blue, solid) and the JCM (red, dotted). Arrows mark transitions of interest, placed at arbitrary locations on the $\eta$-axis, (b) steady state excitation number for incoherent driving (cf.~Fig.~\ref{fig:spectra}), and (c) select transition rates, with colors matching the arrows in (a). On the bottom two panels, solid (dashed) lines are with (without) the gauge correction in the dipole gauge. Note a sudden increase of $N_{\rm cav}$ near $\eta\approx0.4$  when states 2 and 3 cross.}
    \label{fig:evals}
    \vspace{-0.0cm}
\end{figure}

To study the quantum dynamics and spectral resonances, we excite the system with an incoherent pump term, $P_{\rm inc} {\cal D}[x^-_{\rm GC}]/2$, or with a coherent laser drive, $H_{\rm drive}(t) = (\Omega_{\rm d}/2)(x^-_{\rm GC} e^{-i\omega_{\rm L}t} + x^+_{\rm GC}e^{-i\omega_{\rm L}t})$, added to $H_{\rm QR}$, where $\Omega_{\rm d}$ is the Rabi frequency and $\omega_{\rm L}=\omega_{ c}$ is the laser frequency; thus, $H_{\rm S} = H_{\rm QR} + H_{\rm drive}$. Note that the QRM with a coherent drive is time-dependent and oscillates around a pseudo-steady-state. In addition, because of the driving laser, the periodic nature of the system Hamiltonian means that in principle the QRM spectra, already quite rich, are modified further; however, we use $\Omega_{\rm d} \ll g$, and neglect the influence of the coherent drive on the system eigenstates. The first few (lowest) energy eigenvalues are plotted for the QRM (dipole gauge) and JCM in Fig.~\ref{fig:evals}(a) for a range of normalized coupling strengths. 
Three transitions are shown, which we
will refer to below.

\section{Gauge Invariant Observables}

\begin{figure}[thp]
    \centering
    \includegraphics[width=0.7\linewidth]{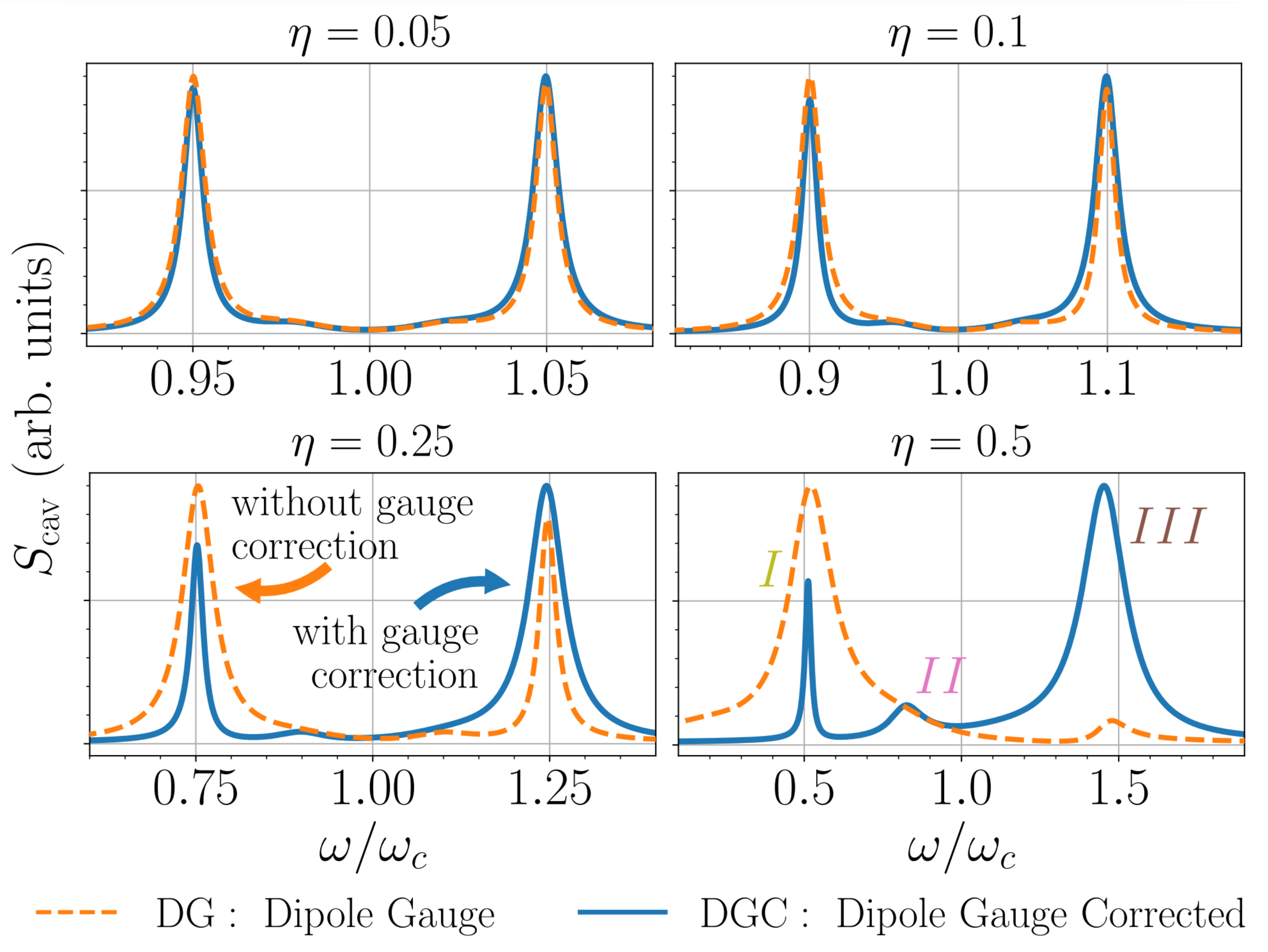}
    \caption{
    Cavity spectra outside the RWA (QRM) with DG model (orange dashed line), and DGC model (with gauge correction, blue line) for varying $\eta$ and weak incoherent driving: $P_{\rm inc} = 0.01 g$. Spectra are normalized to have the same maxima. Other system parameters are $\kappa = 0.25g$, and $\omega_{\rm L}=\omega_c=\omega_0$. Note a small change with the DG corrected model even  below the USC regime ($\eta=0.05$).}
    \label{fig:spectra}
\end{figure}

We first define the system excitation number, 
\begin{equation}
N_{\rm cav}(t){=}\left\langle x_{\rm GC}^-(t)x_{\rm GC}^+(t) \right\rangle,
\end{equation}
and a quadrature operator matrix element squared,
\begin{equation}
\lvert{\cal P}'_{j,k}\rvert^2 {=} \lvert C'_{jk}/\sqrt{2}\rvert^2,
\end{equation}
which is proportional to the photodetection rate of cavity-emitted photons from the $\ket{j}\rightarrow\ket{k}$ transition~\cite{2002.02139}. In  Fig.~\ref{fig:evals}(b), we show $N_{\rm cav}$ versus $\eta$, using incoherent driving (cf.~Fig.~\ref{fig:spectra}), where the solid curves show the effect of gauge corrections. Equivalent gauge-corrected results are obtained in the  Coulomb gauge.
With the correction, the population saturates
, while the uncorrected population continues to increase superlinearly, and jumps when states 2 and 3 cross in energy, potentially related to the photon blockade~\cite{PhysRevA.94.033827}. {With gauge corrections, we  see a strong influence from the TLS operator physics}. In Fig.~\ref{fig:evals}(c), we show $|\mathcal{P}_{jk}'|^2$ for the relevant transitions which are{, for weak excitation,} proportional to the transition linewidths; again, the  solid lines show the gauge corrected results.
{Note that
the corrected dipole gauge quadrature operator 
($\Pi_D' = i(a^{\dagger}-a) + 2\eta \sigma_x$)
causes a major modification of the 
transitions, significantly impacting their behavior in the nonperturbative regime.
}

In Appendix~\ref{sec:AppE}, we give analytical insight into these quadrature matrix elements using a Bloch-Siegert (BS) transformation, which analytically (to lowest order in $\eta$) predicts the following changes with gauge correction:
$|{\cal P}_{\Romannum{1}}|^2{=} {1}/{4}(1{+}3\eta/2)
\rightarrow
{1}/{4}(1{-}5\eta/2)$,
and
$|{\cal P}_{\Romannum{3}}|^2{=} {1}/{4}(1{-}3\eta/2)
\rightarrow
{1}/{4}(1{+}5\eta/2)$,
causing a {\it reversed asymmetry} with gauge corrections. Physically, this asymmetry arises from the BS shift of cavity and TLS resonances giving rise to photon-like and atom-like polariton branches; the  composition of the $\Pi$ operator (which is affected by the gauge correction) ultimately determines which state is more cavity-like, and thus has a greater decay rate (see Appendix~\ref{sec:AppE} for details).

Next, we define the cavity-emitted spectrum,
\begin{equation}\label{eqn:spec}
    S_{\rm{cav}}\propto \text{Re} \left[\int_{0}^{\infty} d\tau e^{i\Omega\tau} \int_{0}^{\infty} \left\langle x_{\rm GC,\Delta}^-(t)x_{\rm GC,\Delta}^+(t+\tau) \right\rangle dt\right],
\end{equation}
where $x_{\rm GC,\Delta}^\pm {=} x_{\rm GC}^\pm {-} \left\langle x_{\rm GC}^\pm\right\rangle$ and $\Omega{=}\omega-\omega_{\rm L}$. Beyond the spectrum, which uses a first-order quantum CF, we also compute the normalized second-order quantum CF,
\begin{equation}\label{eqn:g2(t,tau)}
    g^{(2)}(t,\tau) = 
    \frac{\left\langle x_{\rm GC}^-(t) x_{\rm GC}^-(t+\tau) x_{\rm GC}^+(t+\tau) x_{\rm GC}^+(t)\right\rangle}
    {\left\langle x_{\rm GC}^-(t)x_{\rm GC}^+(t) \right\rangle \left\langle x_{\rm GC}^-(t+\tau)x_{\rm GC}^+(t+\tau) \right\rangle},
\end{equation}
which quantifies the likelihood of a photon being detected at ($t+\tau$) if one was detected at $t$. We also introduce the time-averaged $g^{(2)}(\tau) = {\int_{t_1}^{t_1+T} g^{(2)}(t,\tau) dt}/{T}$, where $t_{\rm end}$ is an arbitrary time point at which the system has reached the pseudo-steady-state and $T$ is the period of oscillation (see Appendix~\ref{sec:AppD}). Note that without the gauge-correction, we use the uncorrected (corresponding to a Coulomb gauge representation) $x^{\pm},x^\pm_\Delta$ for computing the observables, and $x^{\pm}$ for incoherent or coherent driving (see Appendix~\ref{sec:AppA}). All calculations use Python with the QuTiP package~\cite{johansson2012qutip,johansson_qutip_2013}.

\begin{figure}[thp]
    \centering
    \begin{tabular}{c c}
        \includegraphics[width=0.48\linewidth]{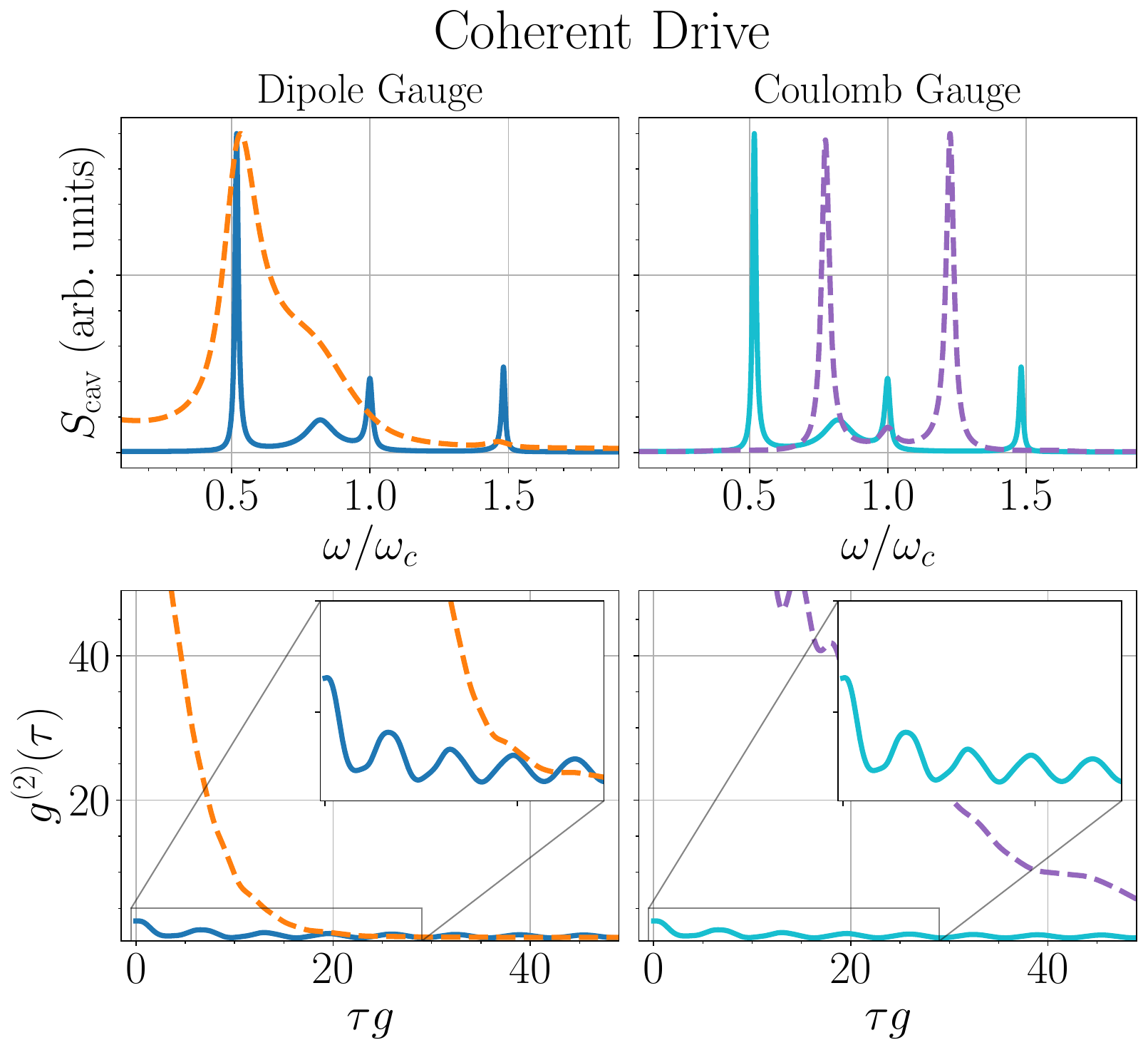}
        &
        \includegraphics[width=0.48\linewidth]{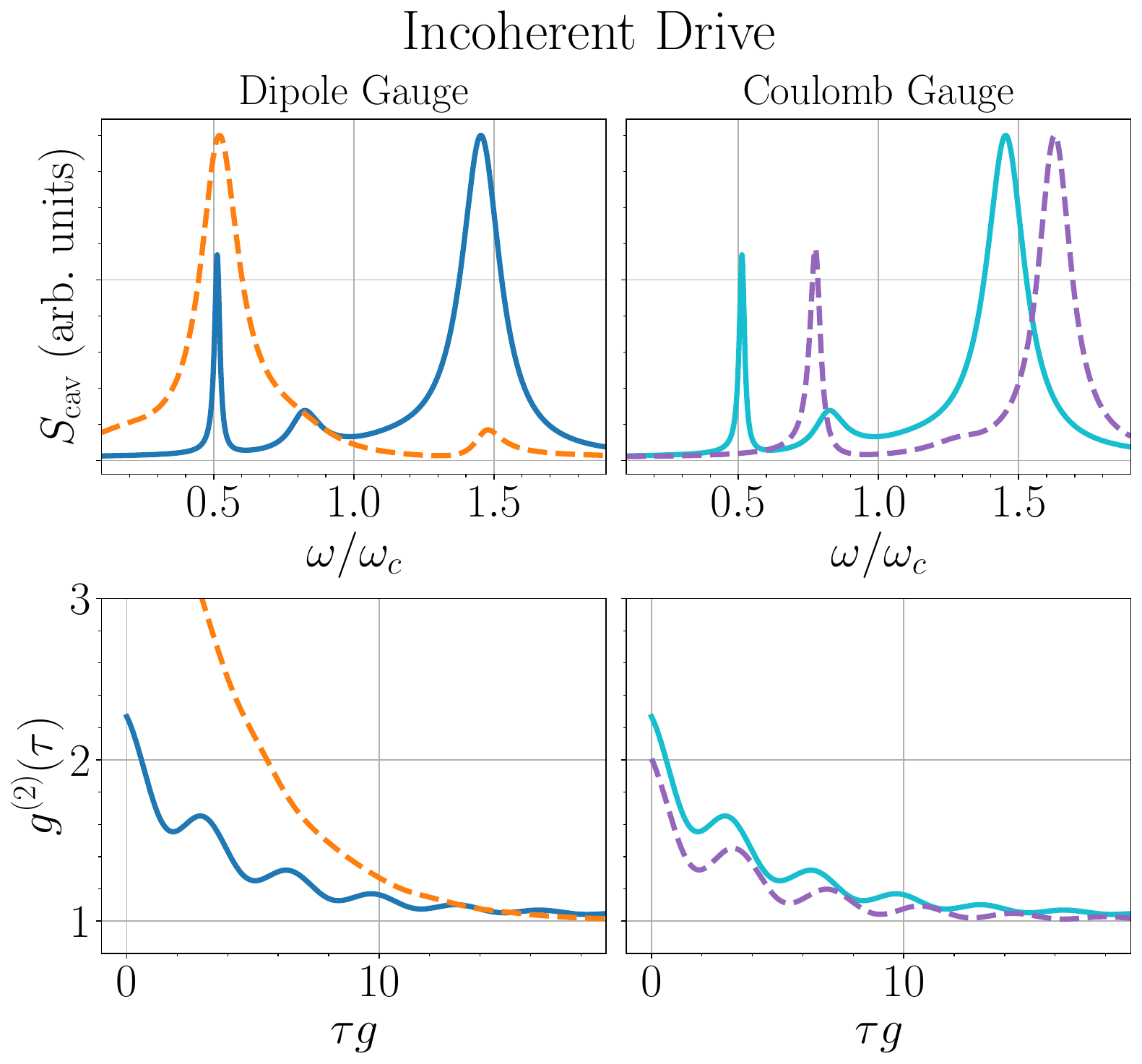}
    \end{tabular}
    \vspace{-0.2cm}
    \caption{Direct comparison between master equation results using the dipole and Coulomb gauges at $\eta=0.5$, for both coherent and incoherent excitation, showing the profound effect of the gauge-correction and how this manifests in identical spectra (top) and $g^{(2)}(\tau)$ correlation functions (bottom). Solid and dashed curves are with and without the gauge correction, respectively. For the coherent drive (left), we use $\Omega_{\rm d} = 0.1g$, and the incoherent pumping (right) is the same as in Fig.~\ref{fig:spectra} ($P_{\rm inc} = 0.01 g$).}
    \label{fig:coh_vs_inc}
\vspace{-0.0cm}
\end{figure}

For weak incoherent pumping, Fig.~\ref{fig:spectra} compares the computed spectra with and without the gauge correction (DGC: dipole-gauge-corrected and DG: dipole-gauge, respectively), for $\eta$ ranging from 0.05 (strong coupling)
to 0.5 (USC). For relatively small $\eta=0.05$, the DGC (with gauge correction) spectra already begin to deviate from the DG spectra (usual QRM master equation solution). 

With increasing  $\eta$,  notably, the DGC and DG spectra are substantially different above $\eta=0.1$: the DGC spectra still show a reversed asymmetry, with a significant narrowing of the lower polariton resonance ($\Romannum{1}$) and a broadening of the upper polariton resonance ($\Romannum{3}$); the ratio of higher-lower polariton peak areas under weak excitation changes from $1 - 3 \eta + \mathcal{O}(\eta^2)$ to $1 + 5 \eta + \mathcal{O}(\eta^2)$ with gauge correction --- a dramatic change even for $\eta < 0.1$ (see Appendix~\ref{sec:AppE}). These peaks can be identified as resulting from the $\ket{1}\rightarrow\ket{0}$ (olive arrow on Fig.~\ref{fig:evals}(a)) and $\ket{2}\rightarrow\ket{0}$ (brown arrow) transitions, respectively. Since $\lvert{\cal P}_{j,k}\rvert^2$ contributes to photon emission directly through the $\kappa$ decay channel~\cite{2002.02139}, the narrowing (broadening) of peak $\Romannum{1}$ ($\Romannum{3}$) with increasing $\eta$ can be explained with Fig.~\ref{fig:evals}(c). Without the correction, the opposite trend is observed, which is again consistent with Fig.~\ref{fig:evals}(c) (dashed lines). At $\eta = 0.5$, there is also a noticeable resonance ($\Romannum{2}$) around $\omega = 0.8g$, showing a {\it deep mixing} of the TLS and cavity dynamics in the USC regime. We can identify this energy difference with the $\ket{3}\rightarrow\ket{1}$ transition, pink arrow on Fig.~\ref{fig:evals}(a), which also has reduced broadening with $\eta$, cf.~Fig.~\ref{fig:evals}(c).

We have shown how the gauge correction manifests in modified linewidths and drastically different spectral weights in comparison to the usual QRM---even so far as to result in a complete reversal of the asymmetry predicted from a non-gauge-corrected model~\cite{cao2011qubit} (Fig.~\ref{fig:spectra2}, $\eta=0.5$). 
We now demonstrate how this gauge correction manifests in the Coulomb gauge. To do this, we display results for the cavity-emitted spectrum and CFs with coherent and incoherent pumping, using the discussed dipole gauge and the Coulomb gauge master equation.

{In the Coulomb gauge, 
the standard system Hamiltonian for the QRM is~\cite{di_stefano_resolution_2019}
\begin{equation}\label{eq:Hi_cg0_2}
    { H}_{\rm QR}^{\rm C} =  \omega_{c} a^\dagger a + 
    \frac{\omega_0}{2}\sigma_z  + 
   g_{\rm C} (a+a^\dagger) \sigma_y + D(a+a^\dagger)^2,
\end{equation}
where $g_{\rm C} = g_{\rm D}\omega_0/\omega_c$ and
$D$ is the strength of the diamagnetic term. Using
 the Thomas-Reiche-Kuhn sum rule~\cite{2002.02139}, then 
$D \geq g_{\rm C}^2/\omega_0$, and
  for our simulations we take $D =  g_{\rm C}^2/\omega_0$. Thus, with
$\omega_0=\omega_c$ and $g_{\rm D}\equiv g$, we have
$D=\eta^2\omega_0$. Unfortunately, this form does not satisfy the gauge principle, and produces the wrong eigenenergies and eigenstates in the USC regime~\cite{de_bernardis_breakdown_2018,di_stefano_resolution_2019,2002.02139}.
Instead, the {\em corrected} Coulomb gauge uses a different system Hamiltonian~\cite{di_stefano_resolution_2019},
\begin{equation}
\label{eq:H_cg}
  {H}_{\rm QR}^{\rm C'} =  \omega_c a^\dagger a
    + \frac{\omega_0}{2}
    \left\{ \sigma_z \cos\left(2 \eta (a+a^\dagger)\right)+ \sigma_y \sin\left(2\eta (a + a^\dagger)\right)\right\} ,
\end{equation}
which  contains field operators to all orders, and the ${\rm C'}$
superscript indicates we are using the corrected form for the system Hamiltonian.} In the Coulomb gauge, the gauge-invariant dissipator term is
(see Appendix~\ref{sec:AppA})
\begin{equation}
    {\cal L}_{\rm dressed}^{\rm C} \rho = \frac{\kappa}{2}{\cal D}[x^+_{\rm C}] \rho,
\end{equation}
where $x^+_{\rm C}= \sum_{j,k>j} C_{jk}^{\rm C} \ket{j} \bra{k}$ with $C_{jk}^{\rm C} = \braket{j|\Pi_{\rm C}|k}$, and we now compute the dressed states in the Coulomb gauge,  using both uncorrected and corrected forms of the system Hamiltonian.

Figure~\ref{fig:coh_vs_inc} (top) shows the coherent and incoherent spectra at $\eta=0.5$, showing that the gauge correction results in a profound effect in either case. For coherent driving, using $\Omega_{\rm d}=0.1g$, there is a significant sharpening of the resonances. The Coulomb gauge result without the gauge correction corresponds to a minimal coupling Hamiltonian naively truncated to a TLS, which results in incorrect energy levels for the dressed-state master equation~\cite{di_stefano_resolution_2019}.
{This effect of having the incorrect energy levels and eigenstates is clearly shown in the uncorrected Coulomb gauge results in
Fig.~\ref{fig:coh_vs_inc}, which is especially {\it wrong} with coherent pumping, since the system is effectively being pumped off resonance (because of the diamagnetic term).}
{
For coherent pumping, additional Rabi field strengths are shown in Appendix \ref{sec:AppC}, where we also show simulations with and without a RWA for the pump field.}

Next, in Fig.~\ref{fig:coh_vs_inc} (bottom), we examine the second-order coherence, which is important for characterising the generation of non-classical light. In all cases shown, we observe photon bunching at short time-delays. With the gauge correction, there is a significant reduction in the level of bunching, and the usual USC master equations significantly overestimate the bunching characteristics. Moreover, the dynamics are qualitatively different, and thus {the non-GC master equations results clearly fail in the USC regime}.
{
In all cases, we confirm full agreement
between the {\it corrected dipole gauge} and {\it corrected Coulomb gauge} results, since these are the correct gauge invariant solutions, and thus produce identical results. }

{
\section{Influence of the Spectral  Bath Function on the Gauge Correction and Gauge-Invariant Spectra with an Ohmic Bath}
\label{sec:G}

 \begin{figure}[h!]
    \centering
    \includegraphics[width=0.7\linewidth]{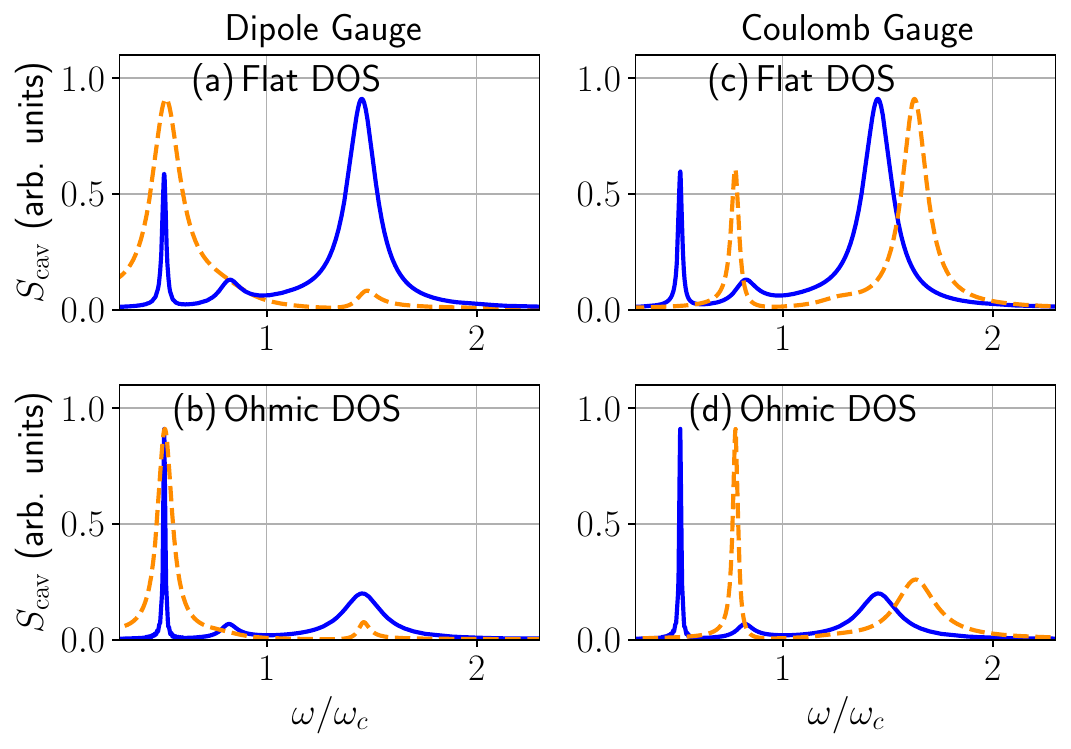}
    \vspace{-0.3cm}
    \caption{{The computed cavity spectra using the dipole gauge (left) and Coulomb gauge (right), with a flat DOS [$\kappa(\omega)=\kappa$, panels (a,c)] and an Ohmic DOS [$\kappa(\omega)= \kappa \omega/\omega_c$, panels (b,d)] using
    the generalized master equation
    [see Eqs.~\eqref{eq:GME1}-\eqref{eq:GME_diss} in Appendix \ref{sec:AppA}].
    In both cases, the effect of the gauge correction (solid lines versus dashed lines) is dramatic. We use the same parameters as in Fig.~2 of the main text, with incoherent driving, and  parameters $\eta=0.5$ and $\kappa=0.25g$. Notably, in all cases, regardless of the spectral function, the corrected dipole gauge and corrected Coulomb gauge results are identical.}}
    \label{fig:GME}
\end{figure}

In the simulations above, for simplicity, we used a flat density of states (DOS) for the spectral bath function; namely, the DOS was assumed to be constant relative to the energy scale of the resonances. This helps to better identify intrinsic spectral asymmetries related to gauge correcting.

For completeness, here we explicitly show an example numerical solution without invoking the approximation that $\kappa(\omega)$ is frequency independent. Specifically, we compute the emitted spectra when $\kappa(\omega)=\kappa$ as well as
$\kappa(\omega)=\kappa \omega/\omega_c$ (Ohmic bath).
We use the same example as in Fig.~\ref{fig:spectra} with incoherent driving at $\eta=0.5$.
These numerical solutions are obtained from the generalized master equation \eqref{eq:GME1}, described in Appendix
\ref{sec:AppA}.

As can be seen in Fig.~\ref{fig:GME}, { clearly the form of the spectral bath function does not affect any of our general conclusions}, as the gauge correction is, in both cases, dramatic, and of course produces exactly the same result for both the dipole gauge and the Coulomb gauge. 
To be clear, if we plot these together, then they are indistinguishable, which also confirms that our numerical results are well converged in terms of basis size and time steps.
}

\section{Conclusions}

We have presented a gauge-invariant master equation approach and calculations for the cavity emission spectra in the USC regime, and shown how the usual QRM in the dipole gauge {\it fails}, yielding effects that are just as pronounced (or even more pronounced) as counter-rotating wave effects in this regime. We have demonstrated how the gauge correction significantly affects the intensity CF and cavity excitation number. We have also shown how the gauge correction modifies results in the Coulomb gauge compared to typically used models. Apart from yielding new insights into the nature of system-bath interactions, and presenting gauge-invariant master equations that can be used to explore a wide range of light-matter interaction in the USC regime, our results show that currently adopted master equations in the USC regime produce ambiguous results since they do not satisfy gauge invariance.

While we have shown explicit results for the cavity spectrum and intensity CF, the gauge correction causes profound effects on {\it any} observable that is computed from the master equations in the same coupling regimes. The nature of the system-bath coupling is also very important, which must also be related to the quadrature coupling to the external fields and the observables to ensure a gauge invariant master equation. For example, it may be more appropriate to use $\Pi_{\rm C}=a+a^\dagger$ (vector potential coupling) rather than $\Pi_{\rm C}=i(a^\dagger -a)$ (electric field coupling) for the interaction (in the Coulomb gauge), or some linear combination of the two; this change affects the dissipators, incoherent pumping, and coherent excitation in a way that still yields gauge-independent results, but the observables are different. 
By unitary equivalence, the form of the quadrature coupling used in the system Hamiltonian is thus also not arbitrary, which is in stark contrast to the JC model, where both these coupling forms yield identical results. These two coupling forms are widely used in the USC literature and are assumed to lead to the same result;  however, they differ significantly, which reinforces the need, highlighted recently~\cite{Bamba2014Feb,Lentrodt2020Jan}, to go beyond the usual phenomenological formulation of system-environment coupling Hamiltonians in the USC regime of cavity QED in favor of a general fundamental microscopic derivation.
Solutions to such problems can likely be rigorously addressed
using quantized quasinormal modes~\cite{PhysRevLett.122.213901,2019_ACS,PhysRevResearch.2.033332,PhysRevResearch.2.033456,Ren2022}, which even apply to cavities and media in the presence
of gain~\cite{PhysRevX.11.041020,PhysRevA.105.023702}.

\vspace{0.2cm}
\section*{Acknowledgements}
We acknowledge funding from 
the Canadian Foundation for Innovation and
the Natural Sciences and Engineering Research Council of Canada.
F.N. is supported in part by: 
Nippon Telegraph and Telephone Corporation (NTT) Research, 
the Japan Science and Technology Agency (JST) [via 
the Quantum Leap Flagship Program (Q-LEAP) program, 
the Moonshot R\&D Grant Number JPMJMS2061, and 
the Centers of Research Excellence in Science and Technology (CREST) Grant No. JPMJCR1676], 
the Japan Society for the Promotion of Science (JSPS) 
[via the Grants-in-Aid for Scientific Research (KAKENHI) Grant No. JP20H00134 and the 
JSPS–RFBR Grant No. JPJSBP120194828],
the Army Research Office (ARO) (Grant No. W911NF-18-1-0358),
the Asian Office of Aerospace Research and Development (AOARD) (via Grant No. FA2386-20-1-4069), and 
the Foundational Questions Institute Fund (FQXi) via Grant No. FQXi-IAF19-06.
S.S. acknowledges the Army Research Office (ARO)
(Grant No. W911NF1910065).

\vspace{1cm}

\appendix

\section{Gauge-Independent Master Equations: Dipole Gauge and Coulomb Gauge Forms}
\label{sec:AppA}

\subsection{Simple Generic Model for Cavity-Bath Leakage}


Let us first consider a general  bath (or reservoir) that  interacts with the system of interest (e.g., the cavity mode) {\em weakly}{, as shown schematically in Fig.~\ref{fig:system} of the main text}. The bath is described in the usual way by a collection of harmonic oscillators ($\hbar=1$),
%
\begin{equation}
    H_{\rm B} = \sum_k \omega_k b_k^\dagger b_k,
\end{equation}
where $b_k$ and $b_k^\dagger$ are bosonic annihilation and creation operators. 

A  simple model for a single cavity interacting with the bath 
can be written as follows:
\begin{equation}\label{eq:SB}
    H_{\rm SB} =
    \sum_{k}
    \lambda_{k}\Pi(b_{k}+b^\dagger_{k}),
\end{equation}
where $\lambda_{k}$ represent the coupling strengths (assumed real), which are model specific, and $\Pi$ is a gauge-dependent system operator linear in the canonical quantization variables, the form of which we specify based on physical considerations in the following sections.
In the interaction picture, we have
\begin{equation}
    \tilde H_{\rm SB} = \sum_{k}
    \lambda_{k} e^{i H_{\rm QR}t} \Pi e^{-i H_{\rm QR}t}(b_{k}e^{-i\omega k t}+b^\dagger_{k}e^{i\omega_k t}).
\end{equation}

In the dressed-state basis, which is necessary to use in the ultrastrong coupling (USC) regime (as the standard dissipator fails, as discussed in the main text), we can express the lowering operators of the system excitations
from 
\begin{equation}
    \tilde S(t) = \sum_{j,k>j}
    C_{jk}\ket{j}\bra{k} e^{i\Delta_{jk} t},
\end{equation}
where 
\begin{equation}
    C_{jk}= \braket{j|\Pi|k},
\end{equation}
and $\Delta_{jk}=\omega_j-\omega_k$, such that
\begin{equation}
\Pi(t) = e^{i H_{\rm QR} t}\Pi e^{-i H_{\rm QR} t} = \tilde{S}(t) + \tilde{S}^{\dagger}(t),
\end{equation}
and any $C_{jj}$ terms uniformly vanish due to the parity symmetry of the quantum Rabi model.
The bath operators can also be written as
\begin{equation}
    \tilde B(t) = \sum_{k}
    \lambda_{k} b_k e^{-i \omega_k t}.
\end{equation}
Thus we can write~\cite{settineri_dissipation_2018},
\begin{equation}
   \tilde H_{\rm SB} = 
    \tilde S(t) \tilde B^\dagger(t) + \tilde S^\dagger(t) \tilde B(t),
\end{equation}
where we have dropped all terms which oscillate at a frequency equal to a \emph{sum} of positive system and reservoir frequency components which do not ultimately contribute to the master equation we will derive.

Applying a Born-Markov approximation,
assuming continuous bath frequencies, a zero temperature approximation (namely, neglecting thermal excitation and taking the bath to be in the vacuum state), and neglecting any Lamb-like renormalization of the quantum Rabi Hamiltonian parameters,
one can derive a generalized master equation~\cite{settineri_dissipation_2018},
which takes into account the dressed-states' coupling to all the relevant baths for each system operator: 
\begin{equation}
\label{eq:GME1}
\frac{\rm{d}}{\rm{dt}}\rho = -\frac{i}{\hbar}[H_{\rm QR} +H_{\rm drive},\rho]  
+ \mathcal{L}_{\rm{G}}\rho
,
\end{equation}
where the cavity dissipator term is
\begin{equation}\label{eq:GME_diss}
    \mathcal{L}_{\rm{G}}\rho = \frac{1}{2} \sum\limits_{\omega,\omega'>0} 
    \Gamma_c(\omega) [ X^+(\omega)\rho  X^-(\omega') -  X^-(\omega') X^+(\omega)\rho] +
    \Gamma_c(\omega')[ X^+(\omega)\rho  X^-(\omega') - \rho X^-(\omega') X^+(\omega)].
\end{equation}    


The dressed-state operators, $X^\pm$,
decomposed in a basis of energy eigenstates with respect to  $H_{\rm QR}$,
are defined through
\begin{equation}
     X^+(\omega)= \braket{j|\Pi|k} \ket{j}\bra{k},   
\end{equation}
where $\omega=\omega_k-\omega_j>0$ and $ X^-=( X^+)^\dagger$. Note we can also derive a similar generalized master equation for other system decay channels (i.e., TLS losses), but below we concentrate on the cavity operators and relevant system-reservoir interactions, though we also briefly discuss the TLS-bath interactions.
    
One can employ any representative bath functions for the cavity reservoir, $J_{ c}(\omega)=  g_{c}(\omega)|\lambda(\omega)|^2$, where $g_{c}(\omega)$ is the bath density of states (DOS), and the decay rates are subsequently defined from
\begin{equation}
    \Gamma_{c}(\omega) = 
    2 \pi J_{c} (\omega ) = 
    2 \pi g_{ c}(\omega)|\lambda(\omega)|^2.
\end{equation}
Thus, for example, in the case of an Ohmic bath ($J_{c}(\omega) \propto \omega$), then
$\Gamma_{c}(\omega) = {\gamma_{ c}\omega}{/\omega_{c}},$
where $\gamma_{c} \equiv \kappa$.

Finally, assuming 
a relatively flat bath function with respect
to the frequency differences of interest (we will relax this approximation later, in Sec.~\ref{sec:G}), so that
$\Gamma_{\rm c}(\Delta_{jk}) \approx \kappa$,
with $\kappa=\kappa(\omega_0)$ (nominal cavity decay rate) over the energy scales of interest, 
we obtain
\begin{equation}\label{eq:dressed1}
    {\cal L}_{\rm dressed} \rho= 
    \frac{\kappa}{2}
    {\cal D}[x^+] \rho,
\end{equation}
where
\begin{equation}\label{eq:15}
    x^+= \sum_{j,k>j}
    C_{jk} \ket{j} \bra{k},
\end{equation}
with
\begin{equation}
    C_{jk} = \braket{j|\Pi|k},
\end{equation}
and the usual
Lindblad superoperator term,
\begin{equation}
{\cal D}[O] \rho =
2 O \rho O^\dagger - \rho O^\dagger O - O^\dagger O \rho.
\end{equation}

Equation \eqref{eq:dressed1}
is the standard dissipator form 
in the USC regime
for the dressed-state master equation. Without any consideration of gauge, one might naively  take $\Pi = i(a^{\dagger} - a)$ (the form we take for the \emph{non gauge-corrected} form of the dipole gauge model in the main text), or perhaps $\Pi = a+a^{\dagger}$; however, in contrast to usual cavity-QED systems outside of the USC regime, these choices 
give rise to different observables, and furthermore,
lead to gauge-dependent results (in the USC regime).
We address this explicitly in the following sections, and review how the breaking of gauge invariance that can be introduced by truncation to a TLS subspace is ``gauge corrected'' in the dipole gauge by modification of the $\Pi$ operators from their naive form, and in the Coulomb gauge by modifying the Hamiltonian~\cite{savasta_gauge_2020,di_stefano_resolution_2019,settineri_gauge_2019}.

\subsection{Gauge-invariant Master Equation in the Dipole Gauge}
To specify our dissipation model, we must assign a specific form to the gauge-dependent system operator $\Pi$, and thus we must consider how relevant physical quantities are represented in each gauge.
In the dipole gauge, it is the displacement field that is expanded in terms of bosonic creation/destruction operators, and not the  transverse electric field as in the Coulomb gauge. The relevant field operator
is ${\bf F}= {\bf D}/\epsilon_0\epsilon_{\rm b}({\bf r})$~\cite{PhysRevA.70.053823,PhysRevB.80.195106,settineri_gauge_2019},
where ${\bf D}$ is the displacement field and $\epsilon_b$ is the dielectric constant that the TLS is embedded (e.g., for free space this is 1). Importantly, ${\bf D}$ also includes a contribution from the TLS dipole field through the polarization.
The single cavity mode field-TLS interaction is then
\begin{equation}
    {\cal V}_{\rm I} = -{\bm \mu} \cdot {\bf F}({\bf r}_0) = \sigma_x(g_c a +g_c^* a^\dagger),
\end{equation}
where $g_c = -i \sqrt{\frac{\omega_c}{2\epsilon_0}} {\bm \mu}\cdot {\bf f}_c({\bf r}_0)$ and
${\bf r}_0$ is the dipole (TLS) location. The cavity mode amplitude is real (corresponding to a normal mode), and defining $g_c=-ig$, where $g$ is real, then
\begin{equation}
    {\cal V}_{\rm I} = i (a^\dagger -a) g \sigma_x , 
\end{equation}
which is projected onto a two level subspace.
Relating ${\bf F}$ to ${\bf E}$ (the electric field operator), and using a
TLS coupling  for the source of the polarization, results in the electric field being expanded in terms of transformed
cavity operators, $a' = a + i \eta \sigma_x$, such that~\cite{savasta_gauge_2020}
\begin{equation}
    {\bf E}_D({\bf r}, t)
    = i \sqrt{\frac{\omega_c}{2 \epsilon_0}}{\bf f}_c({\bf r}) a'(t) + {\rm H.c.}
    = i \omega_c {\bf A}({\bf r}) (a'-a'^\dagger) =
   i \omega_c {\bf A}({\bf r}) (a-a^\dagger + i 2 \eta \sigma_x), 
\end{equation}
where ${\bf A}({\bf r}) =\sqrt{{1}/{2 \epsilon_0 \omega_c}}\, {\bf f}_c({\bf r})$, the amplitude of the vector potential field.

Note that the explicit coupling between
$a$ and $\sigma_x$ here is a direct consequence of a strict single mode approximation.
If we assume a weak coupling between the cavity electric field and reservoir modes, the system-reservoir coupling takes the {\em gauge-corrected} form:
\begin{align}
    \Pi_{\rm D} &= i(a'^{\dagger} - a') \nonumber \\ 
    &= i(a^{\dagger}-a) + 2\eta \sigma_x,
\end{align}
where we let $\Pi_{\rm D}$ denote the system operator to be inserted into Eq.~\eqref{eq:SB} in the dipole gauge, and we assume the
$b_k$ are unchanged. Note as mentioned above we could also consider a linear coupling between the vector potentials of the cavity and reservoir fields, such that $\Pi \propto a+a^{\dagger}$ (which is manifestly gauge invariant in form). Outside of the USC regime, these couplings produce identical results within the rotating wave approximation (RWA), and are often assumed to be interchangable; however, in our simulations, choosing this form of system-reservoir coupling leads to significantly different observables (similar conclusions were drawn in Ref.~\cite{2002.02139}). This is because in the JC model, $x^+ \propto a + \mathcal{O}(\eta)$ in any gauge, and any change in the phase of $a$ is compensated for in the Lindblad term which pairs $a$ and $a^{\dagger}$. In the USC regime, the counter-rotating terms in the QRM ensure that the dissipator is not invariant under such a change. Noting that a coupling of the form $a+a^{\dagger}$ can be transformed into $i(a^{\dagger}-a)$ by the unitary transformation $U = \exp{[-i\frac{\pi}{2}a^{\dagger}a]}$, an important consequence of this is that a coupling in the (dipole gauge) QRM of the form $ig(a^{\dagger}-a)\sigma_x$ is \emph{not equivalent} to one of the form $g(a+a^{\dagger})\sigma_x$ when dissipation is to be considered, despite what is commonly assumed. In the USC regime, the gauge and form of the dissipators must be properly considered in conjunction with the Hamiltonian in order to ensure gauge-invariant observables. The only symmetry in the dissipative QRM is then that of parity symmetry, which ensures that the overall sign of any couplings terms can be changed. For this work, we restrict ourselves to electric field-like couplings such that $\Pi_{\rm C} = i(a^{\dagger}-a)$.


Following the same steps as above,
we obtain the dipole gauge result
for the dressed-state dissipator,
\begin{equation}\label{eq:L_dg}
\boxed{
    {\cal L}_{\rm dressed}^{\rm D} \rho= 
  \frac{\kappa}{2}
    {\cal D}[x^+_{\rm D}] \rho
    }\,,
\end{equation}
where
\begin{equation}
    x^+_{\rm D}= \sum_{j,k>j}
    C_{jk}^{\rm D} \ket{j} \bra{k},
\end{equation}
with
\begin{equation}\label{eq:C_dg}
\boxed{
    C_{jk}^{\rm D} = \braket{j|\Pi_{\rm D}|k}}\,,
\end{equation}
and the QRM system Hamiltonian is 
\begin{equation}\label{eq:Hi_dg}
\boxed{
  {H}_{\rm QR}^{\rm D} =  \omega_c a^\dagger a
    + \frac{\omega_0}{2} \sigma_z
    + i g(a^\dagger -a) \sigma_x}\,,
\end{equation}
where we use ${\omega_0}/{2} \sigma_z$
instead of ${\omega_0} \sigma^+\sigma^-$ (in the main text) to compare with the
Coulomb forms below. {\it The  `boxed'
equations (\eqref{eq:L_dg},\eqref{eq:C_dg},\eqref{eq:Hi_dg}) represent the gauge-corrected dissipator and QRM system Hamiltonian in the dipole gauge.} Note that as in the main text, to compute optical observables emitted from the cavity in this gauge we should also apply the gauge correction (i.e., use the $x_{\rm D}^{\pm}$ operators), to be consistent with input-output theory~\cite{gardiner_1985}.

More formally, before we switch to the Coulomb
gauge, we should also
identify $g \equiv g_{\rm D}$
as being the TLS-cavity coupling rate
in the dipole gauge. In the main text, we let quantities without explicit subscript/superscript refer to the dipole gauge, and in particular, $\Pi = i(a^{\dagger} -a)$ without any gauge correction, and $\Pi = i(a^{\dagger} -a) + 2 \eta \sigma_x$ with the proper gauge correction (subscripts GC in main text), both in the dipole gauge.

\subsection{Gauge-Invariant Master Equation in the Coulomb Gauge}


{As discussed in the main text, in the Coulomb gauge Hamiltonian,
we have the following system Hamiltonian for the QRM~\cite{di_stefano_resolution_2019}
\begin{equation}\label{eq:Hi_cg0_3}
    { H}_{\rm QR}^{\rm C} = 
    \omega_{c} a^\dagger a + 
    \frac{\omega_0}{2}\sigma_z  + 
   g_{\rm C} (a+a^\dagger) \sigma_y + D(a+a^\dagger)^2,
\end{equation}
where $g_{\rm C} = g_{\rm D}\omega_0/\omega_c$ and
$D$ is the diamagnetic term. 
For $\omega_0=\omega_c$ and $g_{\rm D}\equiv g$, we can use
$D=\eta^2\omega_0$ as a lower bound~\cite{2002.02139}.}






In the USC regime, Eq.~\eqref{eq:Hi_cg0_3} does not produce the same eigenenergies
as Eq.~\eqref{eq:Hi_dg}, since it fails to respect the gauge principle.
Instead, the properly gauge-transformed form, which does produce the same eigenenergies, is given by the following system Hamiltonian for the QRM~\cite{di_stefano_resolution_2019}:
\begin{equation}\label{eq:Hi_cg}
\boxed{
  { H}_{\rm QR }^{\rm C'} = \mathcal{U}^{\dagger} H_{\rm QR}^{\rm D} \mathcal{U} = \omega_c a^\dagger a
    + \frac{\omega_0}{2}
    \left \{ \sigma_z \cos(2 \eta (a+a^\dagger))+ \sigma_y \sin(2\eta (a + a^\dagger)) \right \}},
\end{equation}
where $\mathcal{U} = \exp(-i\eta(a+a^\dagger)\sigma_x)$, as in the main text.
Notably, ${ H}_{\rm QR }^{\rm C'}$ contains field operators to all orders.

In the Coulomb gauge, 
the form of the electric field operator
is proportional to $i(a^\dagger -a)$, assuming the same
bath interactions, 
and thus we have $\Pi_C=\mathcal{U}^{\dagger}\Pi_{\rm D} \mathcal{U} = i(a^{\dagger}-a)$, and the
system-bath coupling is
written as
\begin{equation}
    H_{\rm SB}^{\rm C} =
    \sum_{k}
    \lambda_{k}^{\rm C}\Pi_C(b_{k}+b^\dagger_{k}),
\end{equation}
where $\lambda_k^{\rm C}$ is cavity-bath
interaction in the Coulomb gauge.

Following similar steps to before, we obtain the 
gauge-invariant dissipator term:
\begin{equation}
    {\cal L}_{\rm dressed}^{\rm C} \rho= 
 \frac{\kappa}{2}
    {\cal D}[x^+_{\rm C}] \rho,
\end{equation}
where
\begin{equation}\label{eq:C_cg}
\boxed{
    x^+_{\rm C}= \sum_{j,k>j}
    C_{jk}^{\rm C} \ket{j} \bra{k}},
\end{equation}
with
\begin{equation}\label{eq:L_cg}
\boxed{
    C_{jk}^{\rm C} = \braket{j|\Pi_C|k}
    },
\end{equation}
and now one uses the dressed states in the Coulomb gauge, namely using ${H}_{\rm QR}^{\rm C}$.

Note, to include an arbitrary spectral function, then we use 
\begin{equation}
    \Gamma_c(\omega)
    = 2 \pi J_{ c}^{\rm }(\omega)
    = 2\pi g_{ c}(\omega)|\lambda_{\rm }(\omega)|^2,
\end{equation}
and $J_{ c}$ and $\lambda(\omega)$ are identical in the dipole gauge and Coulomb gauge.
Below we show this explicitly for the case of an Ohmic bath.
%

The three boxed equations (\eqref{eq:Hi_cg},\eqref{eq:C_cg},\eqref{eq:L_cg}) represent the
correct 
form for the Coulomb gauge master equation to give equivalent results
to the dipole-gauge forms, which we prove in Sec.~\ref{sec:AppB} and show explicitly in Fig. 3 of the main text.

Note also that the expressions for
$C_{jk}$ can be rewritten via a sum rule~\cite{2002.02139}. For example,
in the Coulomb gauge~\cite{2002.02139}:
\begin{equation}
    \braket{k|a^\dagger-a|j}=
    \frac{\omega_{kj}}{\omega_{c}}\braket{k|a^\dagger+a|j},
\end{equation}
and in the dipole gauge:
\begin{equation}
    \braket{k|a^\dagger-a-2i\eta\sigma_x|j} = 
    \frac{\omega_{kj}}{\omega_{c}}\braket{k|a^\dagger+a|j}.
\end{equation}



\subsection{Two Level System (TLS) Dissipator}

Next, for completeness, we discuss the TLS dissipator (whose contribution is negligible in our simulations) and again show equivalence between the dipole gauge and Coulomb gauge.
The $\sigma_x$ is invariant when transformed through the gauge correction, and thus there is no change; namely we simply have:
\begin{equation}
    {\cal L}_{\rm dressed}^{\rm D/C}|_\gamma \rho= 
    \frac{\gamma}{2}
    {\cal D}[y^+_{D/C}] \rho.
\end{equation}
where
\begin{equation}
    y^+_{\rm D/C}= \sum_{j,k>j}
    C_{jk}^{\rm D/C} \ket{j} \bra{k},
\end{equation}
with
\begin{equation}
    C_{jk}^{\rm D/C} = \braket{j|\sigma_x|k},
\end{equation}
in either gauge.
However, for a specific model for the spontaneous emission decay, a more realistic model would include frequency dependent reservoirs representative of the free space emission channel (such as Ohmic). Later we show an example of how to incorporate such effects for the dominant cavity model decay channel, and it is easy to also do this for the TLS decay, if required.

\subsection{Incoherent Pump Term}
In a standard master equation, the incoherent pumping for the cavity mode is usually written as a reversed Lindblad decay process~\cite{Tian1992,PhysRevB.81.033309},
\begin{equation}
    {\cal L}^{\rm pump}\rho =  \frac{P_{\rm inc}}{2} {\cal D}[a^\dagger] \rho  ,
\end{equation}
which in a dressed-state decomposition is 
\begin{equation}
    {\cal L}^{\rm pump}_{\rm dressed}\rho =  \frac{P_{\rm inc}}{2}  {\cal D}[x^-] \rho  .
\end{equation}

This type of excitation can be derived by input-output theory with (for example) non-vacuum inputs (e.g., a thermal state with $T\neq0$)~\cite{gardiner_1985}. Thus to be consistent with our dissipation channels and the microscopic form of the system-reservoir coupling, we choose
\begin{equation}
\boxed{
    {\cal L}^{\rm pump}_{\rm dressed}\rho =  \frac{P_{\rm inc}}{2}  {\cal D}
    \left [x^-_{\rm D/C}\right ] \rho } \, .
\end{equation}

\subsection{Coherent Pump Term}
Next we present a derivation of the coherent drive term in the Hamiltonian, $H_{\rm drive}(t)$.
We consider the general interaction picture Hamiltonian of
Eq.~\eqref{eq:SB},
\begin{equation}
    \tilde{H}_{\rm SB}(t) = \big(\tilde{S}(t)+\tilde{S}^{\dagger}(t)\big)\big(\tilde{B}(t)+\tilde{B}^{\dagger}(t)\big),
\end{equation}
where the gauge of this interaction is ultimately determined by the form of $\tilde{S}(t)$, which is left general here.
In deriving the master equation models, we formally considered the reservoir to be in a multimode vacuum state $\ket{\mathbf{0}}$ as $t \rightarrow 0$. To model coherent driving at the level of a system-reservoir approach, where the input drive (laser field) is not significantly impacted by the dynamics of the cavity-QED system, we can instead assume the reservoir to be in a multimode (or approximately single mode) coherent state,
with the input condition:
\begin{equation}
    \rho_B(t=0) = U_{\rm c} \ket{\mathbf{0}}\bra{\mathbf{0}}U^{\dagger}_{\rm c},
\end{equation}
where $U_{\rm c}= \prod_{k} D_k(\beta_k)$, $D_{k}(\beta_k) = \exp{\left[\beta_k b_k^{\dagger} - \beta_k^* b_k\right]}$ is the displacement operator,
with $D_{k}(\beta_k) \ket{{\bf 0}}
=\ket{\beta_k}$ (where all $k' \neq k$ remain in the vacuum state),
and $\beta_k$ are substantial only for wavevectors $k$ around the laser resonance. Since $U_{\rm c}$ is unitary, we can apply a unitary transformation to the system plus reservoir density operator and Hamiltonian. Within the Born-Markov approximation, we have $\tilde{\rho}_{\rm S +B} = \tilde{\rho}_{\rm S} \rho_{\rm B}$, thus we apply the unitary transformation $\tilde{\rho}_{\rm S +B} \rightarrow U_{\rm c}^{\dagger} \tilde{\rho}_{\rm S+B}U_{\rm c}$, and $\tilde{H}_{\rm SB} \rightarrow U_{\rm c}^{\dagger} \tilde{H}_{\rm SB}U_{\rm c}$. The effect of this is merely to take $b_k \rightarrow b_k +\beta_k$ within the interaction picture. Thus we have $\tilde{B}(t) \rightarrow \tilde{B}(t) + \sum_k \lambda_k \beta_k e^{-i \omega_k t}$, and
\begin{equation}\label{eq:HS}
    \tilde{H}_{\rm SB}(t) \rightarrow \tilde{H}_{\rm SB}(t) + \sum_k \big(\tilde{S}(t)+\tilde{S}^{\dagger}(t)\big)\big(\lambda_k \beta_k e^{-i \omega_k t} + \rm{c.c.}\big).
\end{equation}

Since the new term in Eq.~\eqref{eq:HS} only depends on the system operators, we can call it $\tilde{H}_{\rm drive}$ and consider it part of the system Hamiltonian. Moving back to the Schr{\"o}dinger picture:
\begin{equation}\label{eq:multimodedrive}
    H_{\rm drive}(t) = \sum_k \big(x^+_{\rm D/C} + x^-_{\rm D/C}\big)\big(\lambda_k \beta_k e^{-i \omega_k t} + {\rm c.c.}\big).
\end{equation}
In this new frame, the bath is in the multimode vacuum state $\ket{\mathbf{0}}$. Thus, the master equation can be derived in exactly the same manner as before, with the only difference in the equations being the addition of $H_{\rm drive}(t)$ in the system Hamiltonian. Typically, we can make a RWA for this term, as we have separated positive and negative frequency components, but in principle we leave this general as the RWA could break down for ultrastrong coherent driving (however, in this regime, a Floquet master equation would be more accurate).

To transform Eq.~\eqref{eq:multimodedrive} into an effective single-mode drive, we move to a continuous frequency representation:
\begin{equation}
    H_{\rm drive}(t) = \big(x^+_{\rm D/C}+x^-_{\rm D/C}\big)\int_0^{\infty} d\omega g_c(\omega) \lambda_c(\omega) \beta_c(\omega) e^{-i \omega t} + {\rm c.c.}
\end{equation}
Since $\beta_c(\omega)$ is only nonzero around a very narrow window around $\omega=\omega_{\rm L}$ (the laser center frequency), we have 
\begin{equation}
    \int_0^{\infty} d\omega g_c(\omega) \lambda_c(\omega) \beta_c(\omega) e^{-i \omega t} \approx g_c(\omega_{\rm L})\lambda_c(\omega_{\rm L}) e^{-i\omega_{\rm L} t} \int_{-\infty}^{\infty}d \delta \beta_c(\omega_{\rm L} + \delta)e^{-i\delta t}.
\end{equation}
The form of $\beta_c(\omega)$ is not important provided it is sharply peaked around $\omega = \omega_{\rm L}$; for concreteness we can assume a Lorentzian form:
\begin{equation}
    \beta_c(\omega) = \beta_c(\omega_{\rm L}) \frac{({ W}_0/2)^2}{({W}_0/2)^2 + \delta^2},
\end{equation}
where ${W}_0$ is the FWHM of the laser beam, and we find:
\begin{equation}
        H_{\rm drive}(t) = \big(x^+_{
        \rm D/C}+x^-_{\rm D/C}\big)\big(\Omega_{\rm d} e^{-i\omega_0t} + {\rm c.c.}\big)e^{-{\omega_0} t/2},
\end{equation}
where we have defined
\begin{equation}
    \Omega_{\rm d} = \pi g_c(\omega_{\rm L})\lambda_c(\omega_{\rm L})\beta_c(\omega_{\rm L}){W}_0.
\end{equation} 

We assume that the laser linewidth $W_0$ is small enough such that the drive remains coherent over any experiment of interest. We can also choose $\Omega_{\rm d}$ to be real without loss of generality, as the phase factor can be absorbed into the initial phase of the drive which is not relevant. Thus we find the form used in the main text:
%
%
\begin{equation}
\boxed{
    H_{\rm pump}^{\rm D/C} = \Omega_{\rm d}\cos{(\omega_{\rm L} t)} \left (x^-_{\rm D/C}  + x^+_{\rm D/C}\right) 
    }\, ,
\end{equation} 
using identical system operators as in the
dissipators and incoherent pump terms. Also note, the coherent drive
should not be too strong to invalidate the dressed-state representation for the system Hamiltonian, namely
$\Omega_{\rm d} \ll g$, and in this regime one could make a RWA for the pump term such that $H_{\rm pump}^{\rm D/C} \approx (\Omega_{\rm d}/2)  (x^-_{\rm D/C}e^{-i \omega_c t}  + x^+_{\rm D/C} e^{i \omega_c t})$. 
In the main text, we use this RWA pumping term and also consider a resonant drive where $\omega_{\rm L} = \omega_c$. For completeness, in Sec.~\ref{sec:AppC} below,
we also show example calculations with and without a RWA on the pump term, and confirm that they yield essentially identical spectra, as expected (i.e., for the stated approximations).

\section{Equivalence between the dipole gauge and Coulomb gauge master equations and gauge {invariant} expectation values}
\label{sec:AppB}

Naturally, any observables from a unitarily transformed quantum master equation should be gauge-independent; we include this section primarily to show that the $x^{\pm}$ operators transform in the way one might expect. {By gauge independent expectation values, we mean expectation values that do not and should not depend on the choice of gauge.}

Ultimately, for any gauge-dependent Hermitian operator $O^{\rm D}$ and $O^{\rm C}$ corresponding to a physical observable, the expectation value should be gauge-independent, so
\begin{equation}
    \dot {\braket{O}} = 
    \Tr{\dot{\rho}^{\rm D} O^{\rm D} } = 
    \Tr{\dot{\rho}^{\rm C} O^{\rm C} }.
\end{equation}

Beginning with the evaluation in the Coulomb gauge,
\begin{align}
    \dot{\braket{O}} & = -i\Tr{O^{\rm C}[H_{\rm QR}^{\rm C},O^{\rm C}]} + \frac{\kappa}{2}\Tr{O^{\rm C}\big(\mathcal{D}[x^+_{\rm C}]\rho^{\rm C}\big)} \nonumber \\ & = -i\Tr{O^{\rm D}[H_{\rm QR}^{\rm D},O^{\rm D}]} + \frac{\kappa}{2}\Tr{O^{\rm D} \mathcal{U}\mathcal{D}[x^+_{\rm C}]\rho^{\rm C}\mathcal{U}^{\dagger}}.
\end{align}

Clearly, the evolution is gauge-invariant if,
\begin{equation}
\mathcal{U}\mathcal{D}[x^+_C]\rho^{\rm C}\mathcal{U}^{\dagger} = \mathcal{D}[x_{\rm D}^+]\rho^{\rm D} = \mathcal{D}[\mathcal{U}x_{\rm C}^+\mathcal{U}^{\dagger}]\rho^{\rm D}.
\end{equation}
 We have (explicitly noting the gauge of each state):
\begin{align}
    &\mathcal{U}x_{\rm C}^+\mathcal{U}^{\dagger} = \sum_{j_{\rm C},k_{\rm C}>j_{\rm C}} C^{\rm C}_{jk} \mathcal{U}\ket{j_{\rm C}}\bra{k_{\rm C}}\mathcal{U}^{\dagger} \nonumber \\ 
    & = \sum_{j_{\rm C},k_{\rm C}>j_{\rm C}} \bra{j_{\rm C}} \Pi_{\rm C}\ket{k_{\rm C}}\ket{j_{\rm D}}\bra{k_{\rm D}} \nonumber \\ 
    & = \sum_{j_{\rm C},k_{\rm C}>j_{\rm C}} \bra{j_{\rm C}} \mathcal{U}^{\dagger}\Pi_{\rm D}\mathcal{U}\ket{k_{\rm C}} \ket{j_{\rm D}}\bra{k_{\rm D}} \nonumber \\ 
    & = \sum_{j_{\rm C},k_{\rm C}>j_{\rm C}} \bra{j_{\rm D}}\Pi_{\rm D}\ket{k_D} \ket{j_{\rm D}}\bra{k_{\rm D}} \nonumber \\ 
    & = \sum_{j_{\rm D},k_{\rm D}>j_{\rm D}} C^{\rm D}_{jk} \ket{j_{\rm D}}\bra{k_{\rm D}} \nonumber \\
    & = x^+_{\rm D},
\end{align}
where we have noted in the second last line that energy eigenvalues are preserved under unitary transformation. This argument can be trivially extended to incoherent excitation master equations with terms like $\mathcal{D}[x^-_{\rm C/D}]\rho^{\rm C/D}$, or time-dependent coherent drive terms, provided they are transformed appropriately between gauges. 

\section{Role of the coherent pump strength and coherent pumping with and without a rotating wave approximation}
\label{sec:AppC}

In the main text, we chose an example coherent pump strength of $\Omega_{\rm d}=0.1g$.
Obviously if we increase this value, then higher order nonlinearities become important, though we cannot increase it arbitrarily or the assumed dressed states are no longer valid. We also note that if this value is too small, then the numerical simulations can become very difficult.
For completeness, here we show two further examples, for the larger pump strengths of $\Omega_{\rm d}=0.2g$ and $\Omega_{\rm d}=0.3g$.

\begin{figure}[t]
    \centering
    \begin{tabular}{c c}
        \includegraphics[width=0.48\linewidth]{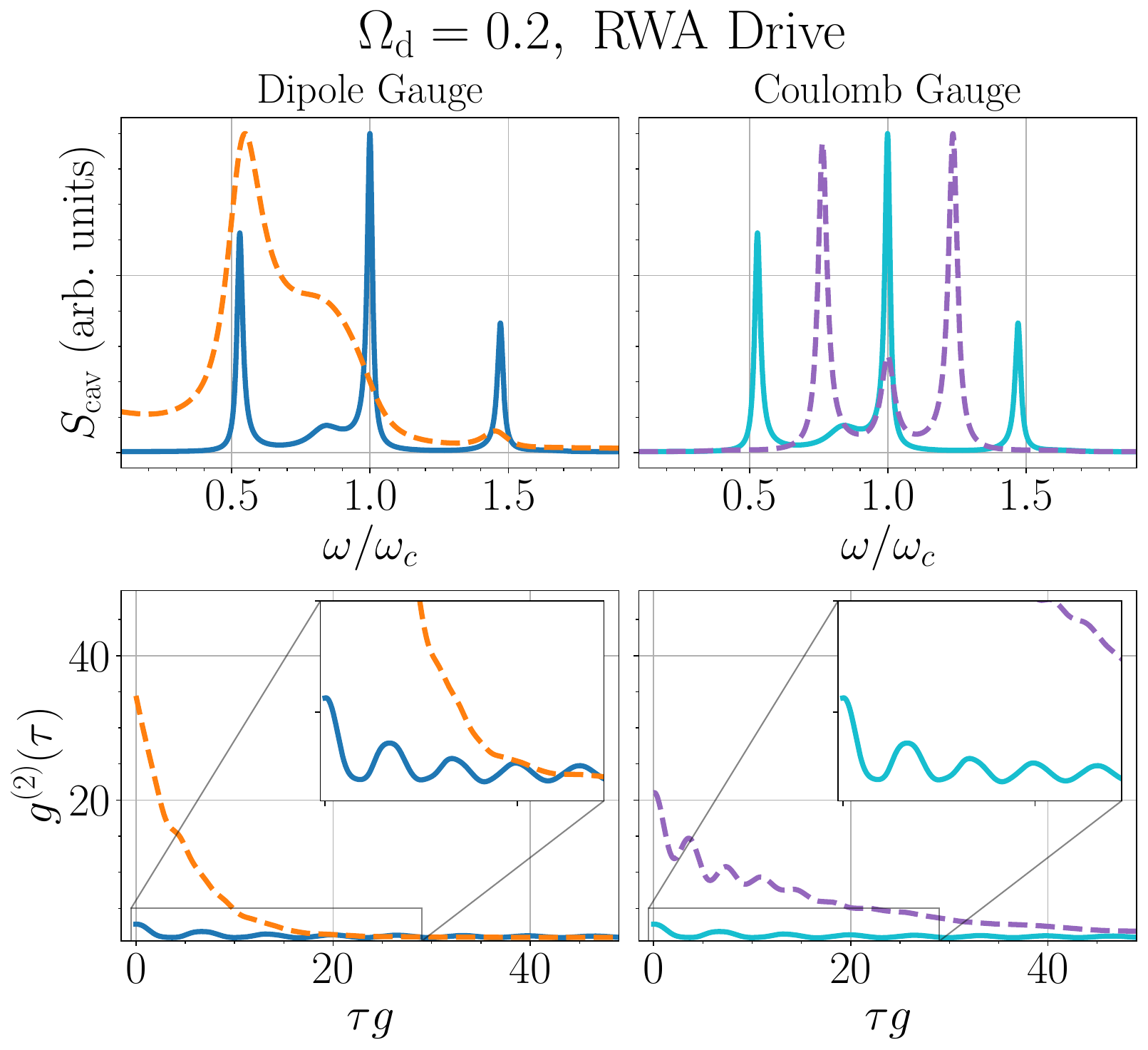}
        &
        \includegraphics[width=0.48\linewidth]{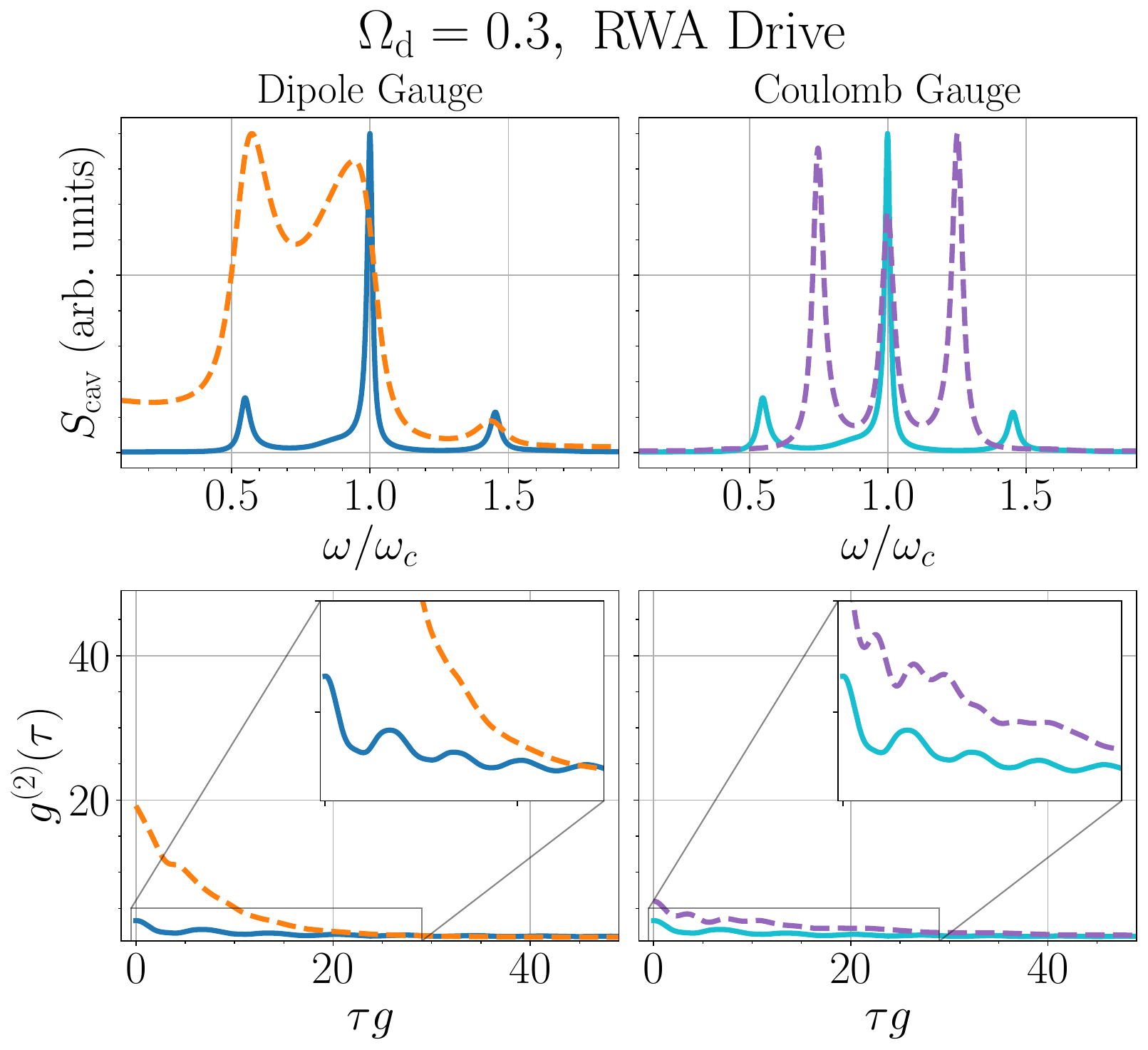}
    \end{tabular}
    \vspace{-0.3cm}
    \caption{
    Left four panels show cavity spectra and $g^{(2)}(\tau)$ for $\Omega_{\rm d}=0.2g$, and the right four panels are for $\Omega_{\rm d}=0.3g$ (cf.~Fig.~\ref{fig:coh_vs_inc} of the main text and also below). Solid lines show the gauge corrected master equation results.}
    \label{fig:pump2}
\vspace{0.2cm}
\end{figure}

Figure~\ref{fig:pump2} shows that in comparison to Fig.~3 if the main text, the center peak increases with larger coherent driving (as expected), and begins to dominate the spectral response when the pump is sufficiently large. The gauge correction is also seen to be even more dramatic for the larger pump strength, especially in the dipole gauge.
In both cases we see a significant influence from the gauge correction, and recognize once again that the corrected dipole gauge and corrected Coulomb gauge master equations yield identical results.

\begin{figure}[t]
    \centering
    \begin{tabular}{c c}
        \includegraphics[width=0.48\linewidth]{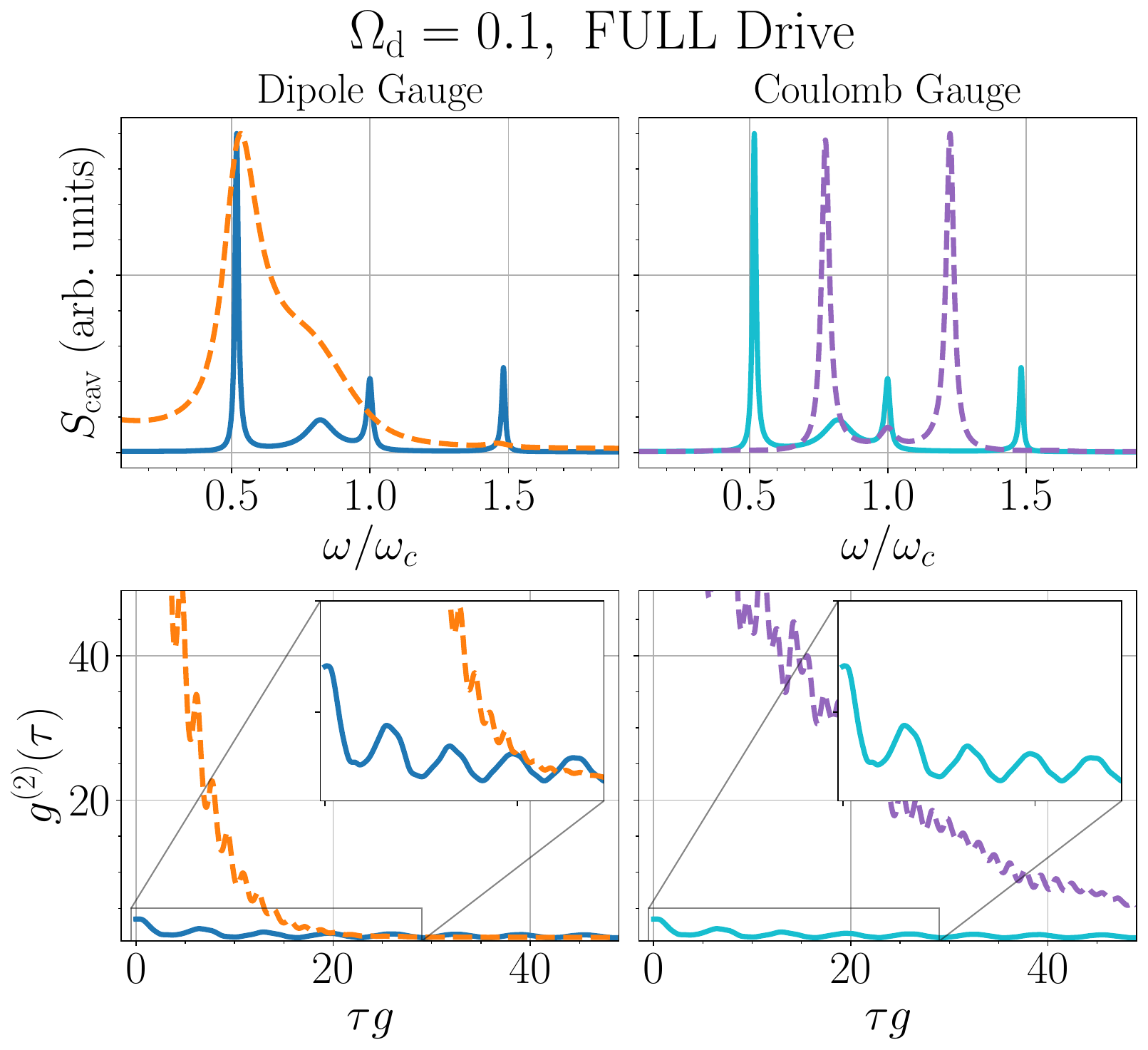}
        &
        \includegraphics[width=0.48\linewidth]{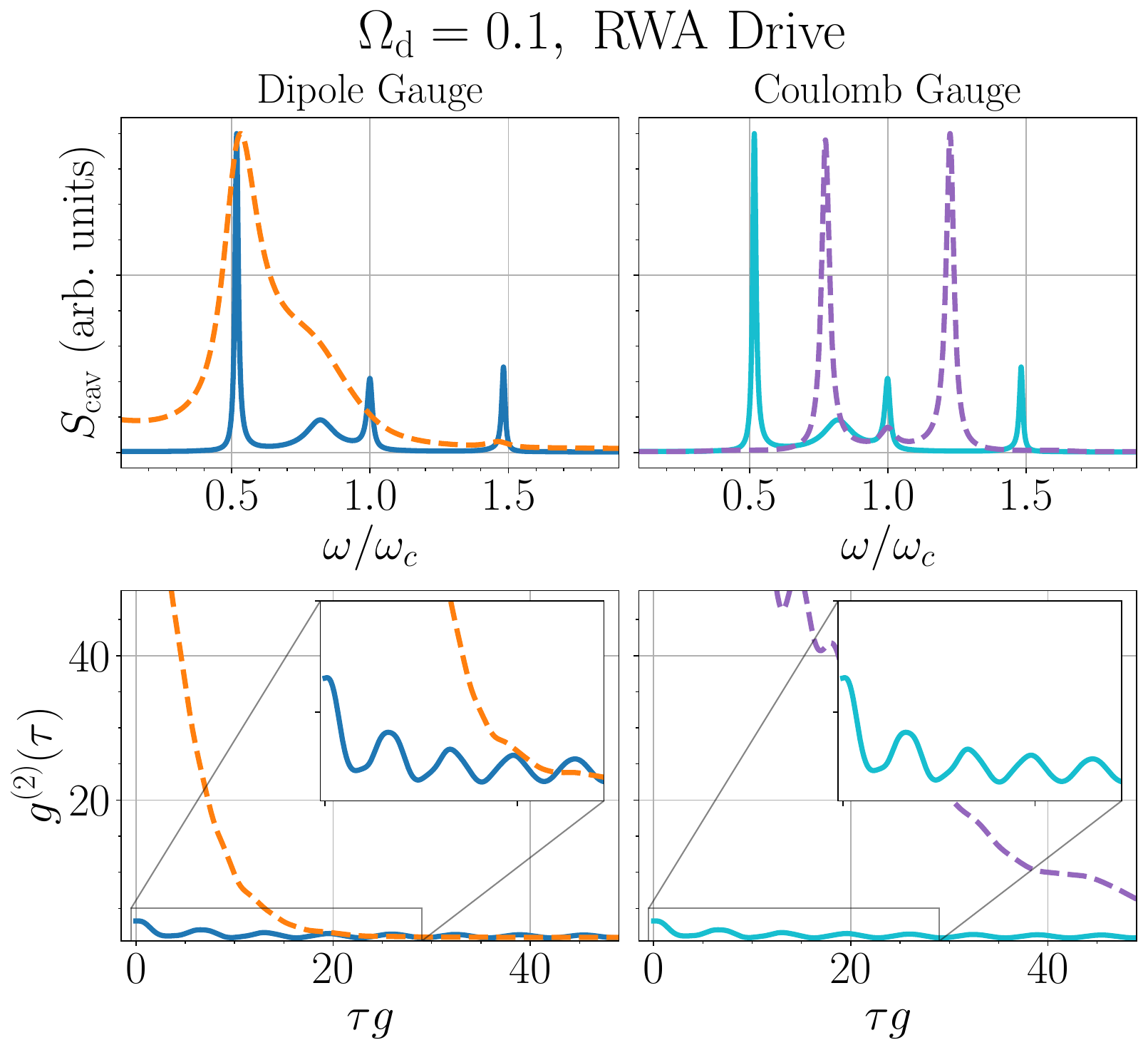}
    \end{tabular}
    \vspace{-0.3cm}
    \caption{
    Cavity spectra and $g^{(2)}$ with $\Omega_{\rm d}=0.1g$ coherent pumping, using a full cosine excitation (left), and a RWA for the pumping term (right, as also shown in Fig.~\ref{fig:coh_vs_inc} of the main text). Solid lines show the gauge corrected master equation results.}
    \label{fig:pumpRWA}
\vspace{0.2cm}
\end{figure}

Next we also investigate the results of coherent driving with and without a RWA on the drive term. With a rotating-wave approximation, as mentioned earlier, we use $H_{\rm drive}(t) = (\Omega_{\rm d}/2)(x^-_{\rm GF} e^{-i\omega_{\rm L}t} + x^+_{\rm GF}e^{-i\omega_{\rm L}t} )$ (as in the main text),
and without this approximation, we use $H_{\rm drive}(t) = \Omega_{\rm d}(x^-_{\rm GF} + x^+_{\rm GF})\cos(\omega_{\rm L}t)$. Figure \ref{fig:pumpRWA} compares these two pump forms for computing the cavity spectrum and $g^{(2)}(\tau)$, which are shown to essentially yield the same behavior, apart from fast oscillations in the correlation functions when a RWA is not made. Since we do not consider the effect of coherent driving on the dressed-states (from which we solve the master equations), then pumping within a RWA should be valid within the same level of approximations, and is arguably more self consistent. 

\section{Further Details on the Numerical Calculations}
\label{sec:AppD}
\subsection{Simulation Parameters}
In our numerical simulations in the main text, a large initial basis size of 50 photon states was used. This ensures that the lowest dressed states, which have a significant chance to become populated, are correct, before computing the spectra in a truncated basis space. With 24 dressed states in the truncated space, we observe negligible excitation in the highest states and numerically converged results (i.e., additional dressed states make no change to our simulations and results). The eigenenergy simulation in Fig.~\ref{fig:evals}(a)
of the main text was conducted with 200 photon states to ensure accurate numerical convergence. Longer times are required for simulations at higher $\eta$ and as such, the simulation time was between $150/g$ and $550/g$ throughout, with 20 time-steps in each period of the pseudo-steady-state oscillation. This was also carefully checked to be sufficient.
Numerical calculations were performed using QuTiP under Python~\cite{johansson_qutip_2013}.

With coherent excitation, numerical calculations of the spectra outside the rotating wave approximation require some care. Specifically, the $t$ integral in the spectrum definition (Eq.~\eqref{eqn:spec} of the main text) was completed over the last $\omega_{\rm L}$ time period so as to ignore turn-on dynamics, after ensuring that the system had reached its pseudo-steady-state (namely, after it evolves to a continuous oscillation dynamic with no change). Consequently, there is a potential issue with computing a Fourier transform of an oscillating function over a finite range. This is commonly done for computing USC spectra but is rarely discussed. Formally, the Fourier transform of a $\sin{\omega_0 t}$ function over a finite range $a$ is proportional to the difference of two $\sinc$ functions at $\pm \omega_0$, shown explicitly below,
\begin{equation}
\begin{split}
    F_{-a\rightarrow a}(\omega) &= \frac{1}{\sqrt{2\pi}}\int^a_{-a}e^{-i\omega t}\sin{\omega_0t}dt\\
    &= \frac{1}{2i\sqrt{2\pi}} \int^a_{-a}\left[e^{i(\omega_0-\omega)t}-e^{-i(\omega_0+\omega)t}\right]dt\\
    &= \frac{1}{2i\sqrt{2\pi}} \left[\frac{e^{ia(\omega_0-\omega)}-e^{-ia(\omega_0-\omega)}}{i(\omega_0-\omega)} + \frac{-e^{-ia(\omega_0+\omega)}+e^{ia(\omega_0+\omega)}}{i(\omega_0+\omega)}\right]\\
    &= \frac{a}{i\sqrt{2\pi}} \left[\frac{\sin{\left(a(\omega_0-\omega)\right)}}{a(\omega_0-\omega)} - \frac{\sin{\left(a(\omega_0+\omega)\right)}}{a(\omega_0+\omega)}\right]\\
    &= \frac{-ia}{\sqrt{2\pi}} \left[{\rm sinc}\left(a(\omega-\omega_0)\right)-{\rm sinc}\left(a(\omega+\omega_0)\right)\right].
\end{split}
\end{equation}

When extending this time sampling range to infinity, the Fourier transform tends towards the sum of two Dirac delta functions,
 \begin{equation}
    F(\omega) = i\sqrt{\frac{\pi}{2}} \left[\delta(\omega+\omega_0)-\delta(\omega-\omega_0)\right].
 \end{equation}

This can have a significant effect on both the total and coherent spectrum, but in all our simulations the {\it incoherent spectrum} with coherent driving is unaffected, as it does decay to zero for large time delays, and performing the quantum regression theorem over only a single period is thus adequate for our case. For example, we have checked that we obtain the same result when integrating over ten periods.

\subsection{Quantum Regression Theorem}
To calculate the two-time correlation function in the spectrum definition (Eq.~\eqref{eqn:spec}  of the main text), we make use of the quantum regression theorem,
\begin{equation}\label{eqn:QRT}
    \Braket{x_\Delta^-(t)x_\Delta^+(t+\tau)} = \Tr{x_\Delta^+(0)U(t+\tau,t)\left[\rho(t)x_\Delta^-(0)\right]U^\dagger(t+\tau,t)},
\end{equation}
where $U(t+\tau,t)$ is the total (system + environment) unitary evolution operator such that $A(t+\tau)=U^\dagger(t+\tau,t)\left[A(t)\right]U(t+\tau,t)$ for an operator $A(t)$. Within the Born-Markov approximation, the implementation of the quantum regression theorem for Eq.~\eqref{eqn:QRT} is as follows: find the reduced density matrix at $t$, multiply on the right by $x_\Delta^-(0)$, evolve this new operator from $t$ to $(t+\tau)$ with the master equation to form the effective density matrix, and finally take the expectation value of $x_\Delta^+(0)$ with respect to this effective density matrix. Note that $x_\Delta^-(0)=x^-(0)-\braket{x^-(t)}$ where the second term must be evaluated at $t$, so $x_\Delta^-(0)$ does implicitly depend on $t$. For the positive frequency operator, we have $x_\Delta^+(0)=x^+(0)-\braket{x^+(t+\tau)}$, but if we substitute this into Eq.~\eqref{eqn:QRT}, we see that the second term (proportional to $\braket{x^+(t+\tau)}$) is exactly zero, so there is no need to find the expectation value of $x^+$ at $(t+\tau)$. In practice, we conduct the quantum regression theorem for every $t$ in the last period (in the simulation) of the pseudo-steady-state.

For the more complex second-order correlation function in Eq.~\eqref{eqn:g2(t,tau)} of the main text, we have a more complicated version of the quantum regression theorem seen in Eq.~\eqref{eqn:QRT} as follows:
\begin{equation}
\begin{split}
    G^{(2)}(t,\tau) &= \left\langle x^-(t) x^-(t+\tau) x^+(t+\tau) x^+(t)\right\rangle\\
    &= \Tr{x^-(t) x^-(t+\tau) x^+(t+\tau) x^+(t) \rho(0)}\\
    &= \Tr{U^\dagger(t)x^-U(t)U^\dagger(t+\tau)x^-x^+U(t+\tau)U^\dagger(t)x^+U(t)\rho(0)}\\
    &= \Tr{x^-U^\dagger(t+\tau,t)x^-x^+U(t+\tau,t)x^+\rho(t)}\\
    &= \Tr{x^-x^+U(t+\tau,t)\left[x^+\rho(t)x^-\right]U^\dagger(t+\tau,t)},
\end{split}
\end{equation}
where we have written $U(t)\equiv U(t,0)$, we use the notation $A(0) = A$ for an operator $A(t)$, and we make use of the identities $U(t+\tau)=U(t+\tau,t)U(t)$ and $U(t)U^\dagger(t) = U^\dagger(t)U(t) = \mathbb{1}$ where $\mathbb{1}$ is the identity matrix. This can be understood simply as the expectation value of the operator $x^-x^+$ with respect to the effective density matrix $\tilde{\rho}(t+\tau) = U(t+\tau,t)\left[x^+\rho(t)x^-\right]U^\dagger(t+\tau,t)$, which is the density matrix at $t$ multiplied on the left by $x^+$ and on the right by $x^-$ and evolved from $t$ to $t+\tau$.

\section{Bloch-Siegert Hamiltonian and perturbative unitary transform to obtain analytical scattering rates and spectra}
\label{sec:AppE}

Here we show the approximate solution to the spectra and the origin of asymmetry using the Bloch-Siegert (BS) Hamiltonian~\cite{beaudoin_dissipation_2011,LeBoite2020Jul}. From the system Hamiltonian in the dipole gauge,
\begin{equation}
    H = \omega_0 \sigma^+\sigma^- + \omega_0 a^{\dagger} a + ig(a^{\dagger}-a)\sigma_x,
\end{equation}
we apply the unitary transformation (``BS transformation'') $H_{\rm BS} = U_{\rm BS}^{\dagger} H U_{\rm BS}$, where 
\begin{equation}
U_{\rm BS} = \exp{\left [-i\frac{\eta}{2}(a^{\dagger}\sigma^+ + a \sigma^-)\right]}\exp{\left[-\frac{\eta^2}{4}\sigma_z(a^2-a^{\dagger 2})\right]},
\end{equation}
which is chosen to eliminate counter-rotating terms in the system Hamiltonian, and retain terms of up to second order in $g$~\cite{LeBoite2020Jul}, finding
\begin{equation}\label{eq:bs}
    H_{\rm BS} = \omega_0(1 + \eta^2/2)\sigma^+\sigma^- + \omega_0(1-\eta^2/2)a^{\dagger} a + ig(a^{\dagger} \sigma^- - a \sigma^+),
\end{equation}
where we are considering a weak excitation approximation (WEA),  and so a term proportional to $a^{\dagger} a \sigma^+ \sigma^-$ has been dropped, as well as terms proportional to the identity so the ground state energy remains zero. The resulting BS Hamiltonian of Eq.~\eqref{eq:bs} conserves excitation number, and thus is easily diagonalized.

\begin{figure}[h]
    \centering
    \includegraphics[width=0.5\linewidth]{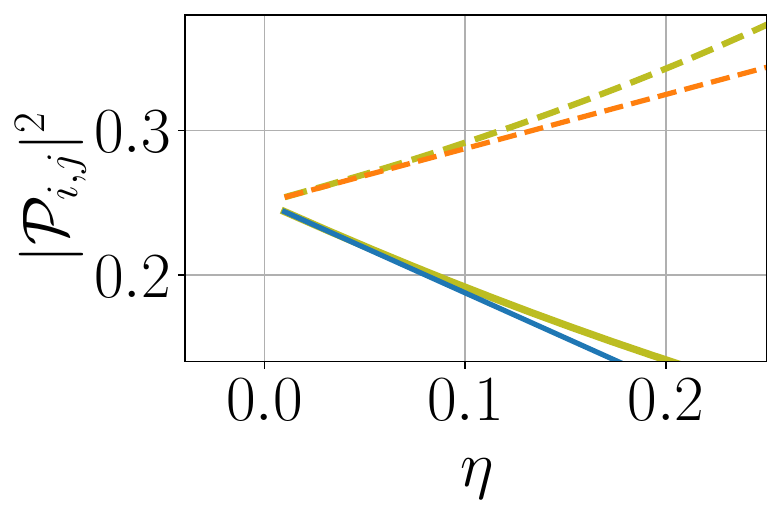}
    \vspace{-0.2cm}
    \caption{Here we show the full numerical calculations as shown in Fig.~\ref{fig:evals}(c) of the main text with (olive solid curve) and without gauge corrections (olive dashed dashed). We also show the BS models, up to first order, again with (blue solid curve, $1/4(1-5\eta/2)$) and without gauge corrections (orange dashed curve, $1/4(1+3\eta/2)$). 
    The general trends are clearly qualitatively very good, especially at lower $\eta$, which is precisely when we expect the BS model to be valid.}
    \label{fig:strengths_vs_approx}
\end{figure}

For resonant bare dipole and cavity frequencies however, the BS Hamiltonian gives no corrections to the JC energies to order $\eta^2$ (and thus often finds more utility in describing the dispersive regime of cavity QED), but does correct the eigenstates. To first order in $\eta$, the corrected JC-like states are 
\begin{equation}
    \ket{+} = \frac{1}{\sqrt{2}}\left[(1+\frac{\eta}{4} )\ket{e,0} + i(1 - \frac{\eta}{4} )\ket{g,1}\right],
\end{equation}
\begin{equation}
    \ket{-} = \frac{1}{\sqrt{2}}\left[(1-\frac{\eta}{4})\ket{e,0} - i(1 + \frac{\eta}{4})\ket{g,1}\right],
\end{equation}
which in conjunction with the ground state $\ket{G} \equiv \ket{g,0}$ gives the three states considered in the WEA. {Note that in this section, we use a notation where the system states are identified by their composition in the \emph{transformed} frame.}


The effect of the counter-rotating terms eliminated in the BS transformation can be quantified by considering the transition matrix elements of the quadrature operator $\Pi$ which we use to couple to the external reservoir fields. As in the main text, we use $\Pi_C$ to refer to the uncorrected operator in the dipole gauge (which is equivalent in form to the Coulomb gauge operator) corresponding to the electric field quadrature mode operator. Performing the BS transformation, we find $\Pi_C = i(a^{\dagger} -a) \rightarrow \mathcal{P} =  \frac{1}{\sqrt{2}}\left[i(a^{\dagger}-a) - \eta/2\sigma_x\right]$, and so $x^+ = -ia - \eta/2\sigma^-$, where $x^+$ is the ``positive frequency'' (taking higher energy states to lower energy ones) component of the transformed operator $\mathcal{P} = \frac{1}{\sqrt{2}}U^{\dagger} \Pi_C U = \frac{1}{\sqrt{2}}\left(x^+ + x^-\right)$. We introduce the notation $\mathcal{P}$ to reiterate that the operator which we assume to couple the system to the reservoir modes is proportional to the momentum quadrature operator of the cavity mode. With gauge corrections, the correct dipole gauge quadrature operator is instead $\Pi_D = i(a^{\dagger}-a) + 2\eta \sigma_x$, and so to first order in $\eta$, we find $\mathcal{P} \rightarrow \mathcal{P'} = \frac{1}{\sqrt{2}}\left[i(a^\dagger -a) +3\eta/2(\sigma^+ +\sigma^-)\right]$, and $x^+ \rightarrow x^+_{\rm GC} = -ia + 3\eta/2\sigma^-$. 

The transition matrix elements (modulus squared) with respect to 
${\cal P}$ are, with no gauge corrections,
\begin{align}
|{\cal P}_{\pm G}|^2 = |\braket{\pm| {\cal P}|G}|^2 &= \frac{1}{4}(1\mp 3\eta/2)+\mathcal{O}(\eta^2). 
\end{align}
However, with gauge corrections, we have
\begin{align}
|{\cal P}'_{\pm G}|^2 = |\braket{\pm| {\cal P}'|G}| &= \frac{1}{4}(1\pm 5\eta/2)+\mathcal{O}(\eta^2),
\end{align}
and we can infer immediately, that the linewidth asymmetry  will change with gauge correction.



As shown in Fig.~\ref{fig:strengths_vs_approx}, 
the leading order effect in $\eta$ of the effect of gauge corrections is in excellent agreement with the numerical solution (Fig.~\ref{fig:evals}(c) of main text), for the perturbative regime ($\eta \lesssim 0.2$).  For higher values
of $\eta$, then the numerically
exact $|\braket{G|{\cal P}|j}|^2$
with and without gauge corrections explain the main features of the spectra, especially the different linewidths as a function of $\eta$,
and how these drastically differ with gauge correction. Below we explain why, to the same order of approximations, that the change in linewidth is directly proportional to the weights of the spectral peaks in the spectra.

Solving the relevant Bloch equation with weak excitations, then the spectral linewidths (full widths at half maxima) of the first two excited states are, in the SC limit $g/\kappa \gg 1$,  simply given by the projections above multiplied
by $2\kappa$.
This is the primary effect
for the observed asymmetry 
for increasing $\eta$ (as we can also see from the full numerical calculations).
Thus, even in the perturbative BS regime, the asymmetry stemming from the
counter rotating wave effects is qualitatively different when one properly accounts for gauge corrections. We justify this assumption in more detail below.

Considering the effect of the BS transformation to first order in in $\eta$,  the relevant Bloch equations are 
\begin{equation}
\dot{\rho}_{\pm} = -\Gamma_{\pm}^{({\rm GC})}\rho_{\pm} + \Gamma_c(\rho_{+-} + \rho_{-+}) + \mathcal{E}\Gamma_{\pm}^{({\rm GC})} \rho_G, 
\end{equation}
\begin{equation}
    \dot{\rho}_{+-} = -2(ig + \Gamma_c)\rho_{+-} + \Gamma_c(\rho_+ + \rho_-)  - 2\mathcal{E}\Gamma_c \rho_G,
\end{equation}
\begin{equation}
    \dot{\rho}_{\pm G} = -\left (\frac{\Gamma_{\pm}^{({\rm GC})}}{2} + i(\omega_c \pm g)\right)\rho_{\pm G} + \Gamma_c \rho_{\mp G},
\end{equation}
where $\rho_m = \bra{m}\rho\ket{m}$, $\rho_{ij} = \bra{i}\rho\ket{j}$, $\Gamma_c = \kappa/4$, and $\mathcal{E} = P_{\rm inc}/\kappa$. The polariton decay rates are $ \Gamma_{\pm} = \frac{\kappa}{2}(1 \mp \frac{3}{2} \eta)$ without considering the dipole gauge correction, and $ \Gamma_{\pm}^{{\rm GC}} = \frac{\kappa}{2}(1 \pm \frac{5}{2} \eta)$ with the correction. We require $\mathcal{E} \ll 1$ for the WEA to be a valid approximation.

Within the WEA, it is possible to find an analytic expression for the emission spectrum $S_{\rm cav}$ that is valid perturbatively up to order $\eta$ by solving the above Bloch equations derived from the BS Hamiltonian. In the strong coupling regime, this spectrum takes on a particularly simple form, which is useful to gain qualitative insight into the spectral asymmetries which are shown to arise in our main results. In particular, to leading order in $\mathcal{E}$, the steady state solutions to the Bloch equations give the very simple solution $\rho_+ = \rho_- = (1 - \rho_G)/2= \mathcal{E}$, with all other matrix elements zero.

The steady-state cavity spectrum with incoherent driving is 
\begin{equation}
    S_{\rm cav}(\omega) \propto \text{Re}\Big[\int_0^{\infty}d\tau e^{i\omega \tau} \langle x^{-} x^+(\tau)\rangle \Big].
\end{equation}
The steady-state correlation function $\langle x^{-} x^+(\tau)\rangle$ can be calculated with the QRT, and the result for the spectrum after Fourier transforming is  
\begin{equation}
S_{\rm cav}(\omega) \propto \text{Re}\Bigg[\sqrt{\Gamma_+^{({\rm GC})}}\tilde{\chi}_{+G}(\omega) - \sqrt{\Gamma_-^{({\rm GC})}}\tilde{\chi}_{-G}(\omega)\Bigg],
\end{equation}
where
\begin{equation}
    \tilde{\chi}_{\pm G}(\omega) = \pm \mathcal{E}\frac{ -\Gamma_c \sqrt{\Gamma_{\mp}^{({\rm GC})}} + \sqrt{\Gamma_{\pm}^{({\rm GC})}}\left[ i(\omega_0 \mp g - \omega) + \frac{\Gamma_{\mp}}{2}\right]}{\left(i(\omega_0+G-\omega) + \frac{\kappa}{4}\right)\left(i(\omega_0-G-\omega) + \frac{\kappa}{4}\right)},
\end{equation}
where 
\begin{equation}
    G = \sqrt{g^2 - \left(\frac{\kappa}{4}\right)^2 - ig\frac{\Gamma_+ - \Gamma_-}{2}}.
\end{equation}


Much simplification can be made if we assume $\alpha \equiv g/\kappa$ is large, and neglect terms of order $1/\alpha^2$ Then, $G \approx g - i(\Gamma_+^{({\rm GC})} - \Gamma_-^{({\rm GC})})/4$, and we find 
\begin{equation}
    \tilde{\chi}_{\pm G}(\omega) \approx \pm \mathcal{E} \frac{ \sqrt{\Gamma_{\pm}^{({\rm GC})}} }{i(\omega_0 \pm g - \omega) + \frac{\Gamma_{\pm}^{({\rm GC})}}{2}},
\end{equation}
and
\begin{equation}\label{eq:spectra}
    S_{\rm cav}(\omega) \approx \frac{\Gamma_+^{2({\rm GC})}}{(\omega-\omega_0 -g)^2 + \frac{\Gamma_+^{2({\rm GC})}}{4}} + \frac{\Gamma_-^{2({\rm GC})}}{(\omega-\omega_0 +g)^2 + \frac{\Gamma_-^{2({\rm GC})}}{4}}.
\end{equation}
%

\begin{figure}[bh]
    \centering
    \includegraphics[width=1\linewidth]{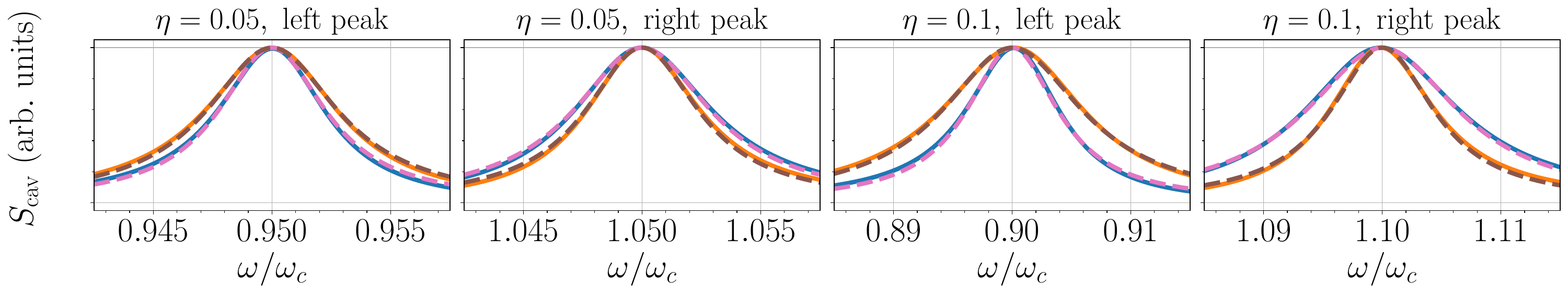}
    \caption{A zoom in of the full numerical spectra (solid curves), with (blue solid curve) and without (orange solid curve) the gauge correction, compared with the analytical solution in Eq.~\eqref{eq:spectra} (dashed curves), again with (pink dashed curve) and without (brown dashed curve) the gauge correction. Parameters are the same as in the main text, with $\kappa=0.25g$, though we use a slightly smaller driving strength to ensure the WEA remains valid, $P_{\rm inc}=0.00025g$.}
    \label{fig:spectra2}
\end{figure}

Within this approximation (SC, first order $\eta$ corrections, and weak excitation), the two polariton peaks have the same height, and have a ratio of peak areas $A_+/A_- = \Gamma_+^{({\rm GC})}/\Gamma_-^{({\rm GC})}$. Without gauge correction, this ratio is $\sim 1 - 3 \eta + \mathcal{O}(\eta^2)$, and with corrections it is $\sim 1 + 5 \eta + \mathcal{O}(\eta^2)$, which quantifies the change in asymmetry to leading order. We can understand this asymmetry on physical grounds as arising from which polariton branch is more cavity-like: In the BS frame,  the BS shift causes a detuning between cavity and TLS resonances, which leads to cavity-like and atom-like polariton branches. In the WEA, both polariton branches become equally populated, and thus, in the SC regime, their spectral weights are determined by the transition matrix elements $\mathcal{P}_{G \pm}$, or in other words, how much the operator which couples the cavity field to decay channel modes also couples the polariton-ground transition. The more cavity-like transition experiences a larger decay rate, but which branch this corresponds to is dependent on both the gauge corrections and the BS frame transformation. The interplay of these effects thus gives the overall asymmetry.

Finally, to confirm the accuracy of the analytical formula using the same material parameters as in the main text, 
we show a zoom in of the two main polariton peaks
(near $\omega_{c}\pm g$) using the full numerical solution
versus the simple analytical formula in Fig.~\ref{fig:spectra}, using $\eta=0.05$
and $\eta=0.1$. Clearly the comparison is qualitatively excellent and the main differences with gauge corrections stem from the changing linewidth.


\begin{thebibliography}{66}%
\makeatletter
\providecommand \@ifxundefined [1]{%
 \@ifx{#1\undefined}
}%
\providecommand \@ifnum [1]{%
 \ifnum #1\expandafter \@firstoftwo
 \else \expandafter \@secondoftwo
 \fi
}%
\providecommand \@ifx [1]{%
 \ifx #1\expandafter \@firstoftwo
 \else \expandafter \@secondoftwo
 \fi
}%
\providecommand \natexlab [1]{#1}%
\providecommand \enquote  [1]{``#1''}%
\providecommand \bibnamefont  [1]{#1}%
\providecommand \bibfnamefont [1]{#1}%
\providecommand \citenamefont [1]{#1}%
\providecommand \href@noop [0]{\@secondoftwo}%
\providecommand \href [0]{\begingroup \@sanitize@url \@href}%
\providecommand \@href[1]{\@@startlink{#1}\@@href}%
\providecommand \@@href[1]{\endgroup#1\@@endlink}%
\providecommand \@sanitize@url [0]{\catcode `\\12\catcode `\$12\catcode
  `\&12\catcode `\#12\catcode `\^12\catcode `\_12\catcode `\%12\relax}%
\providecommand \@@startlink[1]{}%
\providecommand \@@endlink[0]{}%
\providecommand \url  [0]{\begingroup\@sanitize@url \@url }%
\providecommand \@url [1]{\endgroup\@href {#1}{\urlprefix }}%
\providecommand \urlprefix  [0]{URL }%
\providecommand \Eprint [0]{\href }%
\providecommand \doibase [0]{https://doi.org/}%
\providecommand \selectlanguage [0]{\@gobble}%
\providecommand \bibinfo  [0]{\@secondoftwo}%
\providecommand \bibfield  [0]{\@secondoftwo}%
\providecommand \translation [1]{[#1]}%
\providecommand \BibitemOpen [0]{}%
\providecommand \bibitemStop [0]{}%
\providecommand \bibitemNoStop [0]{.\EOS\space}%
\providecommand \EOS [0]{\spacefactor3000\relax}%
\providecommand \BibitemShut  [1]{\csname bibitem#1\endcsname}%
\let\auto@bib@innerbib\@empty
\bibitem [{\citenamefont {Salter}\ \emph {et~al.}(2010)\citenamefont {Salter},
  \citenamefont {Stevenson}, \citenamefont {Farrer}, \citenamefont {Nicoll},
  \citenamefont {Ritchie},\ and\ \citenamefont
  {Shields}}]{salter_entangled-light-emitting_2010}%
  \BibitemOpen
  \bibfield  {author} {\bibinfo {author} {\bibfnamefont {C.~L.}\ \bibnamefont
  {Salter}}, \bibinfo {author} {\bibfnamefont {R.~M.}\ \bibnamefont
  {Stevenson}}, \bibinfo {author} {\bibfnamefont {I.}~\bibnamefont {Farrer}},
  \bibinfo {author} {\bibfnamefont {C.~A.}\ \bibnamefont {Nicoll}}, \bibinfo
  {author} {\bibfnamefont {D.~A.}\ \bibnamefont {Ritchie}},\ and\ \bibinfo
  {author} {\bibfnamefont {A.~J.}\ \bibnamefont {Shields}},\ }\bibfield
  {title} {{\selectlanguage {en}\bibinfo {title} {An entangled-light-emitting
  diode}},\ }\href {https://doi.org/10.1038/nature09078} {\bibfield  {journal}
  {\bibinfo  {journal} {Nature}\ }\textbf {\bibinfo {volume} {465}},\ \bibinfo
  {pages} {594} (\bibinfo {year} {2010})}\BibitemShut {NoStop}%
\bibitem [{\citenamefont {Somaschi}\ \emph {et~al.}(2016)\citenamefont
  {Somaschi}, \citenamefont {Giesz}, \citenamefont {De~Santis}, \citenamefont
  {Loredo}, \citenamefont {Almeida}, \citenamefont {Hornecker}, \citenamefont
  {Portalupi}, \citenamefont {Grange}, \citenamefont {Ant{\'o}n}, \citenamefont
  {Demory}, \citenamefont {G{\'o}mez}, \citenamefont {Sagnes}, \citenamefont
  {Lanzillotti-Kimura}, \citenamefont {Lema{\'i}tre}, \citenamefont {Auffeves},
  \citenamefont {White}, \citenamefont {Lanco},\ and\ \citenamefont
  {Senellart}}]{somaschi_near-optimal_2016}%
  \BibitemOpen
  \bibfield  {author} {\bibinfo {author} {\bibfnamefont {N.}~\bibnamefont
  {Somaschi}}, \bibinfo {author} {\bibfnamefont {V.}~\bibnamefont {Giesz}},
  \bibinfo {author} {\bibfnamefont {L.}~\bibnamefont {De~Santis}}, \bibinfo
  {author} {\bibfnamefont {J.~C.}\ \bibnamefont {Loredo}}, \bibinfo {author}
  {\bibfnamefont {M.~P.}\ \bibnamefont {Almeida}}, \bibinfo {author}
  {\bibfnamefont {G.}~\bibnamefont {Hornecker}}, \bibinfo {author}
  {\bibfnamefont {S.~L.}\ \bibnamefont {Portalupi}}, \bibinfo {author}
  {\bibfnamefont {T.}~\bibnamefont {Grange}}, \bibinfo {author} {\bibfnamefont
  {C.}~\bibnamefont {Ant{\'o}n}}, \bibinfo {author} {\bibfnamefont
  {J.}~\bibnamefont {Demory}}, \bibinfo {author} {\bibfnamefont
  {C.}~\bibnamefont {G{\'o}mez}}, \bibinfo {author} {\bibfnamefont
  {I.}~\bibnamefont {Sagnes}}, \bibinfo {author} {\bibfnamefont {N.~D.}\
  \bibnamefont {Lanzillotti-Kimura}}, \bibinfo {author} {\bibfnamefont
  {A.}~\bibnamefont {Lema{\'i}tre}}, \bibinfo {author} {\bibfnamefont
  {A.}~\bibnamefont {Auffeves}}, \bibinfo {author} {\bibfnamefont {A.~G.}\
  \bibnamefont {White}}, \bibinfo {author} {\bibfnamefont {L.}~\bibnamefont
  {Lanco}},\ and\ \bibinfo {author} {\bibfnamefont {P.}~\bibnamefont
  {Senellart}},\ }\bibfield  {title} {{\selectlanguage {en}\bibinfo {title}
  {Near-optimal single-photon sources in the solid state}},\ }\href
  {https://doi.org/10.1038/nphoton.2016.23} {\bibfield  {journal} {\bibinfo
  {journal} {Nature Photonics}\ }\textbf {\bibinfo {volume} {10}},\ \bibinfo
  {pages} {340} (\bibinfo {year} {2016})}\BibitemShut {NoStop}%
\bibitem [{\citenamefont {Senellart}\ \emph {et~al.}(2017)\citenamefont
  {Senellart}, \citenamefont {Solomon},\ and\ \citenamefont
  {White}}]{senellart_2017}%
  \BibitemOpen
  \bibfield  {author} {\bibinfo {author} {\bibfnamefont {P.}~\bibnamefont
  {Senellart}}, \bibinfo {author} {\bibfnamefont {G.}~\bibnamefont {Solomon}},\
  and\ \bibinfo {author} {\bibfnamefont {G.}~\bibnamefont {White}},\ }\bibfield
   {title} {\bibinfo {title} {High-performance semiconductor quantum-dot
  single-photon sources},\ }\href {https://doi.org/10.1038/nnano.2017.218}
  {\bibfield  {journal} {\bibinfo  {journal} {Nature Nanotech.}\ }\textbf
  {\bibinfo {volume} {12}},\ \bibinfo {pages} {1026} (\bibinfo {year}
  {2017})}\BibitemShut {NoStop}%
\bibitem [{\citenamefont {Tomm}\ \emph {et~al.}(2021)\citenamefont {Tomm},
  \citenamefont {Javadi}, \citenamefont {Antoniadis}, \citenamefont {Najer},
  \citenamefont {L{\" o}bl}, \citenamefont {Korsch}, \citenamefont {Schott},
  \citenamefont {Valentin}, \citenamefont {Wieck}, \citenamefont {Ludwig},\
  and\ \citenamefont {Warburton}}]{tomm_2021}%
  \BibitemOpen
  \bibfield  {author} {\bibinfo {author} {\bibfnamefont {N.}~\bibnamefont
  {Tomm}}, \bibinfo {author} {\bibfnamefont {A.}~\bibnamefont {Javadi}},
  \bibinfo {author} {\bibfnamefont {N.~O.}\ \bibnamefont {Antoniadis}},
  \bibinfo {author} {\bibfnamefont {D.}~\bibnamefont {Najer}}, \bibinfo
  {author} {\bibfnamefont {M.~C.}\ \bibnamefont {L{\" o}bl}}, \bibinfo {author}
  {\bibfnamefont {A.~R.}\ \bibnamefont {Korsch}}, \bibinfo {author}
  {\bibfnamefont {R.}~\bibnamefont {Schott}}, \bibinfo {author} {\bibfnamefont
  {S.~R.}\ \bibnamefont {Valentin}}, \bibinfo {author} {\bibfnamefont {A.~D.}\
  \bibnamefont {Wieck}}, \bibinfo {author} {\bibfnamefont {A.}~\bibnamefont
  {Ludwig}},\ and\ \bibinfo {author} {\bibfnamefont {R.~J.}\ \bibnamefont
  {Warburton}},\ }\bibfield  {title} {\bibinfo {title} {A bright and fast
  source of coherent single photons},\ }\href
  {https://doi.org/10.1038/s41565-020-00831-x} {\bibfield  {journal} {\bibinfo
  {journal} {Nature Nanotech.}\ }\textbf {\bibinfo {volume} {16}},\ \bibinfo
  {pages} {399} (\bibinfo {year} {2021})}\BibitemShut {NoStop}%
\bibitem [{\citenamefont {Buluta}\ \emph {et~al.}(2011)\citenamefont {Buluta},
  \citenamefont {Ashhab},\ and\ \citenamefont {Nori}}]{buluta2011natural}%
  \BibitemOpen
  \bibfield  {author} {\bibinfo {author} {\bibfnamefont {I.}~\bibnamefont
  {Buluta}}, \bibinfo {author} {\bibfnamefont {S.}~\bibnamefont {Ashhab}},\
  and\ \bibinfo {author} {\bibfnamefont {F.}~\bibnamefont {Nori}},\ }\bibfield
  {title} {\bibinfo {title} {Natural and artificial atoms for quantum
  computation},\ }\href@noop {} {\bibfield  {journal} {\bibinfo  {journal}
  {Reports on Progress in Physics}\ }\textbf {\bibinfo {volume} {74}},\
  \bibinfo {pages} {104401} (\bibinfo {year} {2011})}\BibitemShut {NoStop}%
\bibitem [{\citenamefont {Georgescu}\ and\ \citenamefont
  {Nori}(2012)}]{georgescu2012quantum}%
  \BibitemOpen
  \bibfield  {author} {\bibinfo {author} {\bibfnamefont {I.}~\bibnamefont
  {Georgescu}}\ and\ \bibinfo {author} {\bibfnamefont {F.}~\bibnamefont
  {Nori}},\ }\bibfield  {title} {\bibinfo {title} {Quantum technologies: an old
  new story},\ }\href@noop {} {\bibfield  {journal} {\bibinfo  {journal}
  {Physics World}\ }\textbf {\bibinfo {volume} {25}},\ \bibinfo {pages} {16}
  (\bibinfo {year} {2012})}\BibitemShut {NoStop}%
\bibitem [{\citenamefont {Ciuti}\ \emph {et~al.}(2005)\citenamefont {Ciuti},
  \citenamefont {Bastard},\ and\ \citenamefont
  {Carusotto}}]{ciuti_quantum_2005}%
  \BibitemOpen
  \bibfield  {author} {\bibinfo {author} {\bibfnamefont {C.}~\bibnamefont
  {Ciuti}}, \bibinfo {author} {\bibfnamefont {G.}~\bibnamefont {Bastard}},\
  and\ \bibinfo {author} {\bibfnamefont {I.}~\bibnamefont {Carusotto}},\
  }\bibfield  {title} {\bibinfo {title} {Quantum vacuum properties of the
  intersubband cavity polariton field},\ }\href
  {https://doi.org/10.1103/PhysRevB.72.115303} {\bibfield  {journal} {\bibinfo
  {journal} {Physical Review B}\ }\textbf {\bibinfo {volume} {72}},\ \bibinfo
  {pages} {115303} (\bibinfo {year} {2005})}\BibitemShut {NoStop}%
\bibitem [{\citenamefont {Anappara}\ \emph {et~al.}(2009)\citenamefont
  {Anappara}, \citenamefont {De~Liberato}, \citenamefont {Tredicucci},
  \citenamefont {Ciuti}, \citenamefont {Biasiol}, \citenamefont {Sorba},\ and\
  \citenamefont {Beltram}}]{anappara_signatures_2009}%
  \BibitemOpen
  \bibfield  {author} {\bibinfo {author} {\bibfnamefont {A.~A.}\ \bibnamefont
  {Anappara}}, \bibinfo {author} {\bibfnamefont {S.}~\bibnamefont
  {De~Liberato}}, \bibinfo {author} {\bibfnamefont {A.}~\bibnamefont
  {Tredicucci}}, \bibinfo {author} {\bibfnamefont {C.}~\bibnamefont {Ciuti}},
  \bibinfo {author} {\bibfnamefont {G.}~\bibnamefont {Biasiol}}, \bibinfo
  {author} {\bibfnamefont {L.}~\bibnamefont {Sorba}},\ and\ \bibinfo {author}
  {\bibfnamefont {F.}~\bibnamefont {Beltram}},\ }\bibfield  {title} {\bibinfo
  {title} {Signatures of the ultrastrong light-matter coupling regime},\ }\href
  {https://doi.org/10.1103/PhysRevB.79.201303} {\bibfield  {journal} {\bibinfo
  {journal} {Physical Review B}\ }\textbf {\bibinfo {volume} {79}},\ \bibinfo
  {pages} {201303(R)} (\bibinfo {year} {2009})}\BibitemShut {NoStop}%
\bibitem [{\citenamefont {Zaks}\ \emph {et~al.}(2011)\citenamefont {Zaks},
  \citenamefont {Stehr}, \citenamefont {Truong}, \citenamefont {Petroff},
  \citenamefont {Hughes},\ and\ \citenamefont
  {Sherwin}}]{zaks_thz-driven_2011}%
  \BibitemOpen
  \bibfield  {author} {\bibinfo {author} {\bibfnamefont {B.}~\bibnamefont
  {Zaks}}, \bibinfo {author} {\bibfnamefont {D.}~\bibnamefont {Stehr}},
  \bibinfo {author} {\bibfnamefont {T.-A.}\ \bibnamefont {Truong}}, \bibinfo
  {author} {\bibfnamefont {P.~M.}\ \bibnamefont {Petroff}}, \bibinfo {author}
  {\bibfnamefont {S.}~\bibnamefont {Hughes}},\ and\ \bibinfo {author}
  {\bibfnamefont {M.~S.}\ \bibnamefont {Sherwin}},\ }\bibfield  {title}
  {{\selectlanguage {en}\bibinfo {title} {{THz}-driven quantum wells: {Coulomb}
  interactions and {Stark} shifts in the ultrastrong coupling regime}},\ }\href
  {https://doi.org/10.1088/1367-2630/13/8/083009} {\bibfield  {journal}
  {\bibinfo  {journal} {New Journal of Physics}\ }\textbf {\bibinfo {volume}
  {13}},\ \bibinfo {pages} {083009} (\bibinfo {year} {2011})}\BibitemShut
  {NoStop}%
\bibitem [{\citenamefont {Hughes}(1998)}]{hughes_breakdown_1998}%
  \BibitemOpen
  \bibfield  {author} {\bibinfo {author} {\bibfnamefont {S.}~\bibnamefont
  {Hughes}},\ }\bibfield  {title} {\bibinfo {title} {Breakdown of the {Area}
  {Theorem}: {Carrier}-{Wave} {Rabi} {Flopping} of {Femtosecond} {Optical}
  {Pulses}},\ }\href {https://doi.org/10.1103/PhysRevLett.81.3363} {\bibfield
  {journal} {\bibinfo  {journal} {Physical Review Letters}\ }\textbf {\bibinfo
  {volume} {81}},\ \bibinfo {pages} {3363} (\bibinfo {year}
  {1998})}\BibitemShut {NoStop}%
\bibitem [{\citenamefont {M\"{u}cke}\ \emph {et~al.}(2001)\citenamefont
  {M\"{u}cke}, \citenamefont {Tritschler}, \citenamefont {Wegener},
  \citenamefont {Morgner},\ and\ \citenamefont
  {K{\"a}rtner}}]{mucke_signatures_2001}%
  \BibitemOpen
  \bibfield  {author} {\bibinfo {author} {\bibfnamefont {O.~D.}\ \bibnamefont
  {M\"{u}cke}}, \bibinfo {author} {\bibfnamefont {T.}~\bibnamefont
  {Tritschler}}, \bibinfo {author} {\bibfnamefont {M.}~\bibnamefont {Wegener}},
  \bibinfo {author} {\bibfnamefont {U.}~\bibnamefont {Morgner}},\ and\ \bibinfo
  {author} {\bibfnamefont {F.~X.}\ \bibnamefont {K{\"a}rtner}},\ }\bibfield
  {title} {\bibinfo {title} {Signatures of {Carrier}-{Wave} {Rabi} {Flopping}
  in {GaAs}},\ }\href {https://doi.org/10.1103/PhysRevLett.87.057401}
  {\bibfield  {journal} {\bibinfo  {journal} {Physical Review Letters}\
  }\textbf {\bibinfo {volume} {87}},\ \bibinfo {pages} {057401} (\bibinfo
  {year} {2001})}\BibitemShut {NoStop}%
\bibitem [{\citenamefont {Ciappina}\ \emph {et~al.}(2015)\citenamefont
  {Ciappina}, \citenamefont {P{\'e}rez-Hern{\'a}ndez}, \citenamefont
  {Landsman}, \citenamefont {Zimmermann}, \citenamefont {Lewenstein},
  \citenamefont {Roso},\ and\ \citenamefont
  {Krausz}}]{ciappina_carrier-wave_2015}%
  \BibitemOpen
  \bibfield  {author} {\bibinfo {author} {\bibfnamefont {M.~F.}\ \bibnamefont
  {Ciappina}}, \bibinfo {author} {\bibfnamefont {J.~A.}\ \bibnamefont
  {P{\'e}rez-Hern{\'a}ndez}}, \bibinfo {author} {\bibfnamefont {A.~S.}\
  \bibnamefont {Landsman}}, \bibinfo {author} {\bibfnamefont {T.}~\bibnamefont
  {Zimmermann}}, \bibinfo {author} {\bibfnamefont {M.}~\bibnamefont
  {Lewenstein}}, \bibinfo {author} {\bibfnamefont {L.}~\bibnamefont {Roso}},\
  and\ \bibinfo {author} {\bibfnamefont {F.}~\bibnamefont {Krausz}},\
  }\bibfield  {title} {\bibinfo {title} {Carrier-{Wave} {Rabi}-{Flopping}
  {Signatures} in {High}-{Order} {Harmonic} {Generation} for {Alkali}
  {Atoms}},\ }\href {https://doi.org/10.1103/PhysRevLett.114.143902} {\bibfield
   {journal} {\bibinfo  {journal} {Physical Review Letters}\ }\textbf {\bibinfo
  {volume} {114}},\ \bibinfo {pages} {143902} (\bibinfo {year}
  {2015})}\BibitemShut {NoStop}%
\bibitem [{\citenamefont {Frisk~Kockum}\ \emph {et~al.}(2019)\citenamefont
  {Frisk~Kockum}, \citenamefont {Miranowicz}, \citenamefont {De~Liberato},
  \citenamefont {Savasta},\ and\ \citenamefont
  {Nori}}]{frisk_kockum_ultrastrong_2019}%
  \BibitemOpen
  \bibfield  {author} {\bibinfo {author} {\bibfnamefont {A.}~\bibnamefont
  {Frisk~Kockum}}, \bibinfo {author} {\bibfnamefont {A.}~\bibnamefont
  {Miranowicz}}, \bibinfo {author} {\bibfnamefont {S.}~\bibnamefont
  {De~Liberato}}, \bibinfo {author} {\bibfnamefont {S.}~\bibnamefont
  {Savasta}},\ and\ \bibinfo {author} {\bibfnamefont {F.}~\bibnamefont
  {Nori}},\ }\bibfield  {title} {\bibinfo {title} {Ultrastrong coupling between
  light and matter},\ }\href {https://doi.org/10.1038/s42254-018-0006-2}
  {\bibfield  {journal} {\bibinfo  {journal} {Nature Reviews Physics}\ }\textbf
  {\bibinfo {volume} {1}},\ \bibinfo {pages} {19} (\bibinfo {year}
  {2019})}\BibitemShut {NoStop}%
\bibitem [{\citenamefont {Forn-D{\'i}az}\ \emph {et~al.}(2019)\citenamefont
  {Forn-D{\'i}az}, \citenamefont {Lamata}, \citenamefont {Rico}, \citenamefont
  {Kono},\ and\ \citenamefont {Solano}}]{forn-diaz_ultrastrong_2019}%
  \BibitemOpen
  \bibfield  {author} {\bibinfo {author} {\bibfnamefont {P.}~\bibnamefont
  {Forn-D{\'i}az}}, \bibinfo {author} {\bibfnamefont {L.}~\bibnamefont
  {Lamata}}, \bibinfo {author} {\bibfnamefont {E.}~\bibnamefont {Rico}},
  \bibinfo {author} {\bibfnamefont {J.}~\bibnamefont {Kono}},\ and\ \bibinfo
  {author} {\bibfnamefont {E.}~\bibnamefont {Solano}},\ }\bibfield  {title}
  {\bibinfo {title} {Ultrastrong coupling regimes of light-matter
  interaction},\ }\href {https://doi.org/10.1103/RevModPhys.91.025005}
  {\bibfield  {journal} {\bibinfo  {journal} {Reviews of Modern Physics}\
  }\textbf {\bibinfo {volume} {91}},\ \bibinfo {pages} {025005} (\bibinfo
  {year} {2019})}\BibitemShut {NoStop}%
\bibitem [{\citenamefont {Mueller}\ \emph {et~al.}(2020)\citenamefont
  {Mueller}, \citenamefont {Okamura}, \citenamefont {Vieira}, \citenamefont
  {Juergensen}, \citenamefont {Lange}, \citenamefont {Barros}, \citenamefont
  {Schulz},\ and\ \citenamefont {Reich}}]{mueller_deep_2020}%
  \BibitemOpen
  \bibfield  {author} {\bibinfo {author} {\bibfnamefont {N.~S.}\ \bibnamefont
  {Mueller}}, \bibinfo {author} {\bibfnamefont {Y.}~\bibnamefont {Okamura}},
  \bibinfo {author} {\bibfnamefont {B.~G.~M.}\ \bibnamefont {Vieira}}, \bibinfo
  {author} {\bibfnamefont {S.}~\bibnamefont {Juergensen}}, \bibinfo {author}
  {\bibfnamefont {H.}~\bibnamefont {Lange}}, \bibinfo {author} {\bibfnamefont
  {E.~B.}\ \bibnamefont {Barros}}, \bibinfo {author} {\bibfnamefont
  {F.}~\bibnamefont {Schulz}},\ and\ \bibinfo {author} {\bibfnamefont
  {S.}~\bibnamefont {Reich}},\ }\bibfield  {title} {{\selectlanguage
  {en}\bibinfo {title} {Deep strong light–matter coupling in plasmonic
  nanoparticle crystals}},\ }\href {https://doi.org/10.1038/s41586-020-2508-1}
  {\bibfield  {journal} {\bibinfo  {journal} {Nature}\ }\textbf {\bibinfo
  {volume} {583}},\ \bibinfo {pages} {780} (\bibinfo {year}
  {2020})}\BibitemShut {NoStop}%
\bibitem [{\citenamefont {Ashhab}\ and\ \citenamefont
  {Nori}(2010)}]{PhysRevA.81.042311}%
  \BibitemOpen
  \bibfield  {author} {\bibinfo {author} {\bibfnamefont {S.}~\bibnamefont
  {Ashhab}}\ and\ \bibinfo {author} {\bibfnamefont {F.}~\bibnamefont {Nori}},\
  }\bibfield  {title} {\bibinfo {title} {Qubit-oscillator systems in the
  ultrastrong-coupling regime and their potential for preparing nonclassical
  states},\ }\href {https://doi.org/10.1103/PhysRevA.81.042311} {\bibfield
  {journal} {\bibinfo  {journal} {Phys. Rev. A}\ }\textbf {\bibinfo {volume}
  {81}},\ \bibinfo {pages} {042311} (\bibinfo {year} {2010})}\BibitemShut
  {NoStop}%
\bibitem [{\citenamefont {Ashida}\ \emph {et~al.}(2021)\citenamefont {Ashida},
  \citenamefont {\ifmmode \dot{I}\else \.{I}\fi{}mamo\ifmmode~\breve{g}\else
  \u{g}\fi{}lu},\ and\ \citenamefont {Demler}}]{PhysRevLett.126.153603}%
  \BibitemOpen
  \bibfield  {author} {\bibinfo {author} {\bibfnamefont {Y.}~\bibnamefont
  {Ashida}}, \bibinfo {author} {\bibfnamefont {A.}~\bibnamefont {\ifmmode
  \dot{I}\else \.{I}\fi{}mamo\ifmmode~\breve{g}\else \u{g}\fi{}lu}},\ and\
  \bibinfo {author} {\bibfnamefont {E.}~\bibnamefont {Demler}},\ }\bibfield
  {title} {\bibinfo {title} {Cavity quantum electrodynamics at arbitrary
  light-matter coupling strengths},\ }\href
  {https://doi.org/10.1103/PhysRevLett.126.153603} {\bibfield  {journal}
  {\bibinfo  {journal} {Phys. Rev. Lett.}\ }\textbf {\bibinfo {volume} {126}},\
  \bibinfo {pages} {153603} (\bibinfo {year} {2021})}\BibitemShut {NoStop}%
\bibitem [{\citenamefont {Herrera}\ and\ \citenamefont
  {Spano}(2016)}]{herrera_cavity-controlled_2016}%
  \BibitemOpen
  \bibfield  {author} {\bibinfo {author} {\bibfnamefont {F.}~\bibnamefont
  {Herrera}}\ and\ \bibinfo {author} {\bibfnamefont {F.~C.}\ \bibnamefont
  {Spano}},\ }\bibfield  {title} {\bibinfo {title} {Cavity-{Controlled}
  {Chemistry} in {Molecular} {Ensembles}},\ }\href
  {https://doi.org/10.1103/PhysRevLett.116.238301} {\bibfield  {journal}
  {\bibinfo  {journal} {Physical Review Letters}\ }\textbf {\bibinfo {volume}
  {116}},\ \bibinfo {pages} {238301} (\bibinfo {year} {2016})}\BibitemShut
  {NoStop}%
\bibitem [{\citenamefont {Scully}\ and\ \citenamefont
  {Zubairy}(1999)}]{scully1999quantum}%
  \BibitemOpen
  \bibfield  {author} {\bibinfo {author} {\bibfnamefont {M.~O.}\ \bibnamefont
  {Scully}}\ and\ \bibinfo {author} {\bibfnamefont {M.~S.}\ \bibnamefont
  {Zubairy}},\ }\href@noop {} {\bibinfo {title} {Quantum optics}} (\bibinfo
  {year} {1999})\BibitemShut {NoStop}%
\bibitem [{\citenamefont {Miller}\ \emph {et~al.}(2005)\citenamefont {Miller},
  \citenamefont {Northup}, \citenamefont {Birnbaum}, \citenamefont {Boca},
  \citenamefont {Boozer},\ and\ \citenamefont {Kimble}}]{miller_trapped_2005}%
  \BibitemOpen
  \bibfield  {author} {\bibinfo {author} {\bibfnamefont {R.}~\bibnamefont
  {Miller}}, \bibinfo {author} {\bibfnamefont {T.~E.}\ \bibnamefont {Northup}},
  \bibinfo {author} {\bibfnamefont {K.~M.}\ \bibnamefont {Birnbaum}}, \bibinfo
  {author} {\bibfnamefont {A.}~\bibnamefont {Boca}}, \bibinfo {author}
  {\bibfnamefont {A.~D.}\ \bibnamefont {Boozer}},\ and\ \bibinfo {author}
  {\bibfnamefont {H.~J.}\ \bibnamefont {Kimble}},\ }\bibfield  {title}
  {{\selectlanguage {en}\bibinfo {title} {Trapped atoms in cavity {QED}:
  coupling quantized light and matter}},\ }\href
  {https://doi.org/10.1088/0953-4075/38/9/007} {\bibfield  {journal} {\bibinfo
  {journal} {Journal of Physics B: Atomic, Molecular and Optical Physics}\
  }\textbf {\bibinfo {volume} {38}},\ \bibinfo {pages} {S551} (\bibinfo {year}
  {2005})}\BibitemShut {NoStop}%
\bibitem [{\citenamefont {Schuster}\ \emph {et~al.}(2008)\citenamefont
  {Schuster}, \citenamefont {Kubanek}, \citenamefont {Fuhrmanek}, \citenamefont
  {Puppe}, \citenamefont {Pinkse}, \citenamefont {Murr},\ and\ \citenamefont
  {Rempe}}]{schuster_nonlinear_2008}%
  \BibitemOpen
  \bibfield  {author} {\bibinfo {author} {\bibfnamefont {I.}~\bibnamefont
  {Schuster}}, \bibinfo {author} {\bibfnamefont {A.}~\bibnamefont {Kubanek}},
  \bibinfo {author} {\bibfnamefont {A.}~\bibnamefont {Fuhrmanek}}, \bibinfo
  {author} {\bibfnamefont {T.}~\bibnamefont {Puppe}}, \bibinfo {author}
  {\bibfnamefont {P.~W.~H.}\ \bibnamefont {Pinkse}}, \bibinfo {author}
  {\bibfnamefont {K.}~\bibnamefont {Murr}},\ and\ \bibinfo {author}
  {\bibfnamefont {G.}~\bibnamefont {Rempe}},\ }\bibfield  {title}
  {{\selectlanguage {en}\bibinfo {title} {Nonlinear spectroscopy of photons
  bound to one atom}},\ }\href {https://doi.org/10.1038/nphys940} {\bibfield
  {journal} {\bibinfo  {journal} {Nature Physics}\ }\textbf {\bibinfo {volume}
  {4}},\ \bibinfo {pages} {382} (\bibinfo {year} {2008})}\BibitemShut {NoStop}%
\bibitem [{\citenamefont {Flick}\ \emph {et~al.}(2017)\citenamefont {Flick},
  \citenamefont {Ruggenthaler}, \citenamefont {Appel},\ and\ \citenamefont
  {Rubio}}]{flick_atoms_2017}%
  \BibitemOpen
  \bibfield  {author} {\bibinfo {author} {\bibfnamefont {J.}~\bibnamefont
  {Flick}}, \bibinfo {author} {\bibfnamefont {M.}~\bibnamefont {Ruggenthaler}},
  \bibinfo {author} {\bibfnamefont {H.}~\bibnamefont {Appel}},\ and\ \bibinfo
  {author} {\bibfnamefont {A.}~\bibnamefont {Rubio}},\ }\bibfield  {title}
  {{\selectlanguage {en}\bibinfo {title} {Atoms and molecules in cavities, from
  weak to strong coupling in quantum-electrodynamics ({QED}) chemistry}},\
  }\href {https://doi.org/10.1073/pnas.1615509114} {\bibfield  {journal}
  {\bibinfo  {journal} {Proceedings of the National Academy of Sciences}\
  }\textbf {\bibinfo {volume} {114}},\ \bibinfo {pages} {3026} (\bibinfo {year}
  {2017})}\BibitemShut {NoStop}%
\bibitem [{\citenamefont {Hamsen}\ \emph {et~al.}(2017)\citenamefont {Hamsen},
  \citenamefont {Tolazzi}, \citenamefont {Wilk},\ and\ \citenamefont
  {Rempe}}]{hamsen_two-photon_2017}%
  \BibitemOpen
  \bibfield  {author} {\bibinfo {author} {\bibfnamefont {C.}~\bibnamefont
  {Hamsen}}, \bibinfo {author} {\bibfnamefont {K.~N.}\ \bibnamefont {Tolazzi}},
  \bibinfo {author} {\bibfnamefont {T.}~\bibnamefont {Wilk}},\ and\ \bibinfo
  {author} {\bibfnamefont {G.}~\bibnamefont {Rempe}},\ }\bibfield  {title}
  {\bibinfo {title} {Two-{Photon} {Blockade} in an {Atom}-{Driven} {Cavity}
  {QED} {System}},\ }\href {https://doi.org/10.1103/PhysRevLett.118.133604}
  {\bibfield  {journal} {\bibinfo  {journal} {Physical Review Letters}\
  }\textbf {\bibinfo {volume} {118}},\ \bibinfo {pages} {133604} (\bibinfo
  {year} {2017})}\BibitemShut {NoStop}%
\bibitem [{\citenamefont {Yoshie}\ \emph {et~al.}(2004)\citenamefont {Yoshie},
  \citenamefont {Scherer}, \citenamefont {Hendrickson}, \citenamefont
  {Khitrova}, \citenamefont {Gibbs}, \citenamefont {Rupper}, \citenamefont
  {Ell}, \citenamefont {Shchekin},\ and\ \citenamefont
  {Deppe}}]{yoshie_vacuum_2004}%
  \BibitemOpen
  \bibfield  {author} {\bibinfo {author} {\bibfnamefont {T.}~\bibnamefont
  {Yoshie}}, \bibinfo {author} {\bibfnamefont {A.}~\bibnamefont {Scherer}},
  \bibinfo {author} {\bibfnamefont {J.}~\bibnamefont {Hendrickson}}, \bibinfo
  {author} {\bibfnamefont {G.}~\bibnamefont {Khitrova}}, \bibinfo {author}
  {\bibfnamefont {H.~M.}\ \bibnamefont {Gibbs}}, \bibinfo {author}
  {\bibfnamefont {G.}~\bibnamefont {Rupper}}, \bibinfo {author} {\bibfnamefont
  {C.}~\bibnamefont {Ell}}, \bibinfo {author} {\bibfnamefont {O.~B.}\
  \bibnamefont {Shchekin}},\ and\ \bibinfo {author} {\bibfnamefont {D.~G.}\
  \bibnamefont {Deppe}},\ }\bibfield  {title} {{\selectlanguage {en}\bibinfo
  {title} {Vacuum {Rabi} splitting with a single quantum dot in a photonic
  crystal nanocavity}},\ }\href {https://doi.org/10.1038/nature03119}
  {\bibfield  {journal} {\bibinfo  {journal} {Nature}\ }\textbf {\bibinfo
  {volume} {432}},\ \bibinfo {pages} {200} (\bibinfo {year}
  {2004})}\BibitemShut {NoStop}%
\bibitem [{\citenamefont {Reithmaier}\ \emph {et~al.}(2004)\citenamefont
  {Reithmaier}, \citenamefont {S{\k e}k}, \citenamefont {L{\"o}ffler},
  \citenamefont {Hofmann}, \citenamefont {Kuhn}, \citenamefont {Reitzenstein},
  \citenamefont {Keldysh}, \citenamefont {Kulakovskii}, \citenamefont
  {Reinecke},\ and\ \citenamefont {Forchel}}]{reithmaier_strong_2004}%
  \BibitemOpen
  \bibfield  {author} {\bibinfo {author} {\bibfnamefont {J.~P.}\ \bibnamefont
  {Reithmaier}}, \bibinfo {author} {\bibfnamefont {G.}~\bibnamefont {S{\k
  e}k}}, \bibinfo {author} {\bibfnamefont {A.}~\bibnamefont {L{\"o}ffler}},
  \bibinfo {author} {\bibfnamefont {C.}~\bibnamefont {Hofmann}}, \bibinfo
  {author} {\bibfnamefont {S.}~\bibnamefont {Kuhn}}, \bibinfo {author}
  {\bibfnamefont {S.}~\bibnamefont {Reitzenstein}}, \bibinfo {author}
  {\bibfnamefont {L.~V.}\ \bibnamefont {Keldysh}}, \bibinfo {author}
  {\bibfnamefont {V.~D.}\ \bibnamefont {Kulakovskii}}, \bibinfo {author}
  {\bibfnamefont {T.~L.}\ \bibnamefont {Reinecke}},\ and\ \bibinfo {author}
  {\bibfnamefont {A.}~\bibnamefont {Forchel}},\ }\bibfield  {title}
  {{\selectlanguage {en}\bibinfo {title} {Strong coupling in a single quantum
  dot–semiconductor microcavity system}},\ }\href
  {https://doi.org/10.1038/nature02969} {\bibfield  {journal} {\bibinfo
  {journal} {Nature}\ }\textbf {\bibinfo {volume} {432}},\ \bibinfo {pages}
  {197} (\bibinfo {year} {2004})}\BibitemShut {NoStop}%
\bibitem [{\citenamefont {Hennessy}\ \emph {et~al.}(2007)\citenamefont
  {Hennessy}, \citenamefont {Badolato}, \citenamefont {Winger}, \citenamefont
  {Gerace}, \citenamefont {Atat{\"u}re}, \citenamefont {Gulde}, \citenamefont
  {F{\"a}lt}, \citenamefont {Hu},\ and\ \citenamefont {Imamo{\u
  g}lu}}]{hennessy_quantum_2007}%
  \BibitemOpen
  \bibfield  {author} {\bibinfo {author} {\bibfnamefont {K.}~\bibnamefont
  {Hennessy}}, \bibinfo {author} {\bibfnamefont {A.}~\bibnamefont {Badolato}},
  \bibinfo {author} {\bibfnamefont {M.}~\bibnamefont {Winger}}, \bibinfo
  {author} {\bibfnamefont {D.}~\bibnamefont {Gerace}}, \bibinfo {author}
  {\bibfnamefont {M.}~\bibnamefont {Atat{\"u}re}}, \bibinfo {author}
  {\bibfnamefont {S.}~\bibnamefont {Gulde}}, \bibinfo {author} {\bibfnamefont
  {S.}~\bibnamefont {F{\"a}lt}}, \bibinfo {author} {\bibfnamefont {E.~L.}\
  \bibnamefont {Hu}},\ and\ \bibinfo {author} {\bibfnamefont {A.}~\bibnamefont
  {Imamo{\u g}lu}},\ }\bibfield  {title} {{\selectlanguage {en}\bibinfo {title}
  {Quantum nature of a strongly coupled single quantum dot–cavity system}},\
  }\href {https://doi.org/10.1038/nature05586} {\bibfield  {journal} {\bibinfo
  {journal} {Nature}\ }\textbf {\bibinfo {volume} {445}},\ \bibinfo {pages}
  {896} (\bibinfo {year} {2007})}\BibitemShut {NoStop}%
\bibitem [{\citenamefont {Bose}\ \emph {et~al.}(2014)\citenamefont {Bose},
  \citenamefont {Cai}, \citenamefont {Choudhury}, \citenamefont {Solomon},\
  and\ \citenamefont {Waks}}]{bose_all-optical_2014}%
  \BibitemOpen
  \bibfield  {author} {\bibinfo {author} {\bibfnamefont {R.}~\bibnamefont
  {Bose}}, \bibinfo {author} {\bibfnamefont {T.}~\bibnamefont {Cai}}, \bibinfo
  {author} {\bibfnamefont {K.~R.}\ \bibnamefont {Choudhury}}, \bibinfo {author}
  {\bibfnamefont {G.~S.}\ \bibnamefont {Solomon}},\ and\ \bibinfo {author}
  {\bibfnamefont {E.}~\bibnamefont {Waks}},\ }\bibfield  {title}
  {{\selectlanguage {en}\bibinfo {title} {All-optical coherent control of
  vacuum {Rabi} oscillations}},\ }\href
  {https://doi.org/10.1038/nphoton.2014.224} {\bibfield  {journal} {\bibinfo
  {journal} {Nature Photonics}\ }\textbf {\bibinfo {volume} {8}},\ \bibinfo
  {pages} {858} (\bibinfo {year} {2014})}\BibitemShut {NoStop}%
\bibitem [{\citenamefont {You}\ and\ \citenamefont
  {Nori}(2011)}]{you2011atomic}%
  \BibitemOpen
  \bibfield  {author} {\bibinfo {author} {\bibfnamefont {J.~Q.}\ \bibnamefont
  {You}}\ and\ \bibinfo {author} {\bibfnamefont {F.}~\bibnamefont {Nori}},\
  }\bibfield  {title} {\bibinfo {title} {Atomic physics and quantum optics
  using superconducting circuits},\ }\href@noop {} {\bibfield  {journal}
  {\bibinfo  {journal} {Nature}\ }\textbf {\bibinfo {volume} {474}},\ \bibinfo
  {pages} {589} (\bibinfo {year} {2011})}\BibitemShut {NoStop}%
\bibitem [{\citenamefont {Beaudoin}\ \emph {et~al.}(2011)\citenamefont
  {Beaudoin}, \citenamefont {Gambetta},\ and\ \citenamefont
  {Blais}}]{beaudoin_dissipation_2011}%
  \BibitemOpen
  \bibfield  {author} {\bibinfo {author} {\bibfnamefont {F.}~\bibnamefont
  {Beaudoin}}, \bibinfo {author} {\bibfnamefont {J.~M.}\ \bibnamefont
  {Gambetta}},\ and\ \bibinfo {author} {\bibfnamefont {A.}~\bibnamefont
  {Blais}},\ }\bibfield  {title} {\bibinfo {title} {Dissipation and ultrastrong
  coupling in circuit {QED}},\ }\href
  {https://doi.org/10.1103/PhysRevA.84.043832} {\bibfield  {journal} {\bibinfo
  {journal} {Physical Review A}\ }\textbf {\bibinfo {volume} {84}},\ \bibinfo
  {pages} {043832} (\bibinfo {year} {2011})}\BibitemShut {NoStop}%
\bibitem [{\citenamefont {Gu}\ \emph {et~al.}(2017)\citenamefont {Gu},
  \citenamefont {Kockum}, \citenamefont {Miranowicz}, \citenamefont {Liu},\
  and\ \citenamefont {Nori}}]{gu_microwave_2017}%
  \BibitemOpen
  \bibfield  {author} {\bibinfo {author} {\bibfnamefont {X.}~\bibnamefont
  {Gu}}, \bibinfo {author} {\bibfnamefont {A.~F.}\ \bibnamefont {Kockum}},
  \bibinfo {author} {\bibfnamefont {A.}~\bibnamefont {Miranowicz}}, \bibinfo
  {author} {\bibfnamefont {Y.-x.}\ \bibnamefont {Liu}},\ and\ \bibinfo {author}
  {\bibfnamefont {F.}~\bibnamefont {Nori}},\ }\bibfield  {title}
  {{\selectlanguage {en}\bibinfo {title} {Microwave photonics with
  superconducting quantum circuits}},\ }\href
  {https://doi.org/10.1016/j.physrep.2017.10.002} {\bibfield  {journal}
  {\bibinfo  {journal} {Physics Reports}\ }\textbf {\bibinfo {volume}
  {718-719}},\ \bibinfo {pages} {1} (\bibinfo {year} {2017})}\BibitemShut
  {NoStop}%
\bibitem [{\citenamefont {Mirhosseini}\ \emph {et~al.}(2019)\citenamefont
  {Mirhosseini}, \citenamefont {Kim}, \citenamefont {Zhang}, \citenamefont
  {Sipahigil}, \citenamefont {Dieterle}, \citenamefont {Keller}, \citenamefont
  {Asenjo-Garcia}, \citenamefont {Chang},\ and\ \citenamefont
  {Painter}}]{mirhosseini_cavity_2019}%
  \BibitemOpen
  \bibfield  {author} {\bibinfo {author} {\bibfnamefont {M.}~\bibnamefont
  {Mirhosseini}}, \bibinfo {author} {\bibfnamefont {E.}~\bibnamefont {Kim}},
  \bibinfo {author} {\bibfnamefont {X.}~\bibnamefont {Zhang}}, \bibinfo
  {author} {\bibfnamefont {A.}~\bibnamefont {Sipahigil}}, \bibinfo {author}
  {\bibfnamefont {P.~B.}\ \bibnamefont {Dieterle}}, \bibinfo {author}
  {\bibfnamefont {A.~J.}\ \bibnamefont {Keller}}, \bibinfo {author}
  {\bibfnamefont {A.}~\bibnamefont {Asenjo-Garcia}}, \bibinfo {author}
  {\bibfnamefont {D.~E.}\ \bibnamefont {Chang}},\ and\ \bibinfo {author}
  {\bibfnamefont {O.}~\bibnamefont {Painter}},\ }\bibfield  {title}
  {{\selectlanguage {en}\bibinfo {title} {Cavity quantum electrodynamics with
  atom-like mirrors}},\ }\href {https://doi.org/10.1038/s41586-019-1196-1}
  {\bibfield  {journal} {\bibinfo  {journal} {Nature}\ }\textbf {\bibinfo
  {volume} {569}},\ \bibinfo {pages} {692} (\bibinfo {year}
  {2019})}\BibitemShut {NoStop}%
\bibitem [{\citenamefont {Jaynes}\ and\ \citenamefont
  {Cummings}(1963)}]{jaynes_comparison_1963}%
  \BibitemOpen
  \bibfield  {author} {\bibinfo {author} {\bibfnamefont {E.}~\bibnamefont
  {Jaynes}}\ and\ \bibinfo {author} {\bibfnamefont {F.}~\bibnamefont
  {Cummings}},\ }\bibfield  {title} {\bibinfo {title} {Comparison of quantum
  and semiclassical radiation theories with application to the beam maser},\
  }\href {https://doi.org/10.1109/PROC.1963.1664} {\bibfield  {journal}
  {\bibinfo  {journal} {Proceedings of the IEEE}\ }\textbf {\bibinfo {volume}
  {51}},\ \bibinfo {pages} {89} (\bibinfo {year} {1963})}\BibitemShut {NoStop}%
\bibitem [{\citenamefont {Niemczyk}\ \emph {et~al.}(2010)\citenamefont
  {Niemczyk}, \citenamefont {Deppe}, \citenamefont {Huebl}, \citenamefont
  {Menzel}, \citenamefont {Hocke}, \citenamefont {Schwarz}, \citenamefont
  {Garcia-Ripoll}, \citenamefont {Zueco}, \citenamefont {H{\"u}mmer},
  \citenamefont {Solano} \emph {et~al.}}]{niemczyk_circuit_2010}%
  \BibitemOpen
  \bibfield  {author} {\bibinfo {author} {\bibfnamefont {T.}~\bibnamefont
  {Niemczyk}}, \bibinfo {author} {\bibfnamefont {F.}~\bibnamefont {Deppe}},
  \bibinfo {author} {\bibfnamefont {H.}~\bibnamefont {Huebl}}, \bibinfo
  {author} {\bibfnamefont {E.~P.}\ \bibnamefont {Menzel}}, \bibinfo {author}
  {\bibfnamefont {F.}~\bibnamefont {Hocke}}, \bibinfo {author} {\bibfnamefont
  {M.~J.}\ \bibnamefont {Schwarz}}, \bibinfo {author} {\bibfnamefont {J.~J.}\
  \bibnamefont {Garcia-Ripoll}}, \bibinfo {author} {\bibfnamefont
  {D.}~\bibnamefont {Zueco}}, \bibinfo {author} {\bibfnamefont
  {T.}~\bibnamefont {H{\"u}mmer}}, \bibinfo {author} {\bibfnamefont
  {E.}~\bibnamefont {Solano}}, \emph {et~al.},\ }\bibfield  {title} {\bibinfo
  {title} {Circuit quantum electrodynamics in the ultrastrong-coupling
  regime},\ }\href@noop {} {\bibfield  {journal} {\bibinfo  {journal} {Nature
  Physics}\ }\textbf {\bibinfo {volume} {6}},\ \bibinfo {pages} {772} (\bibinfo
  {year} {2010})}\BibitemShut {NoStop}%
\bibitem [{\citenamefont {Settineri}\ \emph
  {et~al.}(2021{\natexlab{a}})\citenamefont {Settineri}, \citenamefont
  {Di~Stefano}, \citenamefont {Zueco}, \citenamefont {Hughes}, \citenamefont
  {Savasta},\ and\ \citenamefont {Nori}}]{Settineri2021Apr}%
  \BibitemOpen
  \bibfield  {author} {\bibinfo {author} {\bibfnamefont {A.}~\bibnamefont
  {Settineri}}, \bibinfo {author} {\bibfnamefont {O.}~\bibnamefont
  {Di~Stefano}}, \bibinfo {author} {\bibfnamefont {D.}~\bibnamefont {Zueco}},
  \bibinfo {author} {\bibfnamefont {S.}~\bibnamefont {Hughes}}, \bibinfo
  {author} {\bibfnamefont {S.}~\bibnamefont {Savasta}},\ and\ \bibinfo {author}
  {\bibfnamefont {F.}~\bibnamefont {Nori}},\ }\bibfield  {title} {\bibinfo
  {title} {{Gauge freedom, quantum measurements, and time-dependent
  interactions in cavity QED}},\ }\href
  {https://doi.org/10.1103/PhysRevResearch.3.023079} {\bibfield  {journal}
  {\bibinfo  {journal} {Phys. Rev. Res.}\ }\textbf {\bibinfo {volume} {3}},\
  \bibinfo {pages} {023079} (\bibinfo {year} {2021}{\natexlab{a}})}\BibitemShut
  {NoStop}%
\bibitem [{\citenamefont {De~Bernardis}\ \emph {et~al.}(2018)\citenamefont
  {De~Bernardis}, \citenamefont {Pilar}, \citenamefont {Jaako}, \citenamefont
  {De~Liberato},\ and\ \citenamefont {Rabl}}]{de_bernardis_breakdown_2018}%
  \BibitemOpen
  \bibfield  {author} {\bibinfo {author} {\bibfnamefont {D.}~\bibnamefont
  {De~Bernardis}}, \bibinfo {author} {\bibfnamefont {P.}~\bibnamefont {Pilar}},
  \bibinfo {author} {\bibfnamefont {T.}~\bibnamefont {Jaako}}, \bibinfo
  {author} {\bibfnamefont {S.}~\bibnamefont {De~Liberato}},\ and\ \bibinfo
  {author} {\bibfnamefont {P.}~\bibnamefont {Rabl}},\ }\bibfield  {title}
  {\bibinfo {title} {Breakdown of gauge invariance in ultrastrong-coupling
  cavity {QED}},\ }\href {https://doi.org/10.1103/PhysRevA.98.053819}
  {\bibfield  {journal} {\bibinfo  {journal} {Physical Review A}\ }\textbf
  {\bibinfo {volume} {98}},\ \bibinfo {pages} {053819} (\bibinfo {year}
  {2018})}\BibitemShut {NoStop}%
\bibitem [{\citenamefont {Stokes}\ and\ \citenamefont
  {Nazir}(2019)}]{adam_stokes_gauge_2019}%
  \BibitemOpen
  \bibfield  {author} {\bibinfo {author} {\bibfnamefont {A.}~\bibnamefont
  {Stokes}}\ and\ \bibinfo {author} {\bibfnamefont {A.}~\bibnamefont {Nazir}},\
  }\bibfield  {title} {\bibinfo {title} {Gauge ambiguities imply
  {Jaynes-Cummings} physics remains valid in ultrastrong coupling qed},\
  }\href@noop {} {\bibfield  {journal} {\bibinfo  {journal} {Nature
  communications}\ }\textbf {\bibinfo {volume} {10}},\ \bibinfo {pages} {499}
  (\bibinfo {year} {2019})}\BibitemShut {NoStop}%
\bibitem [{\citenamefont {Di~Stefano}\ \emph {et~al.}(2019)\citenamefont
  {Di~Stefano}, \citenamefont {Settineri}, \citenamefont {Macr{\`i}},
  \citenamefont {Garziano}, \citenamefont {Stassi}, \citenamefont {Savasta},\
  and\ \citenamefont {Nori}}]{di_stefano_resolution_2019}%
  \BibitemOpen
  \bibfield  {author} {\bibinfo {author} {\bibfnamefont {O.}~\bibnamefont
  {Di~Stefano}}, \bibinfo {author} {\bibfnamefont {A.}~\bibnamefont
  {Settineri}}, \bibinfo {author} {\bibfnamefont {V.}~\bibnamefont
  {Macr{\`i}}}, \bibinfo {author} {\bibfnamefont {L.}~\bibnamefont {Garziano}},
  \bibinfo {author} {\bibfnamefont {R.}~\bibnamefont {Stassi}}, \bibinfo
  {author} {\bibfnamefont {S.}~\bibnamefont {Savasta}},\ and\ \bibinfo {author}
  {\bibfnamefont {F.}~\bibnamefont {Nori}},\ }\bibfield  {title} {\bibinfo
  {title} {Resolution of gauge ambiguities in ultrastrong-coupling cavity
  quantum electrodynamics},\ }\href {https://doi.org/10.1038/s41567-019-0534-4}
  {\bibfield  {journal} {\bibinfo  {journal} {Nature Physics}\ }\textbf
  {\bibinfo {volume} {15}},\ \bibinfo {pages} {803} (\bibinfo {year}
  {2019})}\BibitemShut {NoStop}%
\bibitem [{\citenamefont {Stokes}\ and\ \citenamefont
  {Nazir}(2020)}]{stokes_gauge_2020}%
  \BibitemOpen
  \bibfield  {author} {\bibinfo {author} {\bibfnamefont {A.}~\bibnamefont
  {Stokes}}\ and\ \bibinfo {author} {\bibfnamefont {A.}~\bibnamefont {Nazir}},\
  }\bibfield  {title} {\bibinfo {title} {Gauge non-invariance due to material
  truncation in ultrastrong-coupling {QED}},\ }\href
  {http://arxiv.org/abs/2005.06499} {\bibfield  {journal} {\bibinfo  {journal}
  {arXiv:2005.06499}\ } (\bibinfo {year} {2020})}\BibitemShut {NoStop}%
\bibitem [{\citenamefont {Rouse}\ \emph {et~al.}(2021)\citenamefont {Rouse},
  \citenamefont {Lovett}, \citenamefont {Gauger},\ and\ \citenamefont
  {Westerberg}}]{Rouse2021Feb}%
  \BibitemOpen
  \bibfield  {author} {\bibinfo {author} {\bibfnamefont {D.~M.}\ \bibnamefont
  {Rouse}}, \bibinfo {author} {\bibfnamefont {B.~W.}\ \bibnamefont {Lovett}},
  \bibinfo {author} {\bibfnamefont {E.~M.}\ \bibnamefont {Gauger}},\ and\
  \bibinfo {author} {\bibfnamefont {N.}~\bibnamefont {Westerberg}},\ }\bibfield
   {title} {\bibinfo {title} {{Avoiding gauge ambiguities in cavity quantum
  electrodynamics - Scientific Reports}},\ }\href
  {https://doi.org/10.1038/s41598-021-83214-z} {\bibfield  {journal} {\bibinfo
  {journal} {Sci. Rep.}\ }\textbf {\bibinfo {volume} {11}},\ \bibinfo {pages}
  {1} (\bibinfo {year} {2021})}\BibitemShut {NoStop}%
\bibitem [{\citenamefont {Lamb}\ \emph {et~al.}(1987)\citenamefont {Lamb},
  \citenamefont {Schlicher},\ and\ \citenamefont {Scully}}]{PhysRevA.36.2763}%
  \BibitemOpen
  \bibfield  {author} {\bibinfo {author} {\bibfnamefont {W.~E.}\ \bibnamefont
  {Lamb}}, \bibinfo {author} {\bibfnamefont {R.~R.}\ \bibnamefont
  {Schlicher}},\ and\ \bibinfo {author} {\bibfnamefont {M.~O.}\ \bibnamefont
  {Scully}},\ }\bibfield  {title} {\bibinfo {title} {Matter-field interaction
  in atomic physics and quantum optics},\ }\href
  {https://doi.org/10.1103/PhysRevA.36.2763} {\bibfield  {journal} {\bibinfo
  {journal} {Phys. Rev. A}\ }\textbf {\bibinfo {volume} {36}},\ \bibinfo
  {pages} {2763} (\bibinfo {year} {1987})}\BibitemShut {NoStop}%
\bibitem [{\citenamefont {Savasta}\ \emph {et~al.}(2020)\citenamefont
  {Savasta}, \citenamefont {Stefano},\ and\ \citenamefont {Nori}}]{2002.02139}%
  \BibitemOpen
  \bibfield  {author} {\bibinfo {author} {\bibfnamefont {S.}~\bibnamefont
  {Savasta}}, \bibinfo {author} {\bibfnamefont {O.~D.}\ \bibnamefont
  {Stefano}},\ and\ \bibinfo {author} {\bibfnamefont {F.}~\bibnamefont
  {Nori}},\ }\bibfield  {title} {\bibinfo {title}
  {{Thomas{\textendash}Reiche{\textendash}Kuhn} ({TRK}) sum rule for
  interacting photons},\ }\href {https://doi.org/10.1515/nanoph-2020-0433}
  {\bibfield  {journal} {\bibinfo  {journal} {Nanophotonics}\ }\textbf
  {\bibinfo {volume} {10}},\ \bibinfo {pages} {465} (\bibinfo {year}
  {2020})}\BibitemShut {NoStop}%
\bibitem [{\citenamefont {Carmichael}(2013)}]{carmichael_statistical_2013}%
  \BibitemOpen
  \bibfield  {author} {\bibinfo {author} {\bibfnamefont {H.~J.}\ \bibnamefont
  {Carmichael}},\ }\href@noop {} {\emph {\bibinfo {title} {Statistical
  {Methods} in {Quantum} {Optics} 1: {Master} {Equations} and {Fokker}-{Planck}
  {Equations}}}}\ (\bibinfo  {publisher} {Springer Science \& Business Media},\
  \bibinfo {year} {2013})\BibitemShut {NoStop}%
\bibitem [{\citenamefont {Settineri}\ \emph {et~al.}(2018)\citenamefont
  {Settineri}, \citenamefont {Macr{\'i}}, \citenamefont {Ridolfo},
  \citenamefont {Di~Stefano}, \citenamefont {Kockum}, \citenamefont {Nori},\
  and\ \citenamefont {Savasta}}]{settineri_dissipation_2018}%
  \BibitemOpen
  \bibfield  {author} {\bibinfo {author} {\bibfnamefont {A.}~\bibnamefont
  {Settineri}}, \bibinfo {author} {\bibfnamefont {V.}~\bibnamefont
  {Macr{\'i}}}, \bibinfo {author} {\bibfnamefont {A.}~\bibnamefont {Ridolfo}},
  \bibinfo {author} {\bibfnamefont {O.}~\bibnamefont {Di~Stefano}}, \bibinfo
  {author} {\bibfnamefont {A.~F.}\ \bibnamefont {Kockum}}, \bibinfo {author}
  {\bibfnamefont {F.}~\bibnamefont {Nori}},\ and\ \bibinfo {author}
  {\bibfnamefont {S.}~\bibnamefont {Savasta}},\ }\bibfield  {title} {\bibinfo
  {title} {Dissipation and thermal noise in hybrid quantum systems in the
  ultrastrong-coupling regime},\ }\href
  {https://doi.org/10.1103/PhysRevA.98.053834} {\bibfield  {journal} {\bibinfo
  {journal} {Physical Review A}\ }\textbf {\bibinfo {volume} {98}},\ \bibinfo
  {pages} {053834} (\bibinfo {year} {2018})}\BibitemShut {NoStop}%
\bibitem [{\citenamefont {Zueco}\ and\ \citenamefont
  {Garc{\'i}a-Ripoll}(2019)}]{zueco2019ultrastrongly}%
  \BibitemOpen
  \bibfield  {author} {\bibinfo {author} {\bibfnamefont {D.}~\bibnamefont
  {Zueco}}\ and\ \bibinfo {author} {\bibfnamefont {J.}~\bibnamefont
  {Garc{\'i}a-Ripoll}},\ }\bibfield  {title} {\bibinfo {title} {Ultrastrongly
  dissipative quantum {Rabi} model},\ }\href@noop {} {\bibfield  {journal}
  {\bibinfo  {journal} {Physical Review A}\ }\textbf {\bibinfo {volume} {99}},\
  \bibinfo {pages} {013807} (\bibinfo {year} {2019})}\BibitemShut {NoStop}%
\bibitem [{\citenamefont {Cao}\ \emph {et~al.}(2010)\citenamefont {Cao},
  \citenamefont {You}, \citenamefont {Zheng}, \citenamefont {Kofman},\ and\
  \citenamefont {Nori}}]{cao2010dynamics}%
  \BibitemOpen
  \bibfield  {author} {\bibinfo {author} {\bibfnamefont {X.}~\bibnamefont
  {Cao}}, \bibinfo {author} {\bibfnamefont {J.~Q.}\ \bibnamefont {You}},
  \bibinfo {author} {\bibfnamefont {H.}~\bibnamefont {Zheng}}, \bibinfo
  {author} {\bibfnamefont {A.}~\bibnamefont {Kofman}},\ and\ \bibinfo {author}
  {\bibfnamefont {F.}~\bibnamefont {Nori}},\ }\bibfield  {title} {\bibinfo
  {title} {Dynamics and quantum {Z}eno effect for a qubit in either a low-or
  high-frequency bath beyond the rotating-wave approximation},\ }\href@noop {}
  {\bibfield  {journal} {\bibinfo  {journal} {Physical Review A}\ }\textbf
  {\bibinfo {volume} {82}},\ \bibinfo {pages} {022119} (\bibinfo {year}
  {2010})}\BibitemShut {NoStop}%
\bibitem [{\citenamefont {Le~Boit\'e}\ \emph {et~al.}(2016)\citenamefont
  {Le~Boit\'e}, \citenamefont {Hwang}, \citenamefont {Nha},\ and\ \citenamefont
  {Plenio}}]{PhysRevA.94.033827}%
  \BibitemOpen
  \bibfield  {author} {\bibinfo {author} {\bibfnamefont {A.}~\bibnamefont
  {Le~Boit\'e}}, \bibinfo {author} {\bibfnamefont {M.-J.}\ \bibnamefont
  {Hwang}}, \bibinfo {author} {\bibfnamefont {H.}~\bibnamefont {Nha}},\ and\
  \bibinfo {author} {\bibfnamefont {M.~B.}\ \bibnamefont {Plenio}},\ }\bibfield
   {title} {\bibinfo {title} {Fate of photon blockade in the deep
  strong-coupling regime},\ }\href {https://doi.org/10.1103/PhysRevA.94.033827}
  {\bibfield  {journal} {\bibinfo  {journal} {Phys. Rev. A}\ }\textbf {\bibinfo
  {volume} {94}},\ \bibinfo {pages} {033827} (\bibinfo {year}
  {2016})}\BibitemShut {NoStop}%
\bibitem [{\citenamefont {Johansson}\ \emph {et~al.}(2012)\citenamefont
  {Johansson}, \citenamefont {Nation},\ and\ \citenamefont
  {Nori}}]{johansson2012qutip}%
  \BibitemOpen
  \bibfield  {author} {\bibinfo {author} {\bibfnamefont {J.~R.}\ \bibnamefont
  {Johansson}}, \bibinfo {author} {\bibfnamefont {P.~D.}\ \bibnamefont
  {Nation}},\ and\ \bibinfo {author} {\bibfnamefont {F.}~\bibnamefont {Nori}},\
  }\bibfield  {title} {\bibinfo {title} {Qutip: An open-source python framework
  for the dynamics of open quantum systems},\ }\href@noop {} {\bibfield
  {journal} {\bibinfo  {journal} {Computer Physics Communications}\ }\textbf
  {\bibinfo {volume} {183}},\ \bibinfo {pages} {1760} (\bibinfo {year}
  {2012})}\BibitemShut {NoStop}%
\bibitem [{\citenamefont {Johansson}\ \emph {et~al.}(2013)\citenamefont
  {Johansson}, \citenamefont {Nation},\ and\ \citenamefont
  {Nori}}]{johansson_qutip_2013}%
  \BibitemOpen
  \bibfield  {author} {\bibinfo {author} {\bibfnamefont {J.~R.}\ \bibnamefont
  {Johansson}}, \bibinfo {author} {\bibfnamefont {P.~D.}\ \bibnamefont
  {Nation}},\ and\ \bibinfo {author} {\bibfnamefont {F.}~\bibnamefont {Nori}},\
  }\bibfield  {title} {{\selectlanguage {en}\bibinfo {title} {{QuTiP} 2: {A}
  {Python} framework for the dynamics of open quantum systems}},\ }\href
  {https://doi.org/10.1016/j.cpc.2012.11.019} {\bibfield  {journal} {\bibinfo
  {journal} {Computer Physics Communications}\ }\textbf {\bibinfo {volume}
  {184}},\ \bibinfo {pages} {1234} (\bibinfo {year} {2013})}\BibitemShut
  {NoStop}%
\bibitem [{\citenamefont {Cao}\ \emph {et~al.}(2011)\citenamefont {Cao},
  \citenamefont {You}, \citenamefont {Zheng},\ and\ \citenamefont
  {Nori}}]{cao2011qubit}%
  \BibitemOpen
  \bibfield  {author} {\bibinfo {author} {\bibfnamefont {X.}~\bibnamefont
  {Cao}}, \bibinfo {author} {\bibfnamefont {J.~Q.}\ \bibnamefont {You}},
  \bibinfo {author} {\bibfnamefont {H.}~\bibnamefont {Zheng}},\ and\ \bibinfo
  {author} {\bibfnamefont {F.}~\bibnamefont {Nori}},\ }\bibfield  {title}
  {\bibinfo {title} {A qubit strongly coupled to a resonant cavity: asymmetry
  of the spontaneous emission spectrum beyond the rotating wave
  approximation},\ }\href@noop {} {\bibfield  {journal} {\bibinfo  {journal}
  {New Journal of Physics}\ }\textbf {\bibinfo {volume} {13}},\ \bibinfo
  {pages} {073002} (\bibinfo {year} {2011})}\BibitemShut {NoStop}%
\bibitem [{\citenamefont {Bamba}\ and\ \citenamefont
  {Ogawa}(2014)}]{Bamba2014Feb}%
  \BibitemOpen
  \bibfield  {author} {\bibinfo {author} {\bibfnamefont {M.}~\bibnamefont
  {Bamba}}\ and\ \bibinfo {author} {\bibfnamefont {T.}~\bibnamefont {Ogawa}},\
  }\bibfield  {title} {\bibinfo {title} {{Recipe for the Hamiltonian of
  system-environment coupling applicable to the
  ultrastrong-light-matter-interaction regime}},\ }\href
  {https://doi.org/10.1103/PhysRevA.89.023817} {\bibfield  {journal} {\bibinfo
  {journal} {Phys. Rev. A}\ }\textbf {\bibinfo {volume} {89}},\ \bibinfo
  {pages} {023817} (\bibinfo {year} {2014})}\BibitemShut {NoStop}%
\bibitem [{\citenamefont {Lentrodt}\ and\ \citenamefont
  {Evers}(2020)}]{Lentrodt2020Jan}%
  \BibitemOpen
  \bibfield  {author} {\bibinfo {author} {\bibfnamefont {D.}~\bibnamefont
  {Lentrodt}}\ and\ \bibinfo {author} {\bibfnamefont {J.}~\bibnamefont
  {Evers}},\ }\bibfield  {title} {\bibinfo {title} {{Ab Initio Few-Mode Theory
  for Quantum Potential Scattering Problems}},\ }\href
  {https://doi.org/10.1103/PhysRevX.10.011008} {\bibfield  {journal} {\bibinfo
  {journal} {Phys. Rev. X}\ }\textbf {\bibinfo {volume} {10}},\ \bibinfo
  {pages} {011008} (\bibinfo {year} {2020})}\BibitemShut {NoStop}%
\bibitem [{\citenamefont {Franke}\ \emph {et~al.}(2019)\citenamefont {Franke},
  \citenamefont {Hughes}, \citenamefont {Dezfouli}, \citenamefont {Kristensen},
  \citenamefont {Busch}, \citenamefont {Knorr},\ and\ \citenamefont
  {Richter}}]{PhysRevLett.122.213901}%
  \BibitemOpen
  \bibfield  {author} {\bibinfo {author} {\bibfnamefont {S.}~\bibnamefont
  {Franke}}, \bibinfo {author} {\bibfnamefont {S.}~\bibnamefont {Hughes}},
  \bibinfo {author} {\bibfnamefont {M.~K.}\ \bibnamefont {Dezfouli}}, \bibinfo
  {author} {\bibfnamefont {P.~T.}\ \bibnamefont {Kristensen}}, \bibinfo
  {author} {\bibfnamefont {K.}~\bibnamefont {Busch}}, \bibinfo {author}
  {\bibfnamefont {A.}~\bibnamefont {Knorr}},\ and\ \bibinfo {author}
  {\bibfnamefont {M.}~\bibnamefont {Richter}},\ }\bibfield  {title} {\bibinfo
  {title} {Quantization of quasinormal modes for open cavities and plasmonic
  cavity quantum electrodynamics},\ }\href
  {https://doi.org/10.1103/PhysRevLett.122.213901} {\bibfield  {journal}
  {\bibinfo  {journal} {Phys. Rev. Lett.}\ }\textbf {\bibinfo {volume} {122}},\
  \bibinfo {pages} {213901} (\bibinfo {year} {2019})}\BibitemShut {NoStop}%
\bibitem [{\citenamefont {Hughes}\ \emph {et~al.}(2019)\citenamefont {Hughes},
  \citenamefont {Franke}, \citenamefont {Gustin}, \citenamefont {Dezfouli},
  \citenamefont {Knorr},\ and\ \citenamefont {Richter}}]{2019_ACS}%
  \BibitemOpen
  \bibfield  {author} {\bibinfo {author} {\bibfnamefont {S.}~\bibnamefont
  {Hughes}}, \bibinfo {author} {\bibfnamefont {S.}~\bibnamefont {Franke}},
  \bibinfo {author} {\bibfnamefont {C.}~\bibnamefont {Gustin}}, \bibinfo
  {author} {\bibfnamefont {M.~K.}\ \bibnamefont {Dezfouli}}, \bibinfo {author}
  {\bibfnamefont {A.}~\bibnamefont {Knorr}},\ and\ \bibinfo {author}
  {\bibfnamefont {M.}~\bibnamefont {Richter}},\ }\bibfield  {title} {\bibinfo
  {title} {Theory and limits of on-demand single-photon sources using plasmonic
  resonators: A quantized quasinormal mode approach},\ }\href
  {https://doi.org/10.1021/acsphotonics.9b00849} {\bibfield  {journal}
  {\bibinfo  {journal} {ACS Photonics}\ }\textbf {\bibinfo {volume} {6}},\
  \bibinfo {pages} {2168} (\bibinfo {year} {2019})}\BibitemShut {NoStop}%
\bibitem [{\citenamefont {Franke}\ \emph
  {et~al.}(2020{\natexlab{a}})\citenamefont {Franke}, \citenamefont {Ren},
  \citenamefont {Hughes},\ and\ \citenamefont
  {Richter}}]{PhysRevResearch.2.033332}%
  \BibitemOpen
  \bibfield  {author} {\bibinfo {author} {\bibfnamefont {S.}~\bibnamefont
  {Franke}}, \bibinfo {author} {\bibfnamefont {J.}~\bibnamefont {Ren}},
  \bibinfo {author} {\bibfnamefont {S.}~\bibnamefont {Hughes}},\ and\ \bibinfo
  {author} {\bibfnamefont {M.}~\bibnamefont {Richter}},\ }\bibfield  {title}
  {\bibinfo {title} {Fluctuation-dissipation theorem and fundamental photon
  commutation relations in lossy nanostructures using quasinormal modes},\
  }\href {https://doi.org/10.1103/PhysRevResearch.2.033332} {\bibfield
  {journal} {\bibinfo  {journal} {Phys. Rev. Research}\ }\textbf {\bibinfo
  {volume} {2}},\ \bibinfo {pages} {033332} (\bibinfo {year}
  {2020}{\natexlab{a}})}\BibitemShut {NoStop}%
\bibitem [{\citenamefont {Franke}\ \emph
  {et~al.}(2020{\natexlab{b}})\citenamefont {Franke}, \citenamefont {Richter},
  \citenamefont {Ren}, \citenamefont {Knorr},\ and\ \citenamefont
  {Hughes}}]{PhysRevResearch.2.033456}%
  \BibitemOpen
  \bibfield  {author} {\bibinfo {author} {\bibfnamefont {S.}~\bibnamefont
  {Franke}}, \bibinfo {author} {\bibfnamefont {M.}~\bibnamefont {Richter}},
  \bibinfo {author} {\bibfnamefont {J.}~\bibnamefont {Ren}}, \bibinfo {author}
  {\bibfnamefont {A.}~\bibnamefont {Knorr}},\ and\ \bibinfo {author}
  {\bibfnamefont {S.}~\bibnamefont {Hughes}},\ }\bibfield  {title} {\bibinfo
  {title} {Quantized quasinormal-mode description of nonlinear {cavity-QED}
  effects from coupled resonators with a {Fano-like} resonance},\ }\href
  {https://doi.org/10.1103/PhysRevResearch.2.033456} {\bibfield  {journal}
  {\bibinfo  {journal} {Phys. Rev. Research}\ }\textbf {\bibinfo {volume}
  {2}},\ \bibinfo {pages} {033456} (\bibinfo {year}
  {2020}{\natexlab{b}})}\BibitemShut {NoStop}%
\bibitem [{\citenamefont {Ren}\ \emph {et~al.}(2022)\citenamefont {Ren},
  \citenamefont {Franke},\ and\ \citenamefont {Hughes}}]{Ren2022}%
  \BibitemOpen
  \bibfield  {author} {\bibinfo {author} {\bibfnamefont {J.}~\bibnamefont
  {Ren}}, \bibinfo {author} {\bibfnamefont {S.}~\bibnamefont {Franke}},\ and\
  \bibinfo {author} {\bibfnamefont {S.}~\bibnamefont {Hughes}},\ }\bibfield
  {title} {\bibinfo {title} {Connecting classical and quantum mode theories for
  coupled lossy cavity resonators using quasinormal modes},\ }\href
  {https://doi.org/10.1021/acsphotonics.1c01274} {\bibfield  {journal}
  {\bibinfo  {journal} {{ACS} Photonics}\ }\textbf {\bibinfo {volume} {9}},\
  \bibinfo {pages} {138} (\bibinfo {year} {2022})}\BibitemShut {NoStop}%
\bibitem [{\citenamefont {Ren}\ \emph {et~al.}(2021)\citenamefont {Ren},
  \citenamefont {Franke},\ and\ \citenamefont {Hughes}}]{PhysRevX.11.041020}%
  \BibitemOpen
  \bibfield  {author} {\bibinfo {author} {\bibfnamefont {J.}~\bibnamefont
  {Ren}}, \bibinfo {author} {\bibfnamefont {S.}~\bibnamefont {Franke}},\ and\
  \bibinfo {author} {\bibfnamefont {S.}~\bibnamefont {Hughes}},\ }\bibfield
  {title} {\bibinfo {title} {Quasinormal modes, local density of states, and
  classical {Purcell Factors} for coupled loss-gain resonators},\ }\href
  {https://doi.org/10.1103/PhysRevX.11.041020} {\bibfield  {journal} {\bibinfo
  {journal} {Phys. Rev. X}\ }\textbf {\bibinfo {volume} {11}},\ \bibinfo
  {pages} {041020} (\bibinfo {year} {2021})}\BibitemShut {NoStop}%
\bibitem [{\citenamefont {Franke}\ \emph {et~al.}(2022)\citenamefont {Franke},
  \citenamefont {Ren},\ and\ \citenamefont {Hughes}}]{PhysRevA.105.023702}%
  \BibitemOpen
  \bibfield  {author} {\bibinfo {author} {\bibfnamefont {S.}~\bibnamefont
  {Franke}}, \bibinfo {author} {\bibfnamefont {J.}~\bibnamefont {Ren}},\ and\
  \bibinfo {author} {\bibfnamefont {S.}~\bibnamefont {Hughes}},\ }\bibfield
  {title} {\bibinfo {title} {Quantized quasinormal-mode theory of coupled lossy
  and amplifying resonators},\ }\href
  {https://doi.org/10.1103/PhysRevA.105.023702} {\bibfield  {journal} {\bibinfo
   {journal} {Phys. Rev. A}\ }\textbf {\bibinfo {volume} {105}},\ \bibinfo
  {pages} {023702} (\bibinfo {year} {2022})}\BibitemShut {NoStop}%
\bibitem [{\citenamefont {Savasta}\ \emph {et~al.}(2021)\citenamefont
  {Savasta}, \citenamefont {Di~Stefano}, \citenamefont {Settineri},
  \citenamefont {Zueco}, \citenamefont {Hughes},\ and\ \citenamefont
  {Nori}}]{savasta_gauge_2020}%
  \BibitemOpen
  \bibfield  {author} {\bibinfo {author} {\bibfnamefont {S.}~\bibnamefont
  {Savasta}}, \bibinfo {author} {\bibfnamefont {O.}~\bibnamefont {Di~Stefano}},
  \bibinfo {author} {\bibfnamefont {A.}~\bibnamefont {Settineri}}, \bibinfo
  {author} {\bibfnamefont {D.}~\bibnamefont {Zueco}}, \bibinfo {author}
  {\bibfnamefont {S.}~\bibnamefont {Hughes}},\ and\ \bibinfo {author}
  {\bibfnamefont {F.}~\bibnamefont {Nori}},\ }\bibfield  {title} {\bibinfo
  {title} {Gauge principle and gauge invariance in two-level systems},\ }\href
  {https://doi.org/10.1103/PhysRevA.103.053703} {\bibfield  {journal} {\bibinfo
   {journal} {Phys. Rev. A}\ }\textbf {\bibinfo {volume} {103}},\ \bibinfo
  {pages} {053703} (\bibinfo {year} {2021})}\BibitemShut {NoStop}%
\bibitem [{\citenamefont {Settineri}\ \emph
  {et~al.}(2021{\natexlab{b}})\citenamefont {Settineri}, \citenamefont
  {Di~Stefano}, \citenamefont {Zueco}, \citenamefont {Hughes}, \citenamefont
  {Savasta},\ and\ \citenamefont {Nori}}]{settineri_gauge_2019}%
  \BibitemOpen
  \bibfield  {author} {\bibinfo {author} {\bibfnamefont {A.}~\bibnamefont
  {Settineri}}, \bibinfo {author} {\bibfnamefont {O.}~\bibnamefont
  {Di~Stefano}}, \bibinfo {author} {\bibfnamefont {D.}~\bibnamefont {Zueco}},
  \bibinfo {author} {\bibfnamefont {S.}~\bibnamefont {Hughes}}, \bibinfo
  {author} {\bibfnamefont {S.}~\bibnamefont {Savasta}},\ and\ \bibinfo {author}
  {\bibfnamefont {F.}~\bibnamefont {Nori}},\ }\bibfield  {title} {\bibinfo
  {title} {Gauge freedom, quantum measurements, and time-dependent interactions
  in {cavity QED}},\ }\href {https://doi.org/10.1103/PhysRevResearch.3.023079}
  {\bibfield  {journal} {\bibinfo  {journal} {Phys. Rev. Research}\ }\textbf
  {\bibinfo {volume} {3}},\ \bibinfo {pages} {023079} (\bibinfo {year}
  {2021}{\natexlab{b}})}\BibitemShut {NoStop}%
\bibitem [{\citenamefont {Wubs}\ \emph {et~al.}(2004)\citenamefont {Wubs},
  \citenamefont {Suttorp},\ and\ \citenamefont
  {Lagendijk}}]{PhysRevA.70.053823}%
  \BibitemOpen
  \bibfield  {author} {\bibinfo {author} {\bibfnamefont {M.}~\bibnamefont
  {Wubs}}, \bibinfo {author} {\bibfnamefont {L.~G.}\ \bibnamefont {Suttorp}},\
  and\ \bibinfo {author} {\bibfnamefont {A.}~\bibnamefont {Lagendijk}},\
  }\bibfield  {title} {\bibinfo {title} {Multiple-scattering approach to
  interatomic interactions and superradiance in inhomogeneous dielectrics},\
  }\href {https://doi.org/10.1103/PhysRevA.70.053823} {\bibfield  {journal}
  {\bibinfo  {journal} {Phys. Rev. A}\ }\textbf {\bibinfo {volume} {70}},\
  \bibinfo {pages} {053823} (\bibinfo {year} {2004})}\BibitemShut {NoStop}%
\bibitem [{\citenamefont {Yao}\ \emph {et~al.}(2009)\citenamefont {Yao},
  \citenamefont {Van~Vlack}, \citenamefont {Reza}, \citenamefont {Patterson},
  \citenamefont {Dignam},\ and\ \citenamefont {Hughes}}]{PhysRevB.80.195106}%
  \BibitemOpen
  \bibfield  {author} {\bibinfo {author} {\bibfnamefont {P.}~\bibnamefont
  {Yao}}, \bibinfo {author} {\bibfnamefont {C.}~\bibnamefont {Van~Vlack}},
  \bibinfo {author} {\bibfnamefont {A.}~\bibnamefont {Reza}}, \bibinfo {author}
  {\bibfnamefont {M.}~\bibnamefont {Patterson}}, \bibinfo {author}
  {\bibfnamefont {M.~M.}\ \bibnamefont {Dignam}},\ and\ \bibinfo {author}
  {\bibfnamefont {S.}~\bibnamefont {Hughes}},\ }\bibfield  {title} {\bibinfo
  {title} {Ultrahigh {Purcell} factors and {Lamb} shifts in slow-light
  metamaterial waveguides},\ }\href
  {https://doi.org/10.1103/PhysRevB.80.195106} {\bibfield  {journal} {\bibinfo
  {journal} {Phys. Rev. B}\ }\textbf {\bibinfo {volume} {80}},\ \bibinfo
  {pages} {195106} (\bibinfo {year} {2009})}\BibitemShut {NoStop}%
\bibitem [{\citenamefont {Gardiner}\ and\ \citenamefont
  {Collett}(1985)}]{gardiner_1985}%
  \BibitemOpen
  \bibfield  {author} {\bibinfo {author} {\bibfnamefont {C.~W.}\ \bibnamefont
  {Gardiner}}\ and\ \bibinfo {author} {\bibfnamefont {M.~J.}\ \bibnamefont
  {Collett}},\ }\bibfield  {title} {\bibinfo {title} {Input and output in
  damped quantum systems: Quantum stochastic differential equations and the
  master equation},\ }\href {https://doi.org/10.1103/PhysRevA.31.3761}
  {\bibfield  {journal} {\bibinfo  {journal} {Phys. Rev. A}\ }\textbf {\bibinfo
  {volume} {31}},\ \bibinfo {pages} {3761} (\bibinfo {year}
  {1985})}\BibitemShut {NoStop}%
\bibitem [{\citenamefont {Tian}\ and\ \citenamefont
  {Carmichael}(1992)}]{Tian1992}%
  \BibitemOpen
  \bibfield  {author} {\bibinfo {author} {\bibfnamefont {L.}~\bibnamefont
  {Tian}}\ and\ \bibinfo {author} {\bibfnamefont {H.~J.}\ \bibnamefont
  {Carmichael}},\ }\bibfield  {title} {\bibinfo {title} {Incoherent excitation
  of the {Jaynes-Cummings} system},\ }\href
  {https://doi.org/10.1088/0954-8998/4/2/007} {\bibfield  {journal} {\bibinfo
  {journal} {Quantum Optics: Journal of the European Optical Society Part B}\
  }\textbf {\bibinfo {volume} {4}},\ \bibinfo {pages} {131} (\bibinfo {year}
  {1992})}\BibitemShut {NoStop}%
\bibitem [{\citenamefont {Yao}\ \emph {et~al.}(2010)\citenamefont {Yao},
  \citenamefont {Pathak}, \citenamefont {Illes}, \citenamefont {Hughes},
  \citenamefont {M\"unch}, \citenamefont {Reitzenstein}, \citenamefont
  {Franeck}, \citenamefont {L\"offler}, \citenamefont {Heindel}, \citenamefont
  {H\"ofling}, \citenamefont {Worschech},\ and\ \citenamefont
  {Forchel}}]{PhysRevB.81.033309}%
  \BibitemOpen
  \bibfield  {author} {\bibinfo {author} {\bibfnamefont {P.}~\bibnamefont
  {Yao}}, \bibinfo {author} {\bibfnamefont {P.~K.}\ \bibnamefont {Pathak}},
  \bibinfo {author} {\bibfnamefont {E.}~\bibnamefont {Illes}}, \bibinfo
  {author} {\bibfnamefont {S.}~\bibnamefont {Hughes}}, \bibinfo {author}
  {\bibfnamefont {S.}~\bibnamefont {M\"unch}}, \bibinfo {author} {\bibfnamefont
  {S.}~\bibnamefont {Reitzenstein}}, \bibinfo {author} {\bibfnamefont
  {P.}~\bibnamefont {Franeck}}, \bibinfo {author} {\bibfnamefont
  {A.}~\bibnamefont {L\"offler}}, \bibinfo {author} {\bibfnamefont
  {T.}~\bibnamefont {Heindel}}, \bibinfo {author} {\bibfnamefont
  {S.}~\bibnamefont {H\"ofling}}, \bibinfo {author} {\bibfnamefont
  {L.}~\bibnamefont {Worschech}},\ and\ \bibinfo {author} {\bibfnamefont
  {A.}~\bibnamefont {Forchel}},\ }\bibfield  {title} {\bibinfo {title}
  {Nonlinear photoluminescence spectra from a quantum-dot--cavity system:
  Interplay of pump-induced stimulated emission and anharmonic cavity {QED}},\
  }\href {https://doi.org/10.1103/PhysRevB.81.033309} {\bibfield  {journal}
  {\bibinfo  {journal} {Phys. Rev. B}\ }\textbf {\bibinfo {volume} {81}},\
  \bibinfo {pages} {033309} (\bibinfo {year} {2010})}\BibitemShut {NoStop}%
\bibitem [{\citenamefont
  {Le~Boit{\ifmmode\acute{e}\else\'{e}\fi}}(2020)}]{LeBoite2020Jul}%
  \BibitemOpen
  \bibfield  {author} {\bibinfo {author} {\bibfnamefont {A.}~\bibnamefont
  {Le~Boit{\ifmmode\acute{e}\else\'{e}\fi}}},\ }\bibfield  {title} {\bibinfo
  {title} {{Theoretical Methods for Ultrastrong Light{\textendash}Matter
  Interactions}},\ }\href {https://doi.org/10.1002/qute.201900140} {\bibfield
  {journal} {\bibinfo  {journal} {Adv. Quantum Technol.}\ }\textbf {\bibinfo
  {volume} {3}},\ \bibinfo {pages} {1900140} (\bibinfo {year}
  {2020})}\BibitemShut {NoStop}%
\end{thebibliography}
\end{document}